\documentclass[aps,pre,preprint,groupedaddress,amsmath,amssymb,floatfix,showpacs]{revtex4}
\usepackage{graphicx, verbatim, subfigure}
\usepackage{bm}
\begin{document}
\title{Bulk phase behaviour of binary hard platelet mixtures from density functional theory}
\author{Jonathan Phillips$^{1}$} 
\email{jon.phillips@bristol.ac.uk}
\author{Matthias Schmidt$^{1,2}$}
\begin{abstract}
We investigate isotropic-isotropic, isotropic-nematic and nematic-nematic phase coexistence in binary mixtures of circular platelets with vanishing thickness, continuous rotational degrees of freedom and radial size ratios $\lambda$ up to 5. A fundamental measure density functional theory, previously used for the one-component model, is proposed and results are compared against those from Onsager theory as a benchmark. For $\lambda \leq 1.7$ the system displays isotropic-nematic phase coexistence with a widening of the biphasic region for increasing values of $\lambda$. For size ratios $\lambda \geq 2$, we find demixing into two nematic states becomes stable and an isotropic-nematic-nematic triple point can occur. Fundamental measure theory gives a smaller isotropic-nematic biphasic region than Onsager theory and locates the transition at lower densities. Furthermore, nematic-nematic demixing occurs over a larger range of compositions at a given value of $\lambda$ than found in Onsager theory. Both theories predict the same topologies of the phase diagrams. The partial nematic order parameters vary strongly with composition and indicate that the larger particles are more strongly ordered than the smaller particles.
\end{abstract}
\affiliation{
$^{1}$H.H. Wills Physics Laboratory, Bristol University, Royal Fort, Tyndall Avenue, Bristol BS8 1TL, United Kingdom \\
$^{2}$Theoretische Physik II, Universit\"at Bayreuth, Universit\"atsstra\ss e 30, D-95440 Bayreuth, Germany
}
\date{January 29, 2009}
\maketitle
\section{Introduction}
There is a wide range of colloidal particles with platelet-like shape, including materials such as gibbsite \cite{vanderkooij.f.m:2000.a} and certain clays including montmorillonite, laponite and hydrotalcite \cite{dijkstra.m:1995.a,dijkstra.m:1997.a, pizzey.c:2004.a, vanduijneveldt.js:2005.a, leach.esh:2005.a,mourad.m.c.d.:2008.b}. Clays are some of the most abundant minerals on the Earth's surface and are used as pharmaceuticals, cosmetics and catalysts. There is much current interest in the use of platelets in nanocomposite materials, e.g. the nematic phase of sterically stabilised gibbsite platelets may be used as a template for gibbsite-polymer nanocomposites with nematic order \cite{maurice_thesis}. Interest in platelet dispersions is also present in geophysics \cite{maitland.gc:2000.a}, biomedicine \cite{mason.tg:2002.a} and liquid crystal display (LCD) technology \cite{majumdar.a:2007.a}.

Understanding the liquid crystalline phase behaviour of systems of non-spherical particles \cite{chandrasekhar:1990.a, kumar:2004.a, bushby:2002.a} is an important topic in modern condensed matter physics. One of the most celebrated cases of a phase transition in such systems is the isotropic-nematic (\textit{I}-\textit{N}) transition. For athermal model systems, where the particle interactions are hard core, phase transitions arise purely from entropic contributions to the free energy and the phase behaviour is governed solely by density and is independent of temperature. Such models can be used to describe lyotropic liquid crystals and phase transitions such as the \textit{I}-\textit{N} transition. 

Onsager showed how the formation of liquid crystalline phases can be understood on the basis of pair interactions between the constituents of the material \cite{onsager.l:1949.a,Hansen06,gray.cg:1984.a}. He considered the hard platelet fluid but we know that unlike the case of rod-like particles, his second-virial theory does not produce quantitatively correct results for the equation of state and the \textit{I}-\textit{N} coexistence densities. Onsager himself noted that higher virial contributions are important for obtaining reliable results, estimating the ratio $B_{3}/B_{2}^{2}$ at $\mathcal{O} (1)$, with $B_{2}$ and $B_{3}$ being the second and third virial coefficients, respectively. 

Nevertheless, the second-virial theory has been employed to investigate the monodisperse platelet system \cite{forsyth.pa:1977.a, forsyth.pa:1978.a}. In Ref.~\cite{forsyth.pa:1977.a} a numerical approach was used to calculate the phase diagram of platelets for varying thickness, including the case of zero thickness. This is complemented by a calculation for the equation of state in both the \textit{I} and \textit{N} phases for vanishingly thin platelets in Ref.~\cite{forsyth.pa:1978.a}. The first off-lattice simulation study of the \textit{I}-\textit{N} transition for infinitely thin platelets was carried out in Ref.~\cite{eppenga.r:1982.a}. This showed that the phase diagram differs substantially from the Onsager prediction and that the \textit{I}-\textit{N} transition is actually much more weakly first order and occurs at lower densities than predicted theoretically. The authors also carried out a fifth order virial calculation for the equation of state. More accurate predictions of the higher virial contributions for disks were presented in Ref.~\cite{veerman.jac:1992.a}, where simulation results are reported for hard cut spheres and more recently in Ref.~\cite{fartaria.rps:2009.a}. Later simulation work was carried out on polydisperse platelet systems \cite{bates.ma:1999.a} and systems of platelets with different polygonal shapes (e.g. hexagons, triangles) \cite{bates.ma:1999.b}. Further simulation results of model circular platelets were reported in Ref.~\cite{reich.h:2007.a} and simulations alongside an integral equation approach for mixtures of rods and disks were carried out in Ref.~\cite{cheung.dl:2008.e}. Simulations of binary platelet systems have not yet been carried out. 

Binary mixtures of particles of different shape and/or size are interesting due to the richness of the phase diagrams they may exhibit. Binary rod mixtures form a prominent example. Studies include mixtures of thick and thin rods \cite{vanroij.r:1998.a} and long and short rods \cite{lekkerkerker.hnw:1984.a} using Onsager theory as well as using Parsons-Lee scaling \cite{varga.s:2003.a}. The phase behaviour in binary mixtures can include: the fractionation effect, whereby the larger particles go preferentially into the nematic phase; widening of the biphasic region; a re-entrant $I\rightarrow N \rightarrow I$ phenomenon on increasing density; the possibility of demixing into two different isotropic states and/or two different nematic states and triphasic equilibria (see e.g. Ref.~\cite{vanroij.r:1998.a} for examples of these phenomena). Nematic-nematic (\textit{N}-\textit{N}) demixing, at high enough pressures, can be viewed as a result of competition between orientational entropy of the smaller platelets favouring mixing, and the entropy of mixing \cite{vanroij.r:1998.a}. The \textit{N}-\textit{N} phase separation for binary mixtures of rods, including the high density regime, is studied in detail in Ref.~\cite{varga.s:2005.a}. 

Binary mixtures of thin and thick platelets have been investigated \cite{wensink.hh:2001.a, wensink.hh:2004.a} with the Parsons-Lee scaling of the Onsager functional \cite{parsons.jd:1979.a, lee.sd:1987.a, lee.sd:1989.a}. Studies based on the Zwanzig model for binary hard platelets, where the particles are restricted to occupy only three mutually perpendicular directions, have been carried out for  the bulk and interfacial properties of the demixed phases. Rich phase diagrams, involving isotropic and nematic phases have been reported in Refs. \cite{bier.m:2004.a, harnau.l:2002.c, harnau.l:2002.d}. A recent review \cite{harnau.l:2008.f} of platelet fluids contains a summary of these results. Recently Verhoeff \textit{et al.} \cite{verhoeff.a.a:2009.a} have investigated experimentally and theoretically the phase behaviour of colloidal platelets with bimodal shape distribution. Their theory is based on the Onsager-Parsons free energy and a cell approach for the columnar (\textit{Col}) state \cite{wensink.hh:2004.c}. The authors find agreement between their experimental findings and theoretical predictions for sufficiently large thickness ratios. The phase diagram features an \textit{I}-\textit{N} density inversion and triphasic \textit{I}-\textit{N}-\textit{Col} equilibrium.

Fundamental measure theory (FMT) is an approximate non-perturbative density functional theory (DFT) \cite{evans.r:1979.a}, originally proposed by Rosenfeld for additive hard sphere mixtures \cite{rosenfeld.y:1989.a, rosenfeld.y:1997.a}. The approach was later generalised to other convex shapes \cite{rosenfeld.y:1994.a, rosenfeld.y:1995.a}, which led to subsequent work \cite{cinacchi.g:2002.a, hansen-goos.h:2009.a}. The bulk \textit{I}-\textit{N} coexistence densities (scaled by the cube of the platelet radius) and nematic order parameter at the transition ($c_{I},c_{N}$ and $S_{N}$, respectively) were calculated by Frenkel and Eppenga in Ref.~\cite{eppenga.r:1982.a} by simulation; for more recent simulation results see Ref.~\cite{reich.h:2007.b}. The values previously obtained from FMT are $c_{I}=0.418$, $c_{N}=0.46$ and $S_{N}=0.53$ \cite{reich.h:2007.b}, which are in agreement with the present study. Recently \cite{hendrik_communication} these values were improved using the same method but with increased resolution to $c_{I}=0.419$, $c_{N}=0.469$ and $S_{N}=0.533$ \cite{fmt_note}. The FMT functional for pure platelets was later utilised to study inhomogeneous situations including the \textit{I}-\textit{N} interface and wetting at a hard wall \cite{vanderbeek.d:2006.a, reich.h:2007.a} and capillary nematisation of platelets between two parallel walls \cite{reich.h:2007.b}. 

Generalising the theory for the corresponding one-component system \cite{esztermann.a:2006.a}, we here propose a functional to describe binary mixtures of vanishingly thin circular platelets. Our theory features the exact virial second order term in density and an approximate term of third order in density. We investigate three types of demixing phase behaviour in the case of binary platelets with varying size ratio, finding \textit{I}-\textit{N} and \textit{N}-\textit{N} phase coexistence. We do not find stable \textit{I}-\textit{I} demixing (as could be driven by the depletion effect \cite{vanroij.r:1996.a}) for the regimes considered in the present work. We restrict our attention to uniaxial arrangements of the (uniaxial) platelets, as we do not expect biaxial arrangement of the particles to occur. We study a range of size ratios in this investigation, ranging between $\lambda=1.1$ and $5$. We present the phase diagrams in different representations to facilitate comparison with simulations which may be performed in different ensembles or experiments. We expect the phase diagrams from FMT to be quantitatively more accurate than those from Onsager theory, which we calculate as a reference. The topologies of the phase diagrams are the same in both theories for our chosen values of the size ratio between two species. Although the integral kernel, which represents the pair excluded volume term, is the same for long thin rods as it is for platelets of vanishing thickness, the results from Onsager theory cannot be obtained by simple scaling of literature results for binary mixtures of rods.

This paper is organised as follows. In Sec.\ \ref{dft} we define the model, outline the density functional theory and the conditions for thermodynamic stability and phase coexistence. In Sec.\ \ref{results} we present results for the phase behaviour of binary platelet mixtures. In Sec.\ \ref{outlook} we provide conclusions and an outlook on possible future work.
\section{Density Functional Theory for Binary Hard Platelet Mixtures}
\label{dft}
\subsection{Pair Interactions and Model Parameters}
\label{model}

We consider a binary mixture of hard circular platelets with vanishing thickness and continuous positional and orientational degrees of freedom. Species 1 and species 2 have radii $R_{1}$ and $R_{2}$ respectively, with $R_{2}>R_{1}$. The pair potential $u_{ij}$ between two particles $i$ and $j$, where $i,j=1,2$, is infinite if the geometrical shapes of the two platelets overlap and zero otherwise and is hence given by
\begin{equation}
\label{eq:potential}
{u_{ij}(\bf{r}-\bf{r}', \boldsymbol{\omega},\boldsymbol{\omega}')=\begin{cases}\infty&\text{if particles overlap}  \\0&\text{otherwise, } \end{cases}}
\end{equation}
where $\textbf{r}$ and $\textbf{r}'$ are the positions of the particle centres and $\boldsymbol{\omega}$ and $\boldsymbol{\omega}'$ are unit vectors indicating the particle orientations (normal to the particle surface). The size ratio 
\begin{equation}
\label{eq:size_ratio}
\lambda=\frac{R_{2}}{R_{1}} > 1
\end{equation}
characterises the radial bidispersity and is the only control parameter in the model. We characterise the thermodynamic state by two dimensionless densities $c_{1} = \rho_{1}R_{1}^{3}$  and $c_{2} = \rho_{2}R_{1}^{3}$, where $\rho_{1}$ and $\rho_{2}$ are the number densities of the two species, $\rho_{i}=N_{i}/V$, where $N_{i}$ is the number of particles of species $i=1,2$ and $V$ is the system volume.
The composition (mole fraction) of the (larger) species 2 is $x=\rho_{2}/(\rho_{1}+\rho_{2})$ and the total dimensionless concentration is $c=R_{1}^{3}(\rho_{1}+\rho_{2})=c_{1}+c_{2}$.

\subsection{Grand Potential Functional and Minimisation Principle}

Density functional theory (DFT) is formulated on the one-body level of the density distributions $\rho_{i}(\bf{r},\boldsymbol{\omega})$. The variational principle \cite{evans.r:1979.a} states that the true equilibrium density profile is the one which minimises the grand potential functional $\Omega$ and so obeys
\begin{equation}
\label{eq:minimization_principle}
\frac{\delta\Omega([\rho_{1},\rho_{2}],\mu_{1},\mu_{2},V,T)}{\delta \rho_{i}(\textbf{r},\boldsymbol{\omega})}=0,
\end{equation}
where $\mu_{i}$ is the chemical potential of species $i=1,2$ and $T$ is absolute temperature. The grand potential functional can be decomposed as
\begin{align}
\label{eq:grand_potential}
\Omega([\rho_{1},&\rho_{2}],\mu_{1},\mu_{2},V,T)=  F_{\textrm{id}}([\rho_{1},\rho_{2}],V,T)\notag \\ &+ F_{\textrm{exc}}([\rho_{1},\rho_{2}],V,T) + \sum_{i=1}^{2}\int d \textbf{r} \int d \boldsymbol{\omega} \rho_{i} (V_{\textrm{ext}}^{i}(\textbf{r},\boldsymbol{\omega})-\mu_{i}),
\end{align}
where the spatial integral (over $\textbf{r}$) is over the system volume $V$ and the angular integral (over $\boldsymbol{\omega}$) is over the unit sphere;  $V_{\textrm{ext}}^{i}(\textbf{r},\boldsymbol{\omega})$ is an external potential acting on species $i$; $F_{\textrm{exc}}([\rho_{1},\rho_{2}],V,T)$ is the excess (over ideal gas) contribution to the Helmholtz free energy and describes the inter-particle interactions. The free energy functional for a binary ideal gas of uniaxial rotators is given by
\begin{align}
\label{eq:ideal}
\beta F_{\textrm{id}}([\rho_{1},\rho_{2}],V,T) = \sum_{i=1}^{2}&\int{d\textbf{r}}\int d\boldsymbol{\omega}\rho_{i}(\textbf{r},\boldsymbol{\omega})\notag \\ &\times[\ln(\rho_{i}(\textbf{r},\boldsymbol{\omega})\Lambda_{i}^{3})-1],   
\end{align}
where $\Lambda_{i}$ is the (irrelevant) thermal wavelength of species $i$ and $\beta=1/(k_{B}T)$, where $k_{B}$ is the Boltzmann constant. We let $\Lambda_{i}=R_{1}$, which is equivalent to fixing an additive constant to the chemical potential and hence does not affect any observable properties of the system. One systematic way to express the excess free energy functional is to expand it in a virial series in density \cite{Hansen06}. Onsager theory is based entirely on the second-virial level. FMT (as described in Sec.~\ref{fmt} below) approximates higher order terms using single particle geometries. Nevertheless we find it useful to give the terms in the virial expansion up to third order in density explicitly: the second and third order contributions to the (exact) virial series for the excess free energy $\beta {F}_{\textrm{exc}}([\rho_{1},\rho_{2}],V,T)$ are given respectively by
\begin{eqnarray}
\label{eq:virial_expansion}
-\frac{1}{2}\left[\begin{picture}(10,10)(0,0)
\put(5,-4){\circle*{5}}
\put(3,-15){\small{$1$}}
\put(3,17){\small{$1$}}
\put(5,-1){\line(0,1){12}}
\put(5,12){\circle*{5}}
\end{picture}+
\begin{picture}(10,10)(0,0)
\put(5,-4){\circle*{5}}
\put(3,-15){\small{$2$}}
\put(3,17){\small{$2$}}
\put(5,-1){\line(0,1){12}}
\put(5,12){\circle*{5}}
\end{picture}+
2\begin{picture}(10,10)(0,0)
\put(5,-4){\circle*{5}}
\put(3,-15){\small{$1$}}
\put(3,17){\small{$2$}}
\put(5,-1){\line(0,1){12}}
\put(5,12){\circle*{5}}
\end{picture}
\right],
\end{eqnarray}
\begin{eqnarray}
\label{eq:virial_expansion_2}
-\frac{1}{6}\Big[\begin{picture}(25,10)(0,0)
\put(5,-4){\circle*{5}}
\put(3,-15){\small{$1$}}
\put(3,17){\small{$1$}}
\put(19,-15){\small{$1$}}
\put(5,-2){\line(0,1){12}}
\put(5,12){\circle*{5}}
\put(7,-4){\line(1,0){12}}
\put(19,-4){\circle*{5}}
\put(17.8,-1.5){\line(-1,1){11}}
\end{picture}+
\begin{picture}(25,10)(0,0)
\put(5,-4){\circle*{5}}
\put(3,-15){\small{$2$}}
\put(3,17){\small{$2$}}
\put(19,-15){\small{$2$}}
\put(5,-2){\line(0,1){12}}
\put(5,12){\circle*{5}}
\put(7,-4){\line(1,0){12}}
\put(19,-4){\circle*{5}}
\put(17.8,-1.5){\line(-1,1){11}}
\end{picture}+
3\begin{picture}(25,10)(0,0)
\put(5,-4){\circle*{5}}
\put(3,-15){\small{$1$}}
\put(3,17){\small{$2$}}
\put(19,-15){\small{$1$}}
\put(5,-2){\line(0,1){12}}
\put(5,12){\circle*{5}}
\put(7,-4){\line(1,0){12}}
\put(19,-4){\circle*{5}}
\put(17.8,-1.5){\line(-1,1){11}}
\end{picture}+
3\begin{picture}(25,10)(0,0)
\put(5,-4){\circle*{5}}
\put(3,-15){\small{$1$}}
\put(3,17){\small{$2$}}
\put(19,-15){\small{$2$}}
\put(5,-2){\line(0,1){12}}
\put(5,12){\circle*{5}}
\put(7,-4){\line(1,0){12}}
\put(19,-4){\circle*{5}}
\put(17.8,-1.5){\line(-1,1){11}}
\end{picture}
\Big],
\end{eqnarray}\\
where each line in the diagrams represents a Mayer function $f_{ij}({\textbf r}-{\textbf r }', \boldsymbol{\omega}, \boldsymbol{\omega}')= \exp \left(-\beta u_{ij}({\textbf r}-{\textbf r }', \boldsymbol{\omega}, \boldsymbol{\omega}')\right)-1$, which equals $-1$ if the two particles overlap and zero otherwise. The shaded circles, {\em field points}, indicate multiplication by the one-body density $\rho_{i}({\textbf r},\boldsymbol{\omega})$ and integration over the coordinates $\textbf r$ and $\boldsymbol{\omega}$ \cite{Hansen06}. The number alongside each field point represents the species $i$. Here we consider only spatially homogeneous states, such that the one-body density does not depend on position $\textbf{r}$ and, for the case of uniaxial nematic states considered in this paper, depends only on the polar angle $\theta$ of $\boldsymbol{\omega}$ with respect to the nematic director. Hence the one-body densities factorise as $\rho_{i}(\textbf{r}, \boldsymbol{\omega}) = \rho_{i}\Psi_{i}(\theta)$ where $\Psi_{i}(\theta)$ is the orientational distribution function (ODF) and $\rho_{i}$ the number density of species $i$. Before laying out the FMT we focus on the second-virial level.
\subsection{Onsager Second-Virial Theory for Binary Platelets}
\label{onsager}
The diagrams in Eq.~(\ref{eq:virial_expansion}), using the fact that $\rho_{i}(\textbf{r}, \boldsymbol{\omega}) = \rho_{i}\Psi_{i}(\theta)$ for spatially homogeneous states, become\\
\begin{eqnarray}
\label{eq:virial_expansion_example}
-\frac{1}{2}\begin{picture}(10,10)(0,0)
\put(5,-4){\circle*{5}}
\put(3,-16){\small{$i$}}
\put(3,18){\small{$j$}}
\put(5,-1){\line(0,1){12}}
\put(5,12){\circle*{5}}
\end{picture} =-\frac{1}{2}\rho_{i}\rho_{j}\int d \boldsymbol{\omega} \int d \boldsymbol{\omega}'  \int d\textbf{r} f_{ij}(\textbf{r}, \boldsymbol{\omega}, \boldsymbol{\omega}')  \Psi_{i}(\theta)\Psi_{j}(\theta ') ,
\end{eqnarray}
where we have renamed ${\textbf r}-{\textbf r'}\rightarrow \textbf{r}$. The spatial integral over the Mayer bond yields (minus) the excluded volume $-\mathcal{E}_{ij}(\boldsymbol{\omega},\boldsymbol{\omega}')$ between two particles of species $i$ and $j$, as a function of the angle $\gamma$ between  $\boldsymbol \omega$ and $\boldsymbol \omega '$. Hence
\begin{equation}
\label{eq: second_virial_part}
\mathcal{E}_{ij}(\boldsymbol{\omega},\boldsymbol{\omega}')=-\int d\textbf{r} f_{11}(\textbf{r}, \boldsymbol{\omega}, \boldsymbol{\omega}')=2\pi (R_{i}^{2}R_{j}+R_{j}^{2}R_{i})\sin\gamma ,
\end{equation}
Therefore 
\begin{eqnarray}
\label{eq:virial_expansion_example_1}
-\frac{1}{V}\begin{picture}(10,10)(0,0)
\put(5,-4){\circle*{5}}
\put(3,-16){\small{$i$}}
\put(3,18){\small{$j$}}
\put(5,-1){\line(0,1){12}}
\put(5,12){\circle*{5}}
\end{picture} =16\pi^{2} (R_{i}^{2}R_{j}+R_{j}^{2}R_{i})\rho_{i}\rho_{j} \int_{0}^{\frac{\pi}{2}} d\theta \sin \theta 
\end{eqnarray}
\vspace{-5mm}
$$
\hspace{10mm}\times \int_{0}^{\frac{\pi}{2}} d \theta ' \sin \theta ' K(\theta, \theta') \Psi_{i}(\theta) \Psi_{j}(\theta ').
$$
The integrals in Eq.~(\ref{eq:virial_expansion_example_1}) (omitting the prefactor) can be written as
\begin{align}
\label{eq:virial_expansion_example_2}
\int_{0}^{2\pi} d \phi \int_{0}^{\pi} d \theta \sin \theta &\int_{0}^{2\pi} d \phi '\int_{0}^{\pi} d \theta \sin \theta ' \notag \\ & \times \Psi_{1}(\theta)\Psi_{1}(\theta ')\sin\gamma,
\end{align}
where $\theta$ and $\theta'$ are the polar angles of two platelets with respect to the nematic director and $\phi$ and $\phi '$ are the azimuthal angles. Due to the inversion symmetry of the nematic state $\int_{0}^{\pi} d \theta =2\int_{0}^{\frac{\pi}{2}} d \theta$. In order to deal with the azimuthal integral we introduce the kernel $K(\theta,\theta')$ via
\begin{align}
\label{eq: kernel_k}
&K(\theta,\theta')=\int_{0}^{2\pi}d\phi \sin\gamma=\int_{0}^{2\pi}d\phi \sqrt{1- (\boldsymbol{\omega} \cdot \boldsymbol{\omega}')^{2}}
\notag \\ & =\int_{0}^{2\pi}d\phi \sqrt{1-(\cos\theta \cos\theta' + \sin\theta \sin\theta' \cos\phi)^{2}},
\end{align}
where we have renamed $\phi-\phi ' \rightarrow \phi$ as the difference between the azimuthal angles of the two platelets. 
Adding all three terms and multiplying by $R_{1}^{3}$ yields the Onsager contribution to the excess free energy in the dimensionless form
\begin{align}
\label{eq: free_energy_2}
\frac{\beta F_{\textrm{exc}}^{(2)}}{V}R_{1}^{3} =16 &\pi^{2}c^{2} \int_{0}^{\frac{\pi}{2}} d\theta \sin \theta \int_{0}^{\frac{\pi}{2}} d \theta ' \sin \theta '  K(\theta, \theta') \notag \\ &\times\Big[(1-x)^{2}\Psi_{1}(\theta)\Psi_{1}(\theta ')  +x^{2}\lambda^{3}\Psi_{2}(\theta)\Psi_{2}(\theta ') \notag \\ &\hspace{6mm}+x(1-x)(\lambda^{2}+\lambda)\Psi_{1}(\theta)\Psi_{2}(\theta') \Big],
\end{align}
where the superscript of $F^{(\alpha)}_{\textrm{exc}}$ represents the order in density of the excess free energy.  
\subsection{Fundamental Measure Theory for Binary Platelet Mixtures}
\label{fmt}
We generalise the monodisperse functional to the case of binary mixtures using an approximate term at third order in density which is based on the FMT developed in Ref.~\cite{esztermann.a:2006.a}. Our theory contains the exact second order Onsager term and instead of using any higher order terms from the series, such as the {\em exact} third virial level (\ref{eq:virial_expansion_2}), an approximate term which is of third order in density is used \cite{Sollich_communication}. This term is nonvanishing (and constant) for cases with common triple intersection of the three platelets involved. There are no higher order terms due to the scaled-particle roots \cite{barker.ja:1976.a} of the approach; the vanishing volume of the platelets truncates the series. Global prefactors are used to compensate for \textit{lost cases} \cite{cuesta.ja:2002.a, tarazona.p:1997.a}.  We postulate the excess free energy
\begin{align}
\label{eq: free_energy_fmt}
\beta F_{\textrm{exc}}([\rho_{1},\rho_{2}],V)=\int d& \textbf{r} \int d  \boldsymbol{\omega} \int d  \boldsymbol{\omega}' \Big[n^{\textrm{DD}}_{1}(\textbf{r},\boldsymbol{\omega})  n_{2}^{\textrm{D}}(\textbf{r},\boldsymbol{\omega}') \notag \\ &+ \frac{1}{24 \pi} n_{2}^{\textrm{D}}(\textbf{r},\boldsymbol{\omega})  n_{2}^{\textrm{DDD}}(\textbf{r},\boldsymbol{\omega},\boldsymbol{\omega}')  n_{2}^{\textrm{D}}(\textbf{r},\boldsymbol{\omega}')   \Big],
\end{align}
(where the right hand side is independent of $T$). The first term of the sum in Eq.~(\ref{eq: free_energy_fmt}) is equivalent to the Onsager contribution to the excess free energy and the second is the FMT contribution. The weighted densities are related to the bare one-body densities, $\rho_{i}(\textbf{r}, \boldsymbol{\omega})$, via
\begin{align}
\label{eq: density_1dd}
n_{1}^{\textrm{DD}}(\textbf{r}, \boldsymbol{\omega}) &= \sum_{i=1}^{2} \int d \boldsymbol{\omega}'w_{1}^{\textrm{DD}i}(\textbf{r}, \boldsymbol{\omega}',\boldsymbol{\omega})*\rho_{i}(\textbf{r}, \boldsymbol{\omega}'),\\
\label{eq: density_2d}
n_{2}^{\textrm{D}}(\textbf{r}, \boldsymbol{\omega})&= \sum_{i=1}^{2}w_{2}^{\textrm{D}i}(\textbf{r},\boldsymbol{\omega})*\rho_{i}(\textbf{r}, \boldsymbol{\omega}),\\
\label{eq: density_2ddd}
n_{2}^{\textrm{DDD}}(\textbf{r}, \boldsymbol{\omega},\boldsymbol{\omega}')&= \sum_{i=1}^{2} \int d \boldsymbol{\omega}''w_{2}^{\textrm{DDD}i}(\textbf{r}, \boldsymbol{\omega},\boldsymbol{\omega}',\boldsymbol{\omega}'')*\rho_{i}(\textbf{r}, \boldsymbol{\omega}''),
\end{align}
where $*$ represents the three-dimensional convolution $h(\textbf{r})*g(\textbf{r})=\int d^{3}x h(\textbf{x})g(\textbf{x}-\textbf{r})$. We have kept the notation of Ref.~\cite{esztermann.a:2006.a} where the upper index $\textrm{D}$ (disk) is indicative of the number of particle orientations that appear in the weight function (below) or weighted density. The weight functions for species $i$ are given by
\begin{align}
\label{eq: weight_1d}
w_{1}^{\textrm{D}i}(\textbf{r}, \boldsymbol{\omega})&= \delta(R_{i}-|\textbf{r}|)\delta(\textbf{r} \cdot \boldsymbol{\omega})/8,\\
\label{eq: weight_2d}
w_{2}^{\textrm{D}i}(\textbf{r}, \boldsymbol{\omega}) &= 2 \Theta(R_{i}- |\textbf{r}|) \delta (\textbf{r} \cdot \boldsymbol{\omega}),\\
\label{eq: weight_1dd}
w_{1}^{\textrm{DD}i}(\textbf{r}, \boldsymbol{\omega}, \boldsymbol{\omega}') &= \frac{2}{R_{i}} | \boldsymbol{\omega} \cdot ( \boldsymbol{\omega}' \times \textbf{r})| w_{1}^{\textrm{D}i}(\textbf{r}, \boldsymbol{\omega}),\\
\label{eq: weight_1ddd}
w_{2}^{\textrm{DDD}i}(\textbf{r}, \boldsymbol{\omega}, \boldsymbol{\omega}',\boldsymbol{\omega}'') &= \frac{8}{\pi} | \boldsymbol{\omega} \cdot ( \boldsymbol{\omega}' \times \boldsymbol{\omega}'')| w_{2}^{\textrm{D}i}(\textbf{r}, \boldsymbol{\omega}),
\end{align}
\\
\noindent
where $\delta(\cdot)$ is the Dirac delta distribution, $\Theta(\cdot)$ is the Heaviside step function and $\times$ denotes the vector product, such that $\boldsymbol{\omega} \cdot ( \boldsymbol{\omega}' \times \boldsymbol{\omega}'')$ is the triple scalar product. Note the modulus in Eqs.~(\ref{eq: weight_1dd}) and (\ref{eq: weight_1ddd}). For spatially homogeneous states, Eq.~(\ref{eq: free_energy_fmt}) becomes
\begin{align}
\label{eq: free_energy_fmt_bulk}
\frac{\beta F_{\textrm{exc}} ([\rho_{1},\rho_{2}])}{V} = \int d & \boldsymbol{\omega} \int d  \boldsymbol{\omega}' \Big[n_{1}^{\textrm{DD}}(\boldsymbol{\omega})  n_{2}^{\textrm{D}}(\boldsymbol{\omega}') \notag \\ &+ \frac{1}{24 \pi} n_{2}^{\textrm{D}}(\boldsymbol{\omega})  n_{2}^{\textrm{DDD}}(\boldsymbol{\omega},\boldsymbol{\omega}')  n_{2}^{\textrm{D}}(\boldsymbol{\omega}')   \Big].
\end{align}
Inserting the definitions of the weighted densities (\ref{eq: density_1dd})-(\ref{eq: density_2ddd}) into the excess free energy (\ref{eq: free_energy_fmt}) we obtain
\begin{eqnarray}
\label{eq: free_energy_fmt_alternative}
\beta F_{\textrm{exc}}([\rho_{1},\rho_{2}])=
-\frac{1}{2}\left[\begin{picture}(10,10)(0,0)
\put(5,-4){\circle*{5}}
\put(3,-15){\small{$1$}}
\put(3,17){\small{$1$}}
\put(5,-1){\line(0,1){12}}
\put(5,12){\circle*{5}}
\end{picture}+
\begin{picture}(10,10)(0,0)
\put(5,-4){\circle*{5}}
\put(3,-15){\small{$2$}}
\put(3,17){\small{$2$}}
\put(5,-1){\line(0,1){12}}
\put(5,12){\circle*{5}}
\end{picture}+
2\begin{picture}(10,10)(0,0)
\put(5,-4){\circle*{5}}
\put(3,-15){\small{$1$}}
\put(3,17){\small{$2$}}
\put(5,-1){\line(0,1){12}}
\put(5,12){\circle*{5}}
\end{picture}
\right]
+\int d \textbf{x} \int d \boldsymbol{\omega} \int d \boldsymbol{\omega '} \int d \boldsymbol{\omega ''}
\frac{| \boldsymbol{\omega} \cdot ( \boldsymbol{\omega}' \times \boldsymbol{\omega}'')|}{3\pi^{2}}
\end{eqnarray}
$$
\hspace{55mm}\times n_{2}^{\textrm{D}}(\textbf{x},\boldsymbol{\omega})n_{2}^{\textrm{D}}(\textbf{x},\boldsymbol{\omega}')n_{2}^{\textrm{D}}(\textbf{x},\boldsymbol{\omega}'').
$$
The fundamental measures of a platelet of species $i$ are the integral mean curvature $\xi_{i}^{\textrm{IMC}}=\pi R_{i}/4$ and the surface $\xi_{i}^{\textrm{S}}=2\pi R_{i}^{2}$. The first term of Eq.~(\ref{eq: free_energy_fmt_bulk}) may be expressed as
\begin{align}
\label{eq: measures_interpreation}
A'(\boldsymbol{\omega},\boldsymbol{\omega}')=\frac{4}{\pi}&\Big[\xi_{1}^{\textrm{IMC}}\xi_{1}^{\textrm{S}}\rho_{1}^{2}\Psi_{1}(\theta)\Psi_{1}(\theta') +\xi_{2}^{\textrm{IMC}}\xi_{2}^{\textrm{S}}\rho_{2}^{2}\Psi_{2}(\theta)\Psi_{2}(\theta')\notag \\ &+\left(\xi_{2}^{\textrm{IMC}}\xi_{1}^{\textrm{S}}+\xi_{1}^{\textrm{IMC}}\xi_{2}^{\textrm{S}}\right)\Psi_{1}(\theta)\Psi_{2}(\theta')/2\Big]\sin \gamma,
\end{align}
remembering that $\Psi_{i}(\boldsymbol{\omega})=\Psi_{i}(\theta)$ for uniaxial nematics. Eq.~(\ref{eq: measures_interpreation}) is simply the fundamental measures interpretation \cite{esztermann.a:2006.a} of the Onsager contribution to the free energy.  Note that the term which represents the excluded volume between two particles of different species is given by $\xi_{2}^{\textrm{IMC}}\xi_{1}^{\textrm{S}}+\xi_{1}^{\textrm{IMC}}\xi_{2}^{\textrm{S}}$. This leads to a scaling of the excluded volume by $\lambda^{2} + \lambda$, which is quite different from the scaling that occurs for binary mixtures of rods. The second term in Eq.~(\ref{eq: free_energy_fmt_bulk}) is the FMT contribution to the excess free energy. This is given by
\begin{align}
\frac{\beta F^{(3)}_{\textrm{exc}}}{V}= \int d  \boldsymbol{\omega} \int d  \boldsymbol{\omega}' \int d  \boldsymbol{\omega}'' B'(\boldsymbol{\omega},\boldsymbol{\omega}',\boldsymbol{\omega}'')
\end{align}
\noindent
where 
\begin{align}
B'(\boldsymbol{\omega},\boldsymbol{\omega}',\boldsymbol{\omega}'')=\frac{| \boldsymbol{\omega} \cdot ( \boldsymbol{\omega}' \times \boldsymbol{\omega}'')|}{3\pi^{2}}&\Big[(\xi_{1}^{\textrm{S}})^{3}\rho_{1}^{3}\Psi_{1}(\theta)\Psi_{1}(\theta')\Psi_{1}(\theta'') \notag \\ &+(\xi_{2}^{\textrm{S}})^{3}\rho_{2}^{3}\Psi_{2}(\theta)\Psi_{2}(\theta')\Psi_{2}(\theta'') \notag \\ &+3(\xi_{1}^{\textrm{S}})^{2}\xi_{2}^{\textrm{S}}\rho_{1}^{2}\rho_{2}\Psi_{1}(\theta)\Psi_{1}(\theta')\Psi_{2}(\theta'') \notag \\ &+3\xi_{1}^{\textrm{S}}(\xi_{2}^{\textrm{S}})^{2}\rho_{1}\rho_{2}^{2}\Psi_{1}(\theta)\Psi_{2}(\theta')\Psi_{2}(\theta'')\Big].
\end{align}
We choose coordinates such that $\boldsymbol{\omega}=(\sin\theta,0,\cos\theta)$, $\boldsymbol{\omega}'=(\cos\theta'\sin\theta',\sin\phi'\sin\theta',\cos\theta')$ and $\boldsymbol{\omega}''=(\cos\theta''\sin\theta'',\sin\phi''\sin\theta'',\cos\theta'')$ where $\theta, \theta '$ and $\theta ''$ are the polar angles of three platelets. The third order contribution in density to the FMT excess free energy in these coordinates is given by
\begin{widetext}
\begin{align}
\label{eq: free_energy_3}
\frac{\beta F_{\textrm{exc}}^{(3)}}{V} R_{1}^{3}=  \frac{128\pi^{2}}{3}&c^{3} \int_{0}^{\frac{\pi}{2}} d\theta \sin \theta \int_{0}^{\frac{\pi}{2}} d \theta ' \sin \theta ' \int_{0}^{\frac{\pi}{2}} d\theta'' \sin \theta'' L(\theta, \theta',\theta'') \notag \\ &[(1-x)^{3}\Psi_{1}(\theta)\Psi_{1}(\theta')\Psi_{1}(\theta'') +x^{3}\lambda^{6}\Psi_{2}(\theta)\Psi_{2}(\theta')\Psi_{2}(\theta'') \notag \\ &  +3x(1-x)^{2}\lambda^{2}\Psi_{1}(\theta)\Psi_{1}(\theta')\Psi_{2}(\theta'') +3x^{2}(1-x)\lambda^{4}\Psi_{1}(\theta)\Psi_{2}(\theta')\Psi_{2}(\theta'') ],
\end{align}
\end{widetext}
where the kernel $L(\theta,\theta', \theta'')$ is
\begin{align}
\label{eq: kernel_l}
L(\theta,\theta', \theta'')=&\int_{0}^{2\pi} \int_{0}^{2\pi} d\phi' d\phi''  | \boldsymbol{\omega} \cdot ( \boldsymbol{\omega}' \times \boldsymbol{\omega}'')| \notag \\ &\hspace{-4.2mm}=\int_{0}^{2\pi} \int_{0}^{2\pi} d\phi' d\phi''  |\sin \theta(\sin \phi' \sin \theta' \cos \theta''  \notag \\ &\hspace{5mm}+\cos\theta' \sin \phi'' \sin \theta'' )+\cos\theta(\cos\phi' \sin\theta' \sin\phi '' \sin \theta '' \notag \\ &\hspace{5mm}- \sin \phi' \sin \theta' \cos \phi '' \sin \theta '')|.
\end{align}
The full form of the excess free energy, as used in the calculations described below, is given by the sum of Eqs.~(\ref{eq: free_energy_2}) and (\ref{eq: free_energy_3}). In practice, along with Eqs.~(\ref{eq: kernel_k}) and (\ref{eq: kernel_l}), these require numerical computation on a grid as described in the following.
\subsection{Self-consistency equations for the orientational distribution functions}
The minimisation principle (\ref{eq:minimization_principle}) together with the FMT approximation (\ref{eq: free_energy_fmt_bulk}) for $F_{\textrm{exc}}([\rho_{1},\rho_{2}])$ leads to two coupled Euler-Lagrange equations for the ODFs:
\begin{widetext}
\begin{align}
\label{eq:self_consistency_1}
\Psi_{1}(\theta)  =  \frac{1}{Z_{1}} \exp \bigg[  &-8\pi c \int_{0}^{\frac{\pi}{2}} d \theta '  \sin \theta '  K(\theta, \theta ')[(1-x) \Psi_{1}(\theta ') +  \frac{1}{2}x(\lambda^{2} + \lambda)\Psi_{2}(\theta ')  ] \notag \\ &  -32\pi c^{2} \int_{0}^{\frac{\pi}{2}} d \theta '  \sin \theta ' \int_{0}^{\frac{\pi}{2}} d \theta ''  \sin \theta ''   L(\theta, \theta ', \theta'') \notag \\ &  \times[(1-x)^{2}\Psi_{1}(\theta')\Psi_{1}(\theta'')+2x(1-x)\lambda^{2}\Psi_{1}(\theta')\Psi_{2}(\theta'')+x^{2}\lambda^{4}\Psi_{2}(\theta')\Psi_{2}(\theta'') ] \bigg],
\end{align}
\begin{align}
\label{eq:self_consistency_2}
 \Psi_{2}(\theta)  = \frac{1}{Z_{2}} \exp \bigg[ &-8 \pi c \int_{0}^{\frac{\pi}{2}} d \theta '  \sin \theta ' K(\theta, \theta ')[x \lambda^{3}\Psi_{2}(\theta ') + \frac{1}{2}(1-x) (\lambda^{2} + \lambda)\Psi_{1}(\theta ')] \notag \\ & -32\pi c^{2} \int_{0}^{\frac{\pi}{2}} d \theta '  \sin \theta ' \int_{0}^{\frac{\pi}{2}} d \theta ''  \sin \theta '' L(\theta, \theta ', \theta'') \notag \\ &  \times[x^{2} \lambda^{6}\Psi_{2}(\theta')\Psi_{2}(\theta'')+2x(1-x)\lambda^{4}\Psi_{1}(\theta')\Psi_{2}(\theta'')+(1-x)^{2}\lambda^{2}\Psi_{1}(\theta')\Psi_{1}(\theta'') ] \bigg],
\end{align}
\end{widetext}
where the constants $Z_{1}$ and $Z_{2}$ are such that the normalisation $\int d \boldsymbol{\omega}\Psi_{i}(\boldsymbol{\omega})=1$, for $i=1,2$. Neglecting terms of order $c^{2}$ in the exponentials, the equations for FMT reduce to the Euler-Lagrange equations of Onsager theory. We solve Eqs.~(\ref{eq:self_consistency_1}) and (\ref{eq:self_consistency_2}) numerically with a straightforward extension to the iterative procedure given in Ref.~\cite{herzfeld84} and numerical techniques similar to those described in Ref.~\cite{vanroij.r.:2005.d}. The $\theta$- and $\phi$-grids are defined on $[0,\pi/2]$ and $[0,2\pi]$ respectively. The $\theta$-grid is divided into 200 equal steps on [0,$\pi/8$], where the ODF changes most rapidly in the nematic phase and into 50 equal steps on [$\pi/8$,$\pi/2$] where the ODF is almost zero. The $\phi$-grid is divided into 200 equally spaced intervals on [0,$2\pi$]. We start the iteration with two initial trial distributions, for example a normalized Gaussian $(c/\pi^{2})\exp \left[-2c^{2}\theta^{2}/\pi\right]$ or a constant distribution $1/4\pi$. The choice of the constant distribution is more efficient at low densities where the system is expected to be isotropic. These guesses are substituted into the right hand sides of Eqs.~(\ref{eq:self_consistency_1}) and~(\ref{eq:self_consistency_2}) to obtain a new pair of ODFs, $\Psi_{i,\textrm{new}}(\theta)$. This procedure is repeated until $\max|\Psi_{i,\textrm{new}}(\theta)-\Psi_{i,\textrm{old}}(\theta)|<t$, $i=1,2$, where $t$ is the tolerance given by the magnitude of the largest acceptable deviation to the ODF of the previous step: $\Psi_{i,\textrm{old}}(\theta)$. We set $t=10^{-9}$. For FMT the solutions take a longer time to converge than for Onsager theory, as each step has a slower execution time due to the increased complexity of the coupled equations (\ref{eq:self_consistency_1}) and (\ref{eq:self_consistency_2}).

\subsection{Conditions for Phase Coexistence}
Once we have found the ODFs we solve the phase coexistence equations. The requirements for phase coexistence between two phases \textit{A} and \textit{B} are the mechanical and chemical equilibria between the phases as well as the equality of temperature in the two coexisting phases (which is trivial in hard-body systems). Hence we have the non-trivial conditions
\begin{equation}
\label{eq: p_equality}
p^{\textit{A}}=p^{\textit{B}}
\end{equation}
and
\begin{equation}
\label{eq: mu_equality}
\mu_i^{\textit{A}}=\mu_i^{\textit{B}},
\end{equation}
where $i=1,2$ again labels the species.
We calculate the the total Helmholtz free energy $F$ numerically by inserting $\Psi_{i}(\theta)$ into Eqs.~(\ref{eq:ideal}), (\ref{eq: free_energy_2}) and (\ref{eq: free_energy_3}). Likewise, the pressure can be obtained numerically as
\begin{equation}
\label{eq: pressure}
p = -\frac{F}{V}+\sum_{i=1}^{2}\rho_{i}\frac{\partial(F/V)}{\partial \rho_{i}}
\end{equation}
and the chemical potentials as
\begin{equation}
\label{eq: chem_pot}
\mu_{i}= \frac{\partial (F/V)}{\partial \rho_{i}}.
\end{equation}
We define a reduced pressure $p^{*}=\beta p R_{1}^{3}$ and reduced chemical potentials $\mu_{i}^{*}=\beta \mu_{i}$. Eqs.~(\ref{eq: p_equality}) and (\ref{eq: mu_equality}) are three equations for four unknowns (two state points each characterised by two densities). Therefore, regions of two-phase coexistence depend parametrically on one free parameter. Eqs.~(\ref{eq: pressure}) and (\ref{eq: chem_pot}) are solved numerically with a Newton-Raphson procedure \cite{press.wh:2007.a}. The resulting set of solutions yields the binodal. \textit{I}-\textit{N}-\textit{N} triple points are located where the \textit{I}-\textit{N} and \textit{N}-\textit{N} coexistence curves cross. Therefore, regions of two-phase coexistence depend parametrically on one free parameter (which can be chosen arbitrarily, e.g. as the value of concentration $x$ in one of the phases).

\subsection{Equation of State for the Isotropic Phase}
Analytic expressions for the free energy, pressure and chemical potentials for the isotropic phase may be found on insertion of $\Psi(\theta)=1/(4\pi)$ into the ideal (\ref{eq:ideal}) and excess (\ref{eq: free_energy_fmt_bulk}) parts of free energy functional. For this purpose, we use 
\begin{equation}
\int_{0}^{\frac{\pi}{2}}d\theta \sin \theta\int_{0}^{\frac{\pi}{2}}d\theta' \sin \theta'K(\theta,\theta')= \frac{\mathcal{I}_{1}}{8\pi}=\frac{\pi^{2}}{2}
\end{equation}
and
\begin{equation}
\int_{0}^{\frac{\pi}{2}}d\theta \sin \theta\int_{0}^{\frac{\pi}{2}}d\theta' \sin \theta'\int_{0}^{\frac{\pi}{2}}d\theta'' \sin \theta''  L(\theta,\theta', \theta'')\\= \frac{\mathcal{I}_{2}}{16\pi} =\frac{\pi^{3}}{2}
\end{equation}
where the integrals $\mathcal{I}_{1}$ and $\mathcal{I}_{2}$ are calculated in Appendix A. The expressions for the FMT isotropic free energy, $\beta F_{\textrm{iso}}/V$, pressure, $p_{\textrm{iso}}^{*}$ and chemical potentials $\mu_{i,\textrm{iso}}^{*}$ are 
\begin{align}
\label{eq: isotropic_free_energy}
\frac{\beta F_{\textrm{iso}}}{V}R_{1}^{3} = c&(1-x)\ln \left( \frac{c(1-x)}{4\pi}\right) +  cx\ln \Big( \frac{cx}{4\pi}\Big)-c \notag \\ & +\frac{\pi^{2}}{2}c^{2}\left[(1-x)^{2}+x^{2}\lambda^{3}+x(1-x)(\lambda^{2}+\lambda)   \right] \notag \\ & + \frac{\pi^{2}}{3}c^{3}\left[(1-x)^{3}+x^{3}\lambda^{6}+3x(1-x)^{2}\lambda^{2}+3x^{2}(1-x)\lambda^{4}   \right],
\end{align}
\begin{align}
\label{eq: isotropic_pressure}
p^{*}_{\textrm{iso}}= c &+ \frac{c^{2}\pi^{2}}{2}\big[(1-x)^{2}+x^{2}\lambda^{3}+(\lambda^{2}+\lambda)(x-x^{2})   \big]
\notag \\ & +\frac{2\pi^{2}c^{3}}{3}\Big[x^{3}\lambda^{6}+3(x^{2}-x^{3})\lambda^{4}\notag \\ & +3(x-2x^{2}+x^{3})\lambda^{2}
+(1-x)^{3}      \Big],
\end{align}
\begin{align}
\label{eq: isotropic_mu_1}
\mu^{*}_{1,\textrm{iso}} = \ln& \left(\frac{c(1-x)}{4\pi}\right) +\frac{c\pi^{2}}{2}\big[2(1-x)+x(\lambda^{2}+\lambda)\big] 
\notag \\ &  + c^{2}\pi^{2}[(1-x)^{2}+2x(1-x)\lambda^{2}+x^{2}\lambda^{4}],
\end{align}
\begin{align}
\label{eq: isotropic_mu_2}
\mu^{*}_{2,\textrm{iso}}= \ln& \left(\frac{cx}{4\pi}\right)  +\frac{c\pi^{2}}{2}\big[2x\lambda^{3}+(1-x)(\lambda^{2}+\lambda)\big] \notag \\ & +c^{2}\pi^{2}[x^{2}\lambda^{6}+2x(1-x)\lambda^{4}+(1-x)^{2}\lambda^{2}].
\end{align}
The Onsager versions of these equations are given by the same expressions but without the final bracketed term in $c^{3}$ for $\beta F_{\textrm{iso}}/V$ and $p^{*}_{\textrm{iso}}$ and $c^{2}$ for $\mu^{*}_{i,\textrm{iso}}$. The authors of Ref.~\cite{eppenga.r:1984.a} claim that inclusion of the exact third virial term along with the Onsager term, at least for monodisperse platelets, would give a {\em worse} equation of state in the isotropic phase than the Onsager term alone.

\subsection{Isotropic-Nematic Bifurcation Analysis}
On increasing the density in the isotropic state, a point is reached known as the bifurcation density, where an infinitesimal nematic perturbation destabilises the system. The first \textit{I}-\textit{N} bifurcation analysis for a liquid crystalline system was performed in Ref.~\cite{kayser.rf:1978}. This was extended to a class of liquid crystal models in Ref.~\cite{mulder89}. The bifurcation concentration lies inside the coexistence region for the monodisperse case of platelets \cite{cheung.dl:2008.e} but as we will see, this is not always true in the binary case. We insert $\Psi_{1}(\theta)= [1+\epsilon_{1}P_{2}(\cos \theta)]/4\pi$ and $\Psi_{2}(\theta)= [1+\epsilon_{2}P_{2}(\cos \theta)]/4\pi$ into the free energy (\ref{eq:ideal}) and (\ref{eq: free_energy_fmt_bulk}), where $\epsilon_{1}$ and $\epsilon_{2}$ are small parameters measuring the strengths of the nematic perturbation and $P_{2}(\cos \theta)=(3\cos^{2}(\theta)-1)/2$ is the second degree Legendre polynomial in $\cos\theta$. We then extract all second order terms, i.e. those proportional to $\epsilon_{1}^{2}$, $\epsilon_{2}^{2}$ and  $\epsilon_{1}\epsilon_{2}$, respectively. The coefficients of these terms are denoted by $a_{1}(c)$, $a_{2}(c)$ and $a_{12}(c)$, respectively. These involve integrals of Legendre polynomials and are obtained using the integrals $\mathcal{I}_{3},\mathcal{I}_{4},\mathcal{I}_{5}$ defined in Appendix A. One then solves $\det \textbf{M}=0$ \cite{vanroij.r:1996.a} where
\begin{equation}
\label{eq: bifurcation_matrix}
\textbf{M}=\left( \begin{array}{cc}
a_{1}(c) & a_{12}(c)/2  \\
a_{12}(c)/2 & a_{2}(c) 
 \end{array} \right).
\end{equation}
Here
\begin{align}
\label{eq: matrix_element_11}
a_{1}(c) = & \frac{c}{10}(1-x)-\frac{\pi^{2}}{80}c^{2}(1-x)^{2}\notag \\ &-\frac{\pi^{2}}{40}c^{3}(1-x)^{3}-\frac{\pi^{2}}{40}c^{3}x(1-x)^{2}\lambda^{2},
\end{align}
\begin{align}
\label{eq: matrix_element_22}
a_{2}(c) = & \frac{cx}{10}-\frac{\pi^{2}}{80}c^{2}\lambda^{3}x^{2}\notag \\ &-\frac{\pi^{2}}{40}\lambda^{6}c^{3}x^{3}-\frac{\pi^{2}}{40}c^{3}x^{2}(1-x)\lambda^{4},
\end{align}
\begin{align}
\label{eq: matrix_element_12}
\frac{a_{12}(c)}{2}=&-\frac{\pi^{2}}{160}c^{2}x(1-x)(\lambda^{2}+\lambda)\notag \\ &-\frac{\pi^{2}}{40}c^{3}x(1-x)^{2}\lambda^{2}-\frac{\pi^{2}}{40}c^{3}x^{2}(1-x)\lambda^{4}.
\end{align}
The results for the spinodals were checked by running the self-consistency program for the solutions of the coupled ODFs (\ref{eq:self_consistency_1}) and (\ref{eq:self_consistency_2}) with the trial functions $1/4\pi$; the locus of $c$ values (for a given value of $\lambda$) for which the maximum number of iterations occured was found to agree numerically very well to the \textit{I}-\textit{N} spinodal. To calculate the spinodals in the $(x,p^{*})$ representation we insert the $(x,c)$ values which form the spinodal into the expression for the isotropic pressure (\ref{eq: isotropic_pressure}). In the monodisperse limit, one needs to solve the following cubic polynomial to calculate the bifurcation point $c_{*}$
\begin{equation}
c_{*}-\frac{\pi^{2}}{8}c_{*}^{2}-\frac{\pi^{2}}{4}c_{*}^{3}=0,
\end{equation}
where the bifurcation concentration is $c_{*}=8/\pi^{2}=0.811$ for Onsager theory and $c_{*}=0.434$ for FMT, the latter much closer to the limit of stability of the isotropic phase as observed in simulations \cite{cheung.dl:2008.e}.

\subsection{Symmetry-Conserved Demixing Spinodals}
A thermodynamic phase is locally stable if the determinant of the Hessian matrix of the Helmholtz free energy density with respect to the species densities is positive. The spinodal is the limit of stability, defined by $\det \textbf{N}=0$, where
\begin{equation}
\label{eq: hessian}
\textbf{N}=\left( \begin{array}{cc}
\frac{\partial^{2}(F/V)}{\partial \rho_{1}^{2}} & \frac{\partial^{2}(F/V)}{\partial \rho_{1}\partial \rho_{2}}   \\
\frac{\partial^{2}(F/V)}{\partial \rho_{1}\partial \rho_{2}}  & \frac{\partial^{2}(F/V)}{\partial \rho_{2}^{2}}   \end{array} \right). 
\end{equation}
For the \textit{I}-\textit{I} spinodals, we insert $\Psi_{1}(\theta)=\Psi_{2}(\theta)=1/4\pi$ into the free energy (\ref{eq: isotropic_free_energy}). This yields an analytic solution of Eq.~(\ref{eq: hessian}) for Onsager theory, given by
\begin{equation} 
\left(\pi^{2}+\frac{1}{\rho_{1}} \right)  \left(\pi^{2}\lambda^{3}+\frac{1}{\rho_{2}} \right) - \frac{\pi^{4}}{4}
\left(\lambda^{2}+\lambda \right)^{2}=0
\end{equation}
and for FMT by
\begin{align}
\label{eq:i-i_fmt}
&\Big(\pi^{2}+2\pi^{2}\rho_{1}+2\rho_{2}\pi^{2}{\lambda}^{2}+\frac{1}{\rho_{1}} \Big) \notag \\ & \times\Big(\pi^{2}\lambda^{3}+2\rho_{2}\pi^{2}\lambda^{6}+2\rho_{1}\pi^{2}\lambda^{4}+\frac{1}{\rho_{2}}\Big)\notag \\ & - \left( 2\rho_{1}
\pi^{2}\lambda^{2}+2\rho_{2}\pi^{2}{\lambda}^{4}
 \right) ^{2}=0.
\end{align}
\textit{I}-\textit{I} demixing never occurs for the values of $\lambda$ considered here; indeed solutions of Eq.~(\ref{eq:i-i_fmt}) only begin to exist at about $\lambda=20$. For the \textit{N}-\textit{N} spinodals, there are no such analytic equations (for the Zwanzig model, see Ref.~\cite{harnau.l:2002.d}). For practical reasons we rather solve 
\begin{equation}
\left(\frac{\partial p^{*}}{\partial \rho_{2}}\right)_{\mu_{1}}=0,
\label{eq: spin}
\end{equation}
which is equivalent \cite{vanroij.r:1996.a} to solving $\det \textbf{N}=0$. In order to calculate the \textit{N}-\textit{N} spinodals we have to numerically evaluate the left hand side of Eq.~(\ref{eq: spin}). Exchanging the species labels in Eq.~(\ref{eq: spin}) one obtains the same results. The \textit{N}-\textit{N} spinodals are calculated to give an idea of the location of the \textit{N}-\textit{N} phase boundaries. For cases where there is \textit{N}-\textit{N} coexistence closed by critical point, the spinodal and the binodal coincide at the critical point.
\section{Results}
\label{results}
We have calculated the phase diagrams of binary platelet mixtures for seven different size ratios, $\lambda = 1.1,1.4,1.7,2,2.5,4$ and $5$. Fig.~1a shows the results for the slightly asymmetric case $\lambda = 1.1$ in the $(x,c)$ representation. The \textit{I}-\textit{N} transition is the only type of transition which we find for this size ratio. The FMT results show that the \textit{I}-\textit{N} transition concentrations at $x=0$ are $c_{I}=0.418$ and $c_{N}=0.46$ respectively, in agreement with the monodisperse results \cite{reich.h:2007.a}.  At $x=1$, the coexistence values are $c_{I}=0.418/\lambda^{3}=0.314$ and $c_{N}=0.46/\lambda^{3}=0.346$. The coexistence curves interpolate smoothly from $x=0$ to $x=1$. The tie lines joining coexisting isotropic and nematic phases are naturally vertical at $x=0,1$ (corresponding to the cases of the pure systems of small and big platelets, respectively) whereas between $x=0$ and $x=1$ they vary in gradient, leaning with large positive gradient as $x$ increases from zero composition, less so at about 50\% composition and then more so again on approaching $x=1$. Therefore, there is stronger fractionation at 50\% composition than towards $x=0$ and $x=1$, such that the isotropic phase is dominated by particles of species 1 (smaller species) and the nematic phase by particles of species 2 (larger species). The biphasic region, where there is coexistence between the isotropic and nematic phase, is very narrow in FMT. At zero composition, Onsager theory predicts a density jump of $\thicksim22\%$ whereas this is only $\thicksim9\%$ for FMT, which agrees more closely to simulation results of 8\% \cite{eppenga.r:1984.a}. The bifurcation concentration for FMT is $c_{*}=0.434$ at $x=0$ and $c_{*}=0.434/\lambda^{3}=0.326$ at $x=1$. The spinodal for FMT lies closer to the isotropic phase boundary than the nematic phase boundary (whereas the converse is true for Onsager theory). Fig.~1b shows the same results but in the $(c_{1},c_{2})$ representation. The results for both theories again interpolate smoothly between the two pure limits. The two branches of the binodal in this representation move from the $c_{1}$-axis to the $c_{2}$-axis with increasing composition. Hence the tie lines move from being horizontal on the $c_{1}$-axis to vertical on the $c_{2}$-axis. In the $(x,p^{*})$ representation (Fig.~1c) the tie lines are horizontal due to the requirement of equal pressure in the coexisting phases. The spinodal lies above the nematic branch of the binodal in this representation because it is obtained by inserting the bifurcation densities into the isotropic equation of state, which yields a higher pressure than the coexistence value. The binodal obtained from FMT is located at significantly smaller densities as compared to that from Onsager theory. The isotropic end of the tie line is at a lower composition than the nematic end of the tie line. For Onsager theory at zero composition, the isotropic branch of the binodal intersects the $c$-axis at $c_{I}=0.666$ and the nematic branch intersects at $c_{N}=0.849$, in agreement with the monodisperse limit found in earlier work \cite{esztermann.a:2006.a}. The binodals interpolate smoothly from $x=0$ to $x=1$ where the values of $c$ corresponding to the \textit{I}-\textit{N} coexistence concentrations are $c_{I}=0.5$ and $c_{N}=0.638$.  The nature of the tie lines is similar to FMT. The \textit{I}-\textit{N} spinodal lies between the two branches of the binodal and interpolates smoothly from zero composition, where $c=8/\pi^{2}=0.811$ to $x=1$ where $c_{*}=0.609$. 

The nematic phase of a mixture of two components can be characterised by two partial nematic order parameters, $S_{1}$ and $S_{2}$ defined by
\begin{equation}
\label{eq: order_parameter}
S_{i} = 4\pi \int_{0}^{\frac{\pi}{2}}d\theta\sin\theta\Psi_{i}(\theta)P_{2}(\cos\theta).
\end{equation}
The total nematic order parameter is the weighted average
\begin{equation}
S_{\textrm{tot}}= (1-x)S_{1}+xS_{2}.
\end{equation}
\label{eq: total_order}
Here we investigate the behaviour of these quantities at \textit{I}-\textit{N} and \textit{N}-\textit{N} coexistence. We have chosen $S_{\textrm{tot}}$ to be a simple weighted average of
the $S_{i}$. As such, each particle contributes to $S_{\textrm{tot}}$
independently of its size. Of course there are other
suitable choices which may be appropriate to certain applications; the
$S_{i}$ in the sum for the total order may be weighted by the surface
area of the platelet, for example. Due to the absence of particles of species 2 at $x=0$, $S_{\textrm{tot}}$ and $S_{1}$ take the same value. Similarly at $x=1$, $S_{\textrm{tot}}$ and $S_{2}$ take the same value. As $x$ increases from $0$ to $1$, $S_{\textrm{tot}}$ changes smoothly. In Fig.~1d the FMT values for $S_{1}$, $S_{2}$ and $S_{\textrm{tot}}$ are smaller than those obtained from Onsager theory, which places the values much higher. However FMT predicts that the difference between $S_{1}$ and $S_{2}$ to be bigger than Onsager theory does. $S_{\textrm{tot}}$ at $x=0$ takes the same value $S_{\textrm{tot}}=0.531$ as previous work (and agrees well with simulation \cite{reich.h:2007.a}) in the monodisperse case and the the order parameters vary smoothly in the same manner as for Onsager theory.

In Fig.~2 we show the results for $\lambda=1.4$. There is a widening of the biphasic region between the two pure components, clearly seen in Fig.~2a where we show the phase diagram in the $(x,c)$ representation. FMT again predicts that the mixture undergoes \textit{I}-\textit{N} phase separation at lower densities than Onsager theory. The tie lines become less steep with a smaller positive gradient than for $\lambda=1.1$ for intermediate values of composition indicating that there is larger difference in mole fraction between coexisting isotropic and nematic states. The widening of the biphasic gap is even more noticeable in the ($c_{1},c_{2}$) representation (Fig.~2b), especially for the case of Onsager theory. In the $(x,p^{*})$ representation (Fig.~2c), the spinodal again lies above the nematic binodal for Onsager theory. However, for FMT we see that while the spinodal is above the nematic branch of the binodal close to $x=0$ and $x=1$ the curve enters the biphasic region in between about $x=0.1$ and $x=0.6$.  $S_{1}$ is smaller than $S_{2}$ and their difference has increased for both theories, with FMT still possessing the larger difference suggesting that the particles of species 2 are significantly more ordered than species 1 for a given mole fraction at coexistence. 

In Fig.~3 we plot the graphs for $\lambda=1.7$. The \textit{I}-\textit{N} biphasic gap becomes even more pronounced. The results for FMT (Fig.~3a) show that tracing along the nematic branch of the binodal as $x$ increases leads to an increase in $c$. At approximately $x=0.2$, $c$ then decreases and near $x=0.6$ bends back on itself before approaching $x=1$. This bending of the binodal constitutes a re-entrant phenomenon. There is also a large range of compositions between about $x=0.1$ and $x=0.6$ for which the biphasic \textit{I}-\textit{N} phase overlaps for both theories, which is also seen clearly in the ($c_{1},c_{2}$) representation (Fig.~3b). In the $(x,p^{*})$ representation (Fig.~3c) both theories predict a region of composition values for which the spinodal lies inside the biphasic region. For illustration of the re-entrant part of the phase diagram one should keep a constant fluid composition near $x=0.6$ but increases the pressure from $p^{*}=0$ to $p^{*}=1$: the state changes from \textit{I} $\rightarrow$ \textit{I}+\textit{N$_{2}$} $\rightarrow$ \textit{N} $\rightarrow$ \textit{I}+\textit{N$_{2}$} $\rightarrow$ \textit{N} where \textit{N$_{2}$} is a nematic phase composed mostly of particles of species 2. The order parameters along the nematic branch of the binodal are shown in Fig.~3d. $S_{1}$ remains lower than $S_{2}$ for a given composition, again suggesting that the particles of species 2 are more ordered than those of species 1 at coexistence. When the re-entrant feature occurs at about $x=0.6$, $S_{1}$, $S_{2}$ and $S_{\textrm{tot}}$ drop sharply. The difference between $S_{1}$ and $S_{2}$ along the nematic branch of the binodal has become much larger than for $\lambda=1.4$. 
 
In Fig.~4 we present the phase diagrams for $\lambda=2$. The most striking feature after increasing the size ratio to $\lambda=2$ is the stable \textit{N}-\textit{N} coexistence region. For FMT the $(c_{1},c_{2})$ representation (Fig.~4a) reveals that this region is in the form of an almost symmetric hump suggesting that the fractionation between the two distinct nematic phases becomes less with increasing density. There is also a triple point which is in the form of a triangle connecting a low composition isotropic phase, a nematic phase composed mostly of particles of species 1 (\textit{N$_{1}$}) and a nematic phase composed mostly of particles of species 2 (\textit{N$_{2}$}). The symmetric \textit{N}-\textit{N} coexistence region is more clearly seen in the $(x,p^{*})$ representation (Fig.~4c) between $x$ just greater than 0 and $x=0.45$ ending in a critical point at $(x,p^{*})=(0.19,2.72)$. In the $(x,p^{*})$ representation the triple point collapses onto a line and the pressure at the triple point is approximately $p^{*}=1.75$. The \textit{I}-\textit{N} biphasic region has become increasingly pronounced. The topology of the phase diagrams is the same as that obtained in Onsager theory results (Fig.~4b and Fig.~4d). The re-entrant feature obtained from Onsager theory is less pronounced than in FMT. The \textit{N}-\textit{N} phase separation occurs over a larger range of composition values (up to near $x=0.5$) and the triple point occurs at approximately a unit of reduced pressure higher than the triple point predicted by FMT, nevertheless, as we emphasise, the topologies predicted by both theories are the same. The \textit{I}-\textit{N} spinodal enters the biphasic region for both theories. We postpone the discussion of the order parameters for $\lambda=2$ and higher size ratios until the end of this section.

In Fig.~5 we show the results for $\lambda=2.5$. There is \textit{N}-\textit{N} demixing in the system, as is observed for $\lambda=2$, however there is a big difference in the toplogy of the phase diagrams in that the demixing now does not end in a critical point. For FMT in the ($c_{1},c_{2}$) representation (Fig.~5a) the \textit{N}-\textit{N} region opens up and the two branches of the binodal extend outwards suggesting that the demixing extends to arbitrarily high density. No critical point is observed up to the densities we examine; we follow the phase boundaries up to $c=1$ for FMT and $c=1.9$ for Onsager theory. A similar splaying of the phase boundaries is observed in Fig.~5b in the ($c_{1},c_{2}$) for Onsager theory, suggesting that the topology obtained with FMT is correct. Also, the \textit{I}-\textit{N}$_{1}$ region has shrunk to a tiny region close to zero composition, making the triple region mostly dominated by the nematic state rich in species 2. In the $(x,p^{*})$ representation for FMT (Fig.~5c) the re-entrant feature has become extremely pronounced, extending as low as approximately $x=0.5$ in the $(x,p^{*})$ representation and the triple point occurs at approximately $p^{*}=1.8$. In the ($x,p^{*}$) representation for Onsager theory (Fig.~5d), the re-entrant feature has also become more pronounced but less so than for FMT. The triple line extends over approximately the same composition range ($x \leq 0.8$) as for FMT but occurs at just over a unit of reduced pressure higher than for FMT. 

In Fig.~6 we present the phase diagrams for $\lambda = 4$. In the $(c_{1},c_{2})$ representation for FMT (Fig.~6a) the strength of the \textit{I}-\textit{N}$_{2}$ fractionation effect becomes very large. The \textit{N}-\textit{N} separation is very wide giving a huge immiscibility gap, hence we do not show it. Fig.~6b also shows a large immiscibility gap in the ($c_{1},c_{2}$) representation for Onsager theory. We again have confidence that the results from FMT are quantitatively more accurate than those from Onsager theory. At concentrations $c_{1},c_{2}<0.1$ the coexisting compositions already approach closely the $c_{1},c_{2}$ axes highlighting that coexisting compositions are close to $x=0,1$ as is clearly seen in the $(x,p^{*})$ representation (Fig.~6c for FMT and Fig.~6d for Onsager theory). The general trend of the re-entrant bend moving to lower composition with increasing $\lambda$ has continued here, reaching as low as about $x=0.3$ for FMT and near $x=0.5$ for Onsager theory. 

In Fig.~7 we plot the results for $\lambda=5$. In the ($c_{1},c_{2}$) representation for FMT (Fig.~7a) the \textit{I}-\textit{N}$_{2}$ coexistence region is very pronounced. The larger platelets have an area twenty-five times that of the smaller platelet so the asymmetry of the mixture is large. The \textit{I}-\textit{N} spinodal remains close to the isotropic branch of the binodal; more so than in Fig.~7b, the ($c_{1},c_{2}$) representation of Onsager theory. The $(x,p^{*})$ representation for FMT (Fig.~5c) shows that the general trend of the re-entrant bend moving to lower composition has continued, here reaching as low as about $x=0.2$ for FMT. There is a very narrow region of coexistence between the isotropic and nematic state of species 2 up to about $p^{*}=0.06$, suggesting that up to this pressure there is only a small fractionation effect. This fractionation becomes wider on increasing the pressure beyond this point, with the phase boundaries almost reaching $x=0,1$ by $p^{*}=0.1$. The re-entrant bend is also a dominating feature of the phase diagram in Fig.~7d for Onsager theory suggesting the FMT phase behaviour is correct, however it is not as pronounced as for FMT and the narrow \textit{I}-\textit{N}$_{2}$ \textit{handle} present at low pressures for FMT is wider for Onsager theory.

In Fig.~8 we show the partial nematic order parameters $S_i$
        in the coexisting nematic phase(s) for $\lambda = 2, 2.5, 4$ and 
        $5$, as obtained from FMT. For $\lambda=2$ (Fig. 8a) $S_1$ and
        $S_2$ both rise rapidly near zero composition to near
        unity. $S_1$ takes on smaller values than $S_2$ for a given
        mole fraction as has already been observed for the lower size ratios. Hence the (smaller)   
        particles of species 1 are much less ordered than particles of species 2 at a given
        composition, since the particles of species 2 are high in
        number and there is more freedom for the smaller particles to
        rotate in the dense system. Both partial order parameters
        remain close to unity as the composition $x$ is increased from
        0 to about 0.5.  The system is at \textit{N}-\textit{N}
        coexistence over this range (see Fig. 4c for the corresponding
        phase diagram). $S_{\textrm{tot}}$ reaches as low as about 0.5
        and $S_{1}$ becomes as small as 0.2 at $x=1$. For $\lambda=
        2.5$ (Fig. 8b) the \textit{N}-\textit{N} coexistence does not
        end at a critical point, leaving an interval (in composition)
        that the curves $S_i(x)$ do not enter. For $\lambda=4$
        (Fig. 8c) $S_{\textrm{tot}}$ varies between about 0.3 and
        almost 1. This range of $S_{\textrm{tot}}$ increases further
        for $\lambda=5$ (Fig. 8d) where it reaches as low as about
        0.2. For $\lambda=4,5$, $S_{\textrm{tot}}$ also increases more
        rapidly as the composition increases than is the case for the
        lower size ratios. At high compositions, the (small) particles
        of species 1 possess very low nematic order, reaching less
        than 0.1 due to the larger size difference between the two
        species.
        
Our phase diagrams share several features with those reported in
     previous studies of binary mixtures of anisometric hard core
     particles. Widening of the \textit{I}-\textit{N} phase coexistence region upon
     increasing the bidispersity parameter was found in binary
     mixtures of thick and thin rods \cite{vanroij.r:1998.a}, long and
     short rods \cite{lekkerkerker.hnw:1984.a}, mixtures of rods and
     platelets \cite{wensink.hh:2001.b}, as well as in binary mixtures
     of platelets using both the Parons-Lee scaling of the Onsager
     functional \cite{wensink.hh:2001.a, wensink.hh:2004.a} and the Zwanzig model
     \cite{harnau.l:2002.c, harnau.l:2002.d}. \textit{N}-\textit{N} phase coexistence
     ending in an upper critical point, as we find for an intermediate
     range of values of size ratios $\lambda$, occurs in certain regimes for binary mixtures of rods and
     platelets. A notable difference to other studies of platelet
     mixtures \cite{wensink.hh:2004.a} is that we do not find \textit{N}-\textit{N} coexistence
     ending in a lower critical point, at least not for the range of
     densities and size ratios that we explored. Similar results to
     ours, where \textit{N}-\textit{N} coexistence ends in an upper critical point were
     obtained using Onsager theory for mixtures of thick and thin rods
     \cite{vanroij.r:1998.a}. This system of rods also displays
     isotropic-isotropic phase coexistence at high enough values of
     the diameter ratio, which we do not find to be stable in the
     present study.
     
\section{Conclusions and Outlook}

\label{outlook}
We have studied the bulk phase behaviour of binary mixtures of hard platelets, including isotropic and nematic states. The platelets are assumed to have circular shape and vanishing thickness. We have not considered positionally ordered phases such as columnar and crystalline phases, which are not expected to occur for the present model of particles with zero volume and zero packing fraction. Note that a first order \textit{N}-\textit{Col} transition was found at non-zero packing fraction in the limit of vanishing thickness \cite{bates.m:1998.a}; simulations in this limit are possible because the model can be mapped onto a system of particles with finite volume but variable shape. Also, platelets with non-zero thickness exhibit a \textit{N}-\textit{Col} transition \cite{wensink.h:2009.a}; this was also reported for the Zwanzig model \cite{bier.m:2004.a}.

For a variety of size ratios, we have compared the results of Onsager theory and FMT. In the monodisperse limit we find that Onsager theory overestimates the \textit{I}-\textit{N} transition densities and predicts a larger biphasic gap than FMT; the results for FMT in this limit compare quantitatively well with those from simulation. Both theories predict a first order \textit{I}-\textit{N} phase transition. We expect that Onsager theory overestimates the density jump at coexistence and also overestimates the size of the biphasic gap. The FMT results show the appearance of re-entrant phenomena at a lower size ratio than for Onsager theory. FMT results also show a larger range of compositions for which \textit{N}-\textit{N} demixing at a given size ratio occurs. The \textit{N}-\textit{N} demixing occurs at $\lambda = 2$, where it is closed by an upper critical point. For $\lambda \geq 2.5$ there is no critical point up to the densities we consider. In an experimental system of platelets with nonzero volume, one would expect a positionally ordered phase to be favoured before any possible remixing into a homogeneous nematic phase. We also examine the degree of nematic ordering along the phase boundaries for a selection of size ratios up to $\lambda = 2$ where \textit{N}-\textit{N} demixing ends in an upper critical point. The partial nematic order parameters $S_{1}$, $S_{2}$ vary smoothly with increasing composition and hence so does $S_{\textrm{tot}}$. Where a re-entrant feature occurs, all three order parameters along the nematic branch of the binodal drop sharply. As $\lambda$ increases, $S_{2}$ becomes smaller at $x=1$, reaching $<0.2$ for $\lambda = 2$. We do not find any stable \textit{I}-\textit{I} demixing for the range of $\lambda$ in this investigation.

Through the range of $\lambda$ values considered, a striking observation is that whilst FMT predicts the occurence of phase boundaries at locations quantitatively different from Onsager theory, the topology of the phase diagrams for a particular choice of $\lambda$ is the same as for Onsager theory. Coupled with the fact that the \textit{I}-\textit{N} transition for the monodisperse case agrees well with simulation results \cite{reich.h:2007.a}, we gain confidence that the phase diagrams predicted by FMT are quantitatively more reliable since we assume that Onsager theory predicts the correct physics qualitatively. Furthermore, we have confidence that the order parameter profiles predicted by FMT are closer to those that would be observed from simulation studies, since the monodisperse limit in the present theory yields results for the \textit{I}-\textit{N} transition and nematic order parameter that is in good agreement with simulation results. Whether the differences to the true transition densities for the case of binary mixtures increases remains to be seen. 

The results from Onsager theory cannot be obtained by some scaling of the results that exist for binary mixtures of rods (thick and thin \cite{vanroij.r:1998.a} or long and short \cite{lekkerkerker.hnw:1984.a}). For example, for there to be a mapping between thick and thin rods \cite{vanroij.r:1998.a} and the present system, we would require simultaneously that $(1+d)=\lambda^{2}+\lambda$ and $d=\lambda^{3}$, where $d=D_{2}/D_{1}>1$ is the diameter ratio of the thick and thin rods. These requirements are obtained from consideration of the free energy for thick and thin rods (compare Eq.~(2) of Ref.~\cite{vanroij.r:1998.a} with Eq.~(\ref{eq: free_energy_2}) of the present investigation). Clearly these conditions cannot be satisfied simultaneously except for the trivial case $\lambda=d=1$. In the monodisperse limit, the mapping of concentrations from rods to platelets is $c_{\textrm{rod}}=(\pi^{2}/2)c_{\textrm{platelet}}$ where $c_{\textrm{rod}}=(\pi/4)L^{2}D$ is the conventional dimensionless concentration for rods ($L$ is the rod length, $D$ is the rod diameter). Clearly, simulation results for the systems studied in the present paper are most desirable. The predictions of the present work could also be tested experimentally. For example, for gibbsite (having typical experimental radius of approximately 100nm) and hydrotalcite (typically between about 25 and 75 nm) \cite{maurice_thesis} one has values of $\lambda$ between $1.3$ and $4$, which are in the range of the present work. We have taken mean radii here. Of course in experiments there will be effects due to polydispersity and our model ignores effects of finite platelet thickness.

Interesting future work could involve examining the phase behaviour of binary mixtures of polarizable platelets, that is particles that interact with some external applied magnetic field, as has been investigated for binary rod mixtures \cite{varga.s.:2000.a, dobra.s:2006.a}. Also, binary mixtures of rods and platelets could be investigated \cite{vanderkooij.f.m.:2000.b, varga.s.:2002.a, varga.s.:2002.b, varga.s.:2002.c, wensink.h.h.:2002.b}. Another line of investigation would be to incorporate polydispersity into the model since many platelet systems often have a significant polydispersity, see for example \cite{vanderkooij.f.m:2000.a}. This would lead to an extension of the work by Speranza and Sollich and others on rod-like particles, \cite{speranza.a:2002.a, speranza.a:2003.a, speranza.a:2003.b} and by Wensink and Vroege on thickness-polydisperse platelets \cite{wensink.h.h:2002.a}.
\\
\appendix{
\section{Calculation of Relevant Integrals}}
Here we evaluate some integrals that are important in the bifurcation analysis. The first, $\mathcal{I}_{1}$, is a standard integral in excluded volume calculations
\begin{equation}
\mathcal{I}_{1}=\int d \boldsymbol{\omega} \int d \boldsymbol{\omega}' \sin \gamma=\int d \boldsymbol{\omega} \int d \boldsymbol{\omega}' \sqrt{1-(\boldsymbol{\omega}\cdot\boldsymbol{\omega}')^{2}}.
\end{equation}
Without loss of generality, let $\boldsymbol{\omega}'=\boldsymbol{e_{z}}$ in the integrand, where $\boldsymbol{e_{z}}$ is the unit vector in the $z$-direction (the symmetry axis for the system). Now we have
\begin{equation}
\mathcal{I}_{1}=\int d \boldsymbol{\omega}' \int d \boldsymbol{\omega} \sqrt{1-(\boldsymbol{\omega}\cdot \boldsymbol{e_{z}})^{2}}.
\end{equation}
Remembering that in our coordinate system $\boldsymbol{\omega}'=(\cos\theta'\sin\theta',\sin\phi'\sin\theta',\cos\theta')$,
\begin{align}
\mathcal{I}_{1}&=\int d \boldsymbol{\omega}'  \int d \boldsymbol{\omega}\sqrt{1-\cos^{2}\theta} 
\notag \\ &=\int d \boldsymbol{\omega}'  \int_{0}^{2\pi}d\phi \int_{0}^{\pi}d \theta\sin^{2} \theta\notag \\ &=\pi^{2}\int d \boldsymbol{\omega}' = 4\pi^{3}
\end{align}
as required.
The second integral arises in FMT. We wish to integrate the triple scalar product over three unit orientation vectors, $\boldsymbol{\omega},\boldsymbol{\omega}',\boldsymbol{\omega}''.$
\begin{equation}
\mathcal{I}_{2}=\int d \boldsymbol{\omega} \int d \boldsymbol{\omega}' \int d \boldsymbol{\omega}'' | \boldsymbol{\omega} \cdot ( \boldsymbol{\omega}' \times \boldsymbol{\omega}'')|.
\end{equation}
Similar to the $\mathcal{I}_{1}$ calculation, let $\boldsymbol{\omega}''=\boldsymbol{e_{z}}$ in the integrand.
\begin{equation}
\mathcal{I}_{2}=\int d \boldsymbol{\omega}'' \int d \boldsymbol{\omega} \int d \boldsymbol{\omega}' | \boldsymbol{\omega} \cdot ( \boldsymbol{\omega}' \times \boldsymbol{e_{z}})|.
\end{equation}
Remembering that in our coordinate system $\boldsymbol{\omega} = (\sin \theta,0,\cos\theta)$ and $\boldsymbol{\omega}'=(\cos\theta'\sin\theta',\sin\phi'\sin\theta',\cos\theta')$,
\begin{align}
\mathcal{I}_{2}&=\int d \boldsymbol{\omega}'' \int d \boldsymbol{\omega} \int d \boldsymbol{\omega}' \left| \left( \begin{array}{ccc}\sin\theta\\0\\\cos\theta\end{array}\right)  \cdot  \left( \begin{array}{ccc}\sin\phi' \sin\theta '\\-\cos\theta' \sin \theta '\\0\end{array}\right)  \right|
\notag \\ &=\int d \boldsymbol{\omega}''  \int d \boldsymbol{\omega} \int d \boldsymbol{\omega}' |\sin\theta \sin\theta ' \sin \phi '|
\notag \\ &=\int d \boldsymbol{\omega}'' \cdot 2 \pi^{3}=8\pi^{4}
\end{align}
as required. (Here the inner two integrals alone give $\pi^{3}/2$ which can be found straightforwardly).
There are three further integrals required for the bifurcation analysis.
$\mathcal{I}_{3}= \int d \boldsymbol{\omega} \left[ P_{2}(\cos\theta)    \right]^{2}=4\pi/5$ trivially.
Let us turn our attention to 
\begin{equation}
\label{eq:I4}
\mathcal{I}_{4}= \int d \boldsymbol{\omega} \int d \boldsymbol{\omega}' \sin\gamma  \left[ P_{2}(\cos\theta)P_{2}(\cos\theta')    \right].
\end{equation}
We expand $\sin \gamma$ as
\begin{equation}
\sin \gamma =\sum_{n=0}^{\infty}c_{2n}P_{2n}(\cos \gamma)
\end{equation}
with coefficients
\begin{equation}
c_{2n}=\frac{-\pi(4n+1)(2n-3)!!(2n-1)!!}{2^{n+1}(n+1)!},
\end{equation}
where the double factorial is defined by
\begin{equation}
\label{eq:double_factorial}
{n!!\equiv\begin{cases}n\cdot(n-2)\cdots5\cdot3\cdot1&\text{$n>0$ odd}  \\n\cdot(n-2)\cdots6\cdot4\cdot2&\text{$n>0$ even} \\1&\text{$n=-1,0$.} \end{cases}}
\end{equation}
Therefore the inner integral of (\ref{eq:I4}) becomes
\begin{align}
\int d \boldsymbol{\omega}' & \sin\gamma P_{2}(\cos\theta ') = \int d \boldsymbol{\omega}'\sum_{n=0}^{\infty}c_{2n}P_{2n}(\cos\gamma) P_{2}(\cos \theta ')
\notag \\ &=\int_{0}^{2\pi}d \phi ' \int_{0}^{\pi}d \theta ' \sin \theta '\sum_{n=0}^{\infty}c_{2n}P_{2n}(\cos\gamma) P_{2}(\cos \theta ').
\end{align}
We now utilise the addition formula for Legendre polynomials,
\begin{align}
P_{2n}(\cos \gamma)& =P_{2n}(\cos \theta)P_{2n}(\cos \theta ')  +2\sum_{m=1}^{2n}\frac{(2n-m)!}{(2n+m)!}\notag \\ &\times P_{2n}^{m}(\cos \theta) P_{2n}^{m}(\cos \theta ')\cos m(\phi-\phi ').
\end{align}
Hence
\begin{align}
\int d \boldsymbol{\omega}' \sin \gamma & P_{2}(\cos \theta ') =  2\pi \int_{0}^{2\pi}\sin \theta ' d \theta ' \notag \\ &\times \sum_{n=0}^{\infty}c_{2n}P_{2n}(\cos \theta)P_{2n}(\cos \theta ')P_{2}(\cos \theta)
\end{align}
where the sum involving associated Legendre functions of the first kind vanish in the integration over $\phi '$.
Now we introduce the notation
\begin{equation}
\langle P_{2n}\rangle_{f} = \int_{0}^{\pi}d \theta ' f(\theta ')P_{2n}(\cos \theta ').
\end{equation}
So
\begin{equation}
\int d \boldsymbol{\omega}' \sin \gamma P_{2}(\cos \theta ') =  2\pi \sum_{n=0}^{\infty} c_{2n}P_{2n}(\cos \theta)\langle P_{2n}\rangle_{f}   \end{equation}
with $f=P_{2}(\cos \theta ')$. From the orhogonality conditions of Legendre polynomials \cite{abramowitz.m:1965.a} we have that $\langle P_{2n}\rangle_{f}=2/(2m+1)\delta_{mn}$ where $\delta_{mn}$ is the Kronecker-$\delta$ symbol.
Here, $m=2$ and, given that $c_{2} = -5\pi/32$, we have
\begin{equation}
\int d \boldsymbol{\omega}' \sin\gamma P_{2}(\cos\theta ')= 2\pi c_{2} P_{2}(\cos \theta)\cdot \frac{2}{5}=-\frac{\pi^{2}}{2} P_{2}(\cos \theta).
\end{equation}
Hence
\begin{align}
\mathcal{I}_{4}&=\int d \boldsymbol{\omega}P_{2}(\cos \theta)\left[-\frac{\pi^{2}}{2}P_{2}(\cos \theta) \right]
\notag \\ & =-2\pi^{3}\int_{0}^{\frac{\pi}{2}}d \theta \sin \theta [P_{2}(\cos \theta)]^{2} = -2\pi^{3}\cdot \frac{1}{5} = -\frac{\pi^{3}}{10}
\end{align}
as required.
The last integral to consider is
\begin{align}
\label{eq:I5}
\mathcal{I}_{5} &= \int d \boldsymbol{\omega} \int d \boldsymbol{\omega}' \int d \boldsymbol{\omega}'' | \boldsymbol{\omega} \cdot ( \boldsymbol{\omega}' \times \boldsymbol{\omega}'')|\big[ P_{2}(\cos\theta)P_{2}(\cos\theta') \notag \\ &+ P_{2}(\cos\theta)P_{2}(\cos\theta'')+P_{2}(\cos\theta')P_{2}(\cos\theta'')   \big].
\end{align}
We consider integrating the first term of (\ref{eq:I5}), namely
\begin{align}
\label{eq:I5_tilde}
\tilde{\mathcal{I}_{5}} &= \int d \boldsymbol{\omega} \int d \boldsymbol{\omega}' \int d \boldsymbol{\omega}'' | \boldsymbol{\omega} \cdot ( \boldsymbol{\omega}' \times \boldsymbol{\omega}'')| P_{2}(\cos\theta)P_{2}(\cos\theta').
\end{align}
We let $\boldsymbol{\omega}'' = \boldsymbol{e_{z}}$ in the integrand so that 
\begin{align}
\tilde{\mathcal{I}_{5}} &= \int d \boldsymbol{\omega}'' \int d \boldsymbol{\omega} \int d \boldsymbol{\omega}' | \sin \gamma  (\hat{\textbf{n}} \cdot \boldsymbol{e_{z}})| P_{2}(\cos\theta)P_{2}(\cos\theta'),
\end{align}
where $\gamma$ is the angle between $\boldsymbol{\omega}$ and  $\boldsymbol{\omega}'$ and $\hat{\textbf{n}}$ is a unit vector perpendicular to both $\boldsymbol{\omega}$ and  $\boldsymbol{\omega}'$. Since $\boldsymbol{e_{z}}$ is fixed, $\hat{\textbf{n}} \cdot \boldsymbol{e_{z}}= \sin\theta ''$. Hence
\begin{align}
\tilde{\mathcal{I}_{5}} &= \int d \boldsymbol{\omega}'' \sin\theta'' \int d \boldsymbol{\omega} \int d \boldsymbol{\omega}'  \sin \gamma P_{2}(\cos\theta)P_{2}(\cos\theta') \notag \\ & = \int d \boldsymbol{\omega}'' \sin\theta'' \mathcal{I}_{4}  = 4\pi \cdot \left(-\frac{\pi^{3}}{10}\right)=-\frac{\pi^{4}}{5}.
\end{align}
$\mathcal{I}_{5}$ comprises two other similar terms so by symmetry, $\mathcal{I}_{5}=3\tilde{\mathcal{I}_{5}}$. Hence $\mathcal{I}_{5} =-3\pi^{4}/5$ as required.
\newline
\begin{acknowledgments}

We thank Chris Newton and Susanne Klein and the Liquid Crystals group of HP Labs, Bristol for useful discussions. We thank Peter Sollich for useful discussion regarding the structure of the third order term in the binary functional. Bob Evans, Paul Hopkins, Hendrik Reich, Robert Hales and Tom Smith are thanked for valuable comments on the manuscript. Financial support from the EPSRC and HP Labs, Bristol is gratefully acknowledged as well as through the SFB840/A3 of the DFG.
\end{acknowledgments}

\begin{widetext}

\begin{figure}
\centering 
\subfigure{\includegraphics[width=0.45\textwidth,clip]{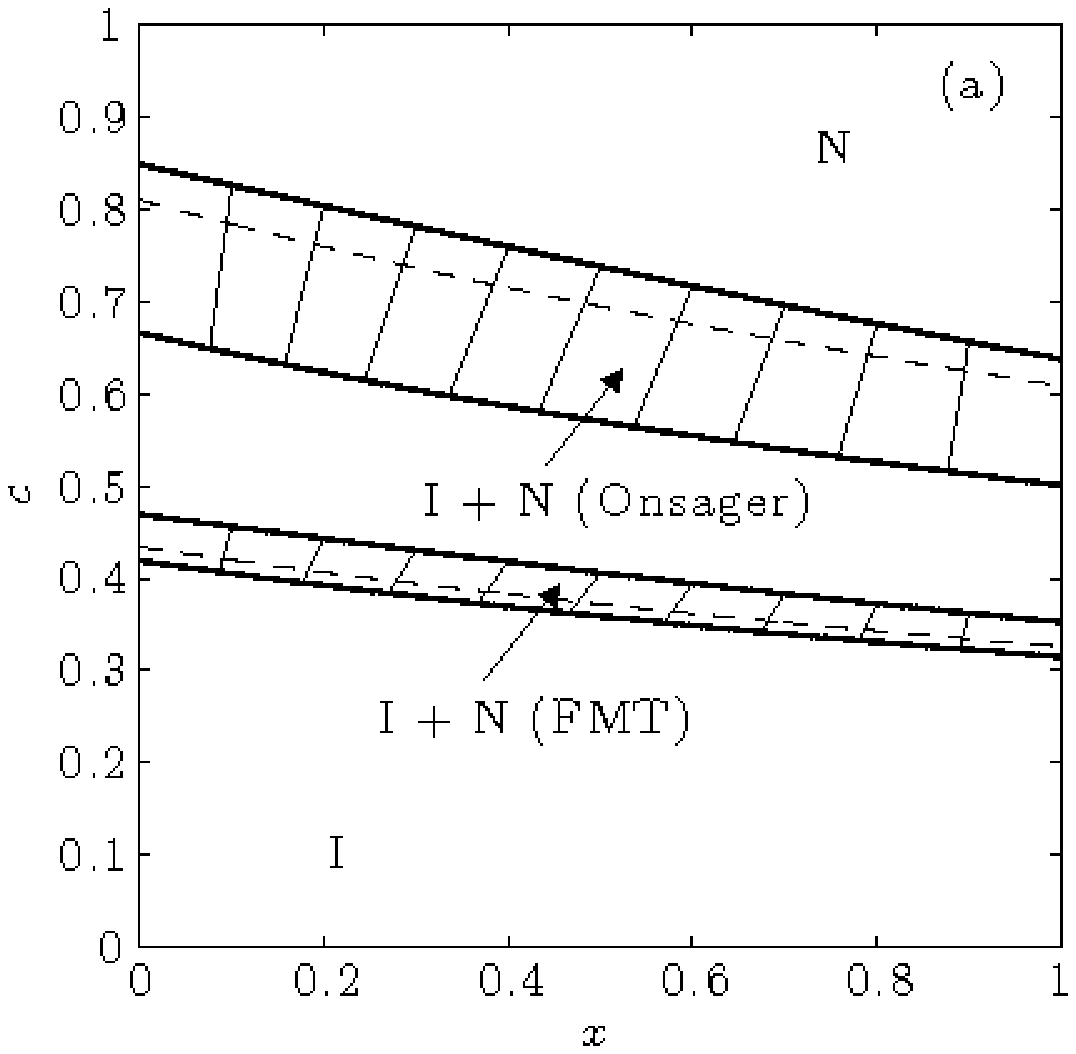}}
\subfigure{\includegraphics[width=0.45\textwidth,clip]{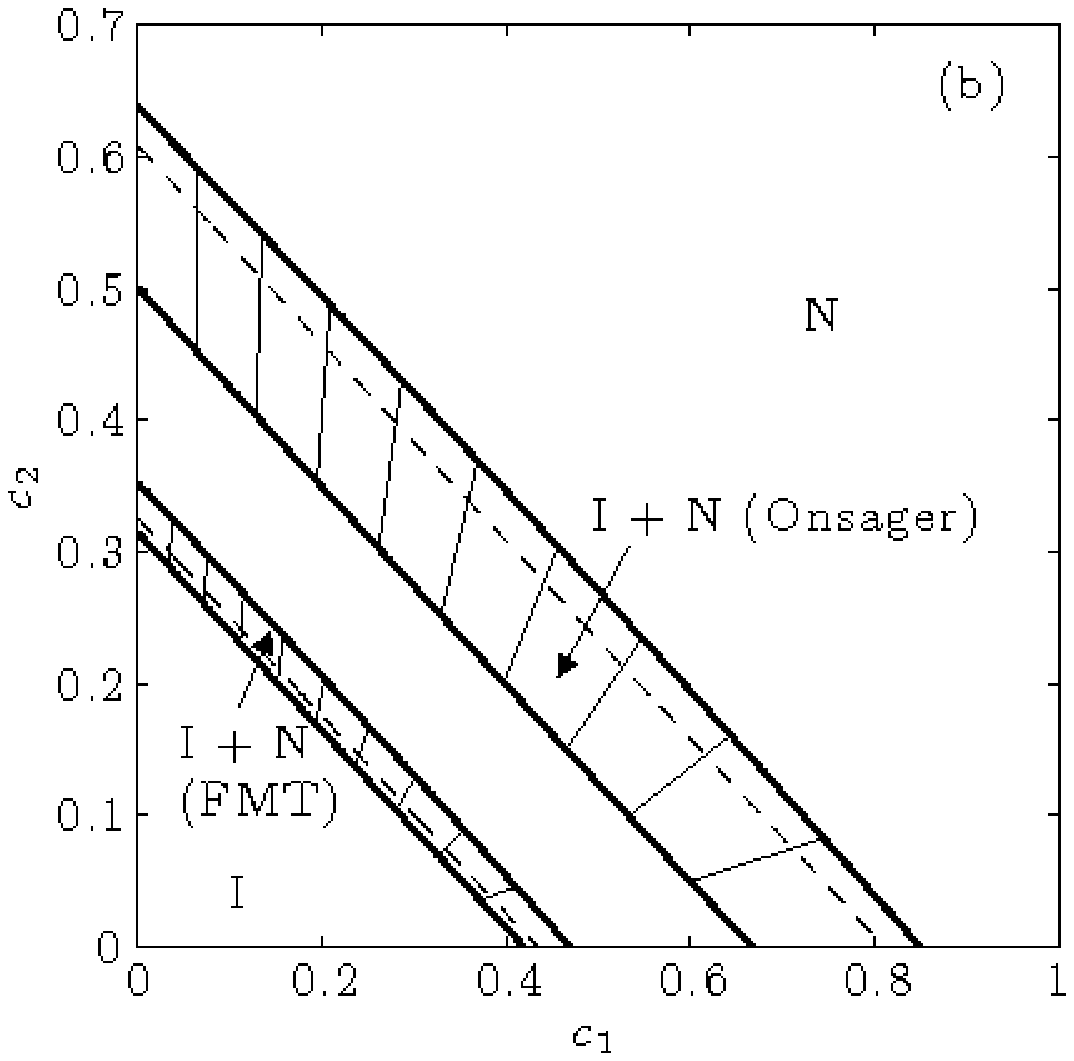}}
\subfigure{\includegraphics[width=0.45\textwidth,clip]{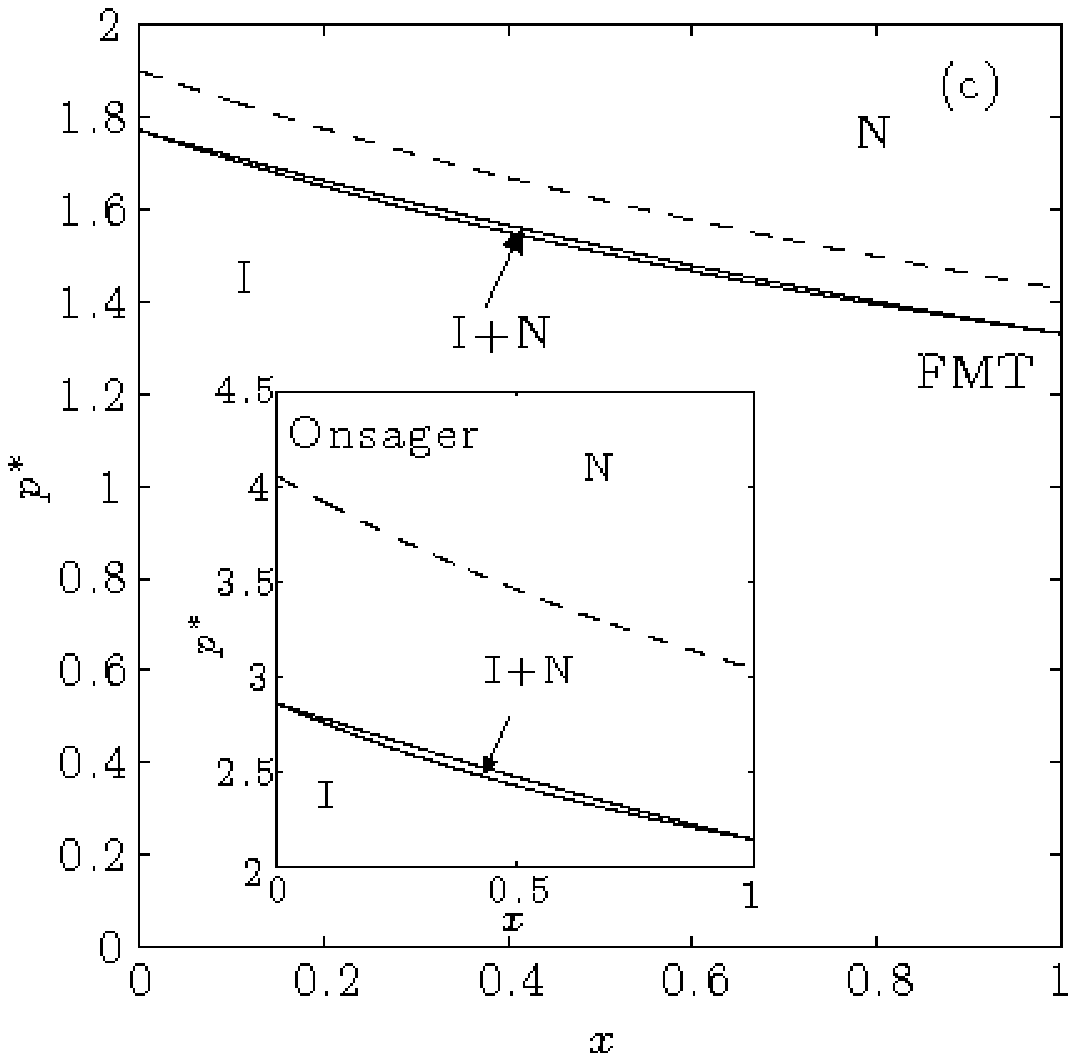}}
\subfigure{\includegraphics[width=0.45\textwidth,clip]{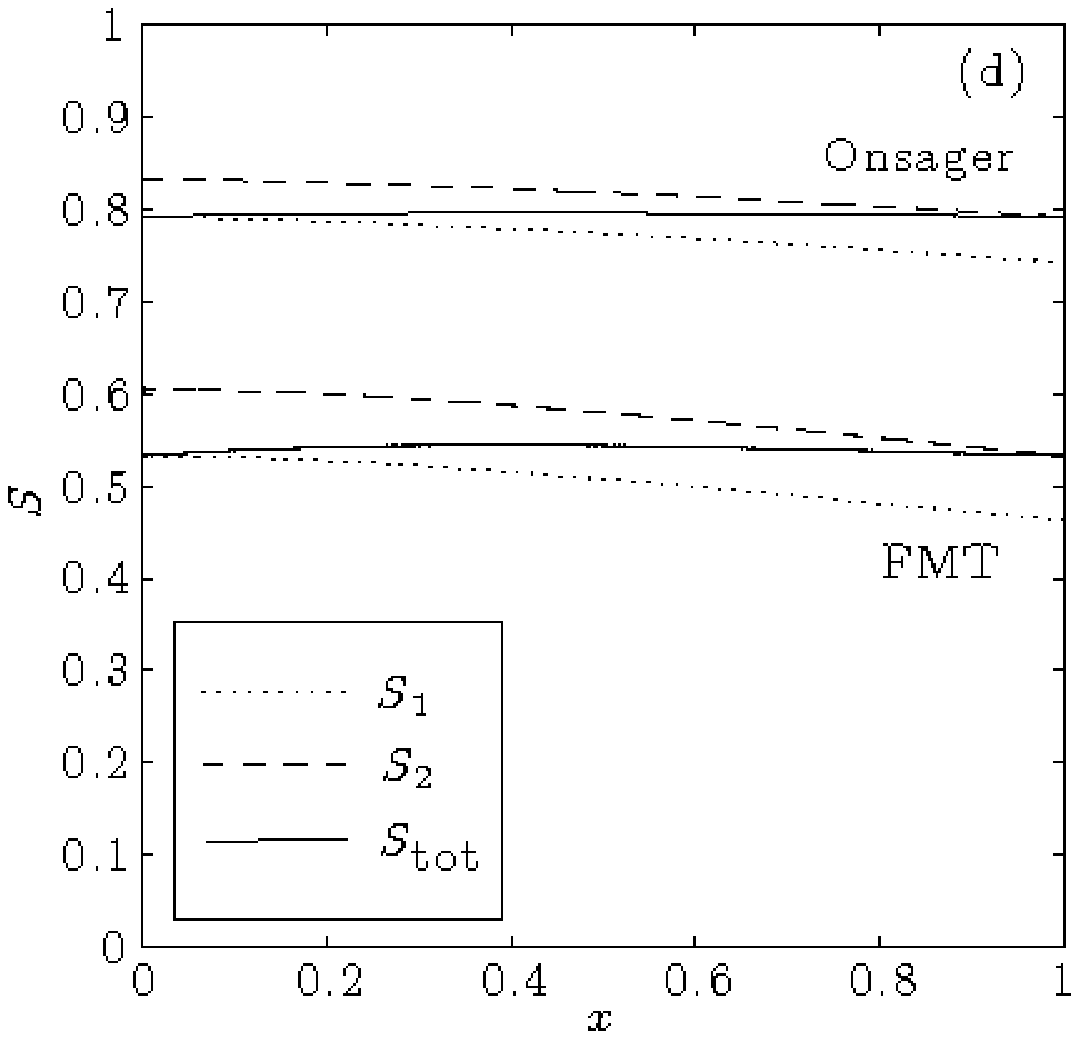}}
\caption{\label{fig:fig1} Results for $\lambda=1.1$. (a) shows the phase behaviour in the $(x,c)$ representation, (b) in the $(c_{1},c_{2})$ representation and (c) in the $(x,p^{*})$ representation. The results for Onsager theory and FMT are plotted on the same graph. In (a)-(c) the upper pairs of solid lines are the binodals according to Onsager theory, the lower pairs of solid lines are the binodals according to FMT and the dashed lines indicate the \textit{I}-\textit{N} spinodals. An isotropic phase (\textit{I}), isotropic and nematic coexistence (\textit{I} + \textit{N}) and a nematic phase (\textit{N}) are present. In (c) the inset is for Onsager theory. In (a) and (b) the thiner solid lines are selected tie lines connecting coexisting state points.  Note that the two theories are plotted on the same graph for comparative purposes; if one is interested in the FMT results for example, then the phase behaviour above the nematic binodal is nematic only and one should ignore the results from Onsager theory. In (d) we plot the nematic order parameters along the nematic binodal for Onsager theory and FMT. The dotted curve is $S_{1}$, the dashed curve is $S_{2}$ and the solid line is $S_{\textrm{tot}}$.}
\label{fig:figure1} 
\end{figure}

\begin{figure}
\centering 
\subfigure{\includegraphics[width=0.45\textwidth,clip]{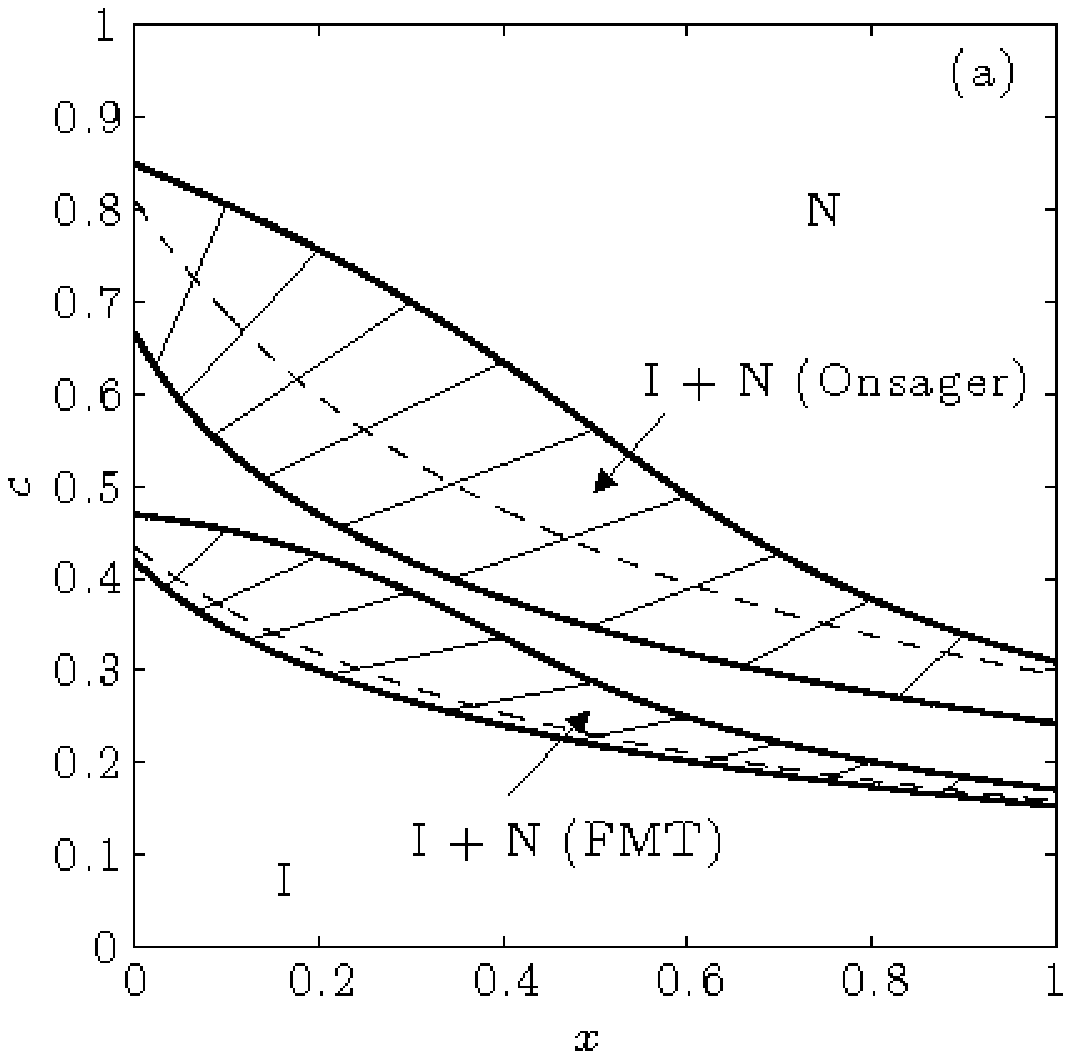}}
\subfigure{\includegraphics[width=0.45\textwidth,clip]{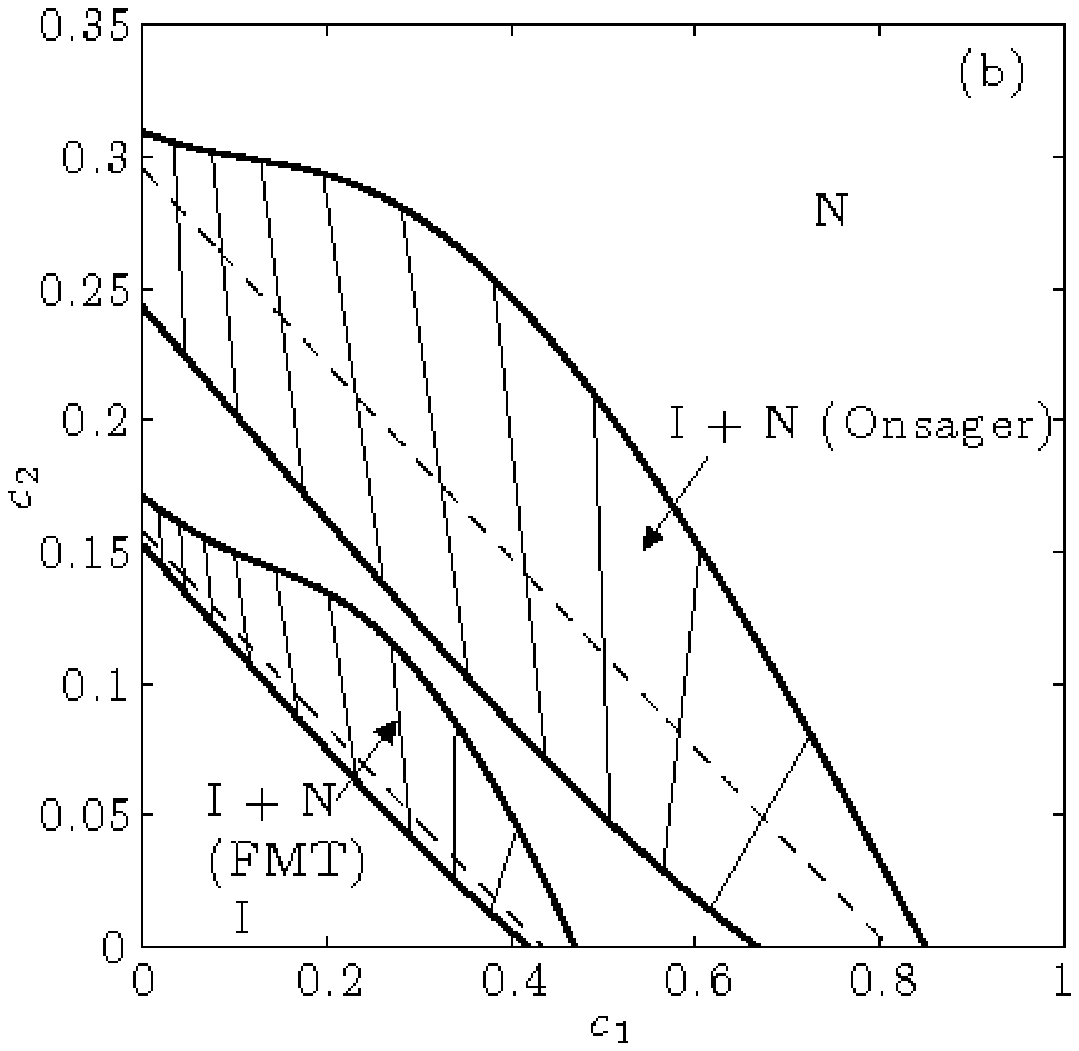}}
\subfigure{\includegraphics[width=0.45\textwidth,clip]{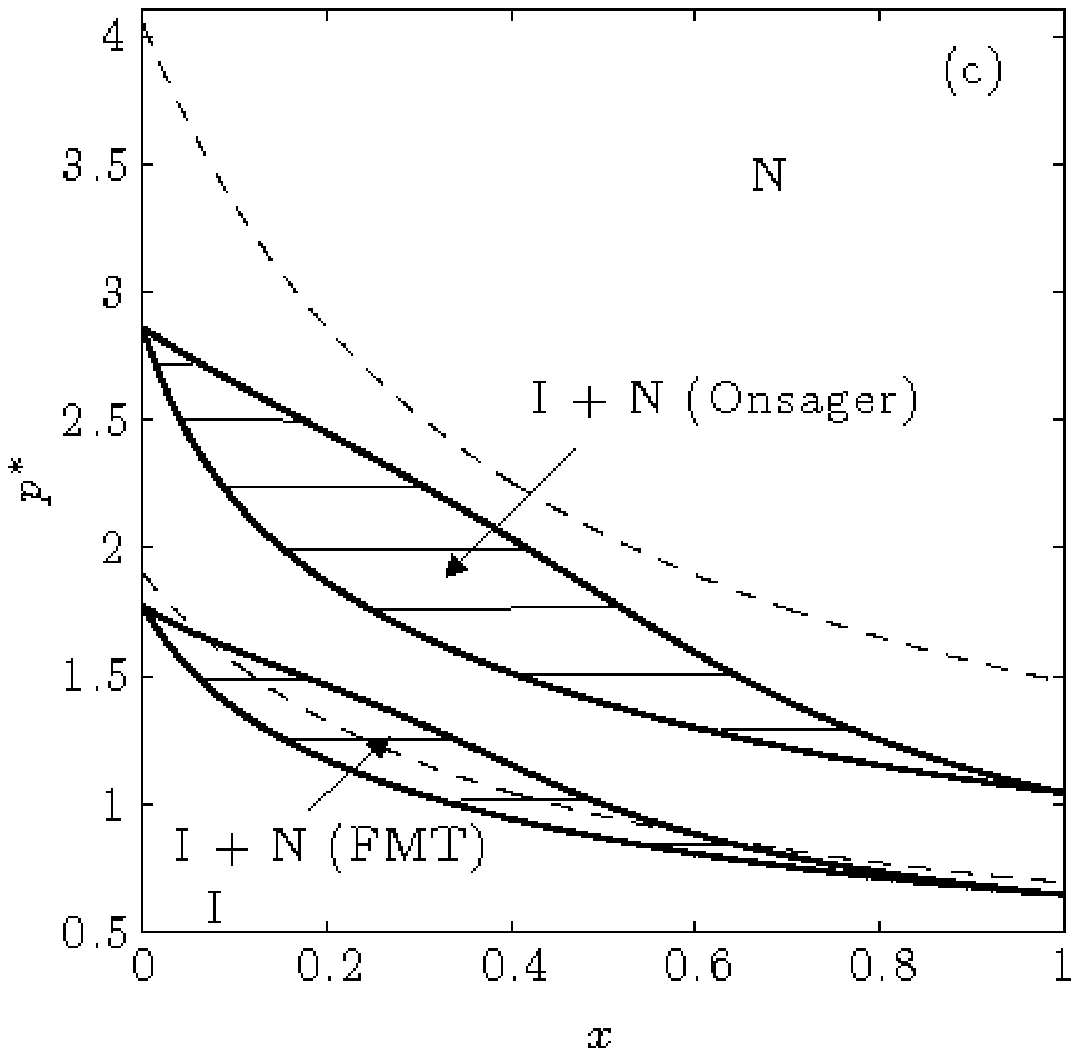}}
\subfigure{\includegraphics[width=0.45\textwidth,clip]{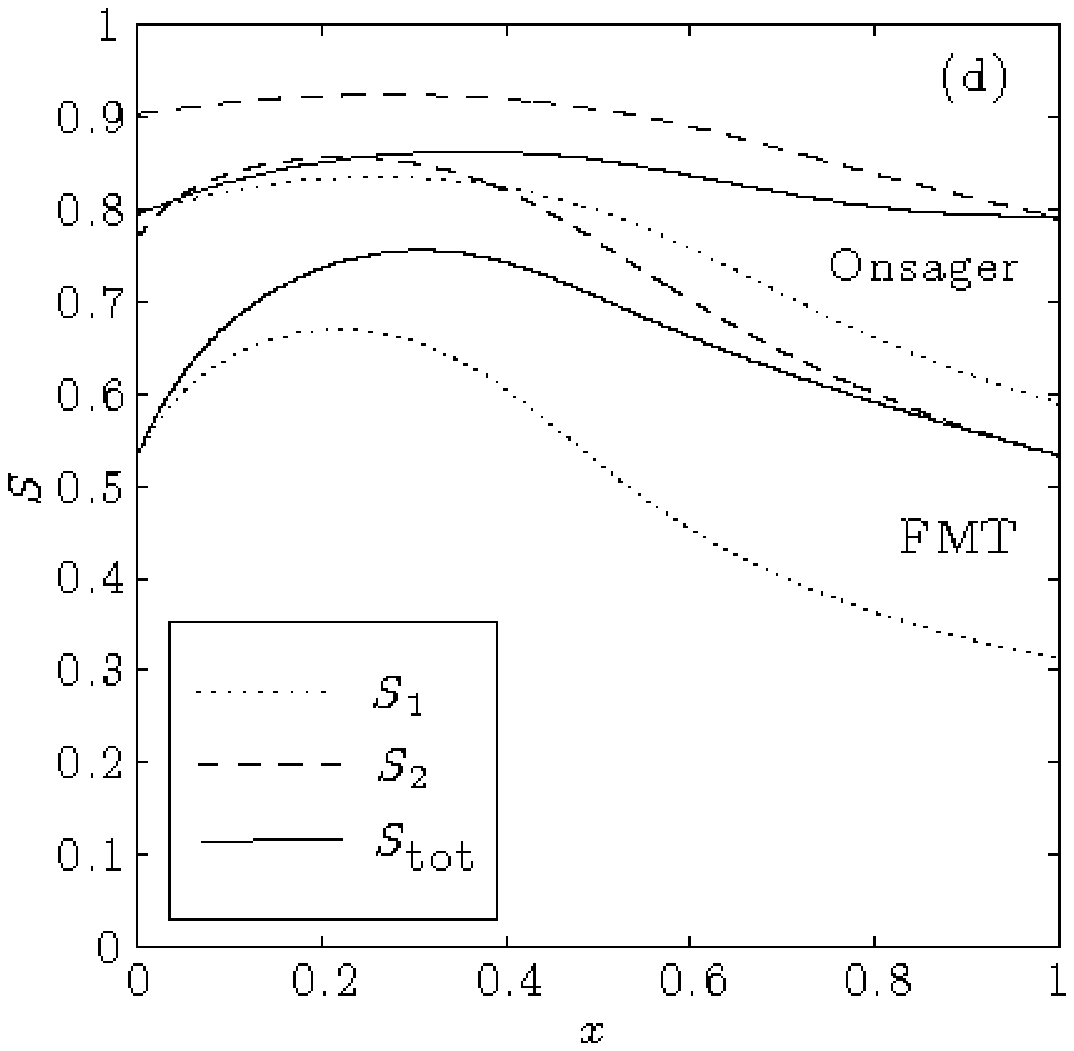}}
\caption{\label{fig:fig2}  Results for $\lambda=1.4$. The notation is the same as in Fig.~1. Note the widening of the \textit{I}-\textit{N} biphasic gap as predicted by both theories.}
\label{figure2} 
\end{figure}

\begin{figure}
\centering 
\subfigure{\includegraphics[width=0.45\textwidth,clip]{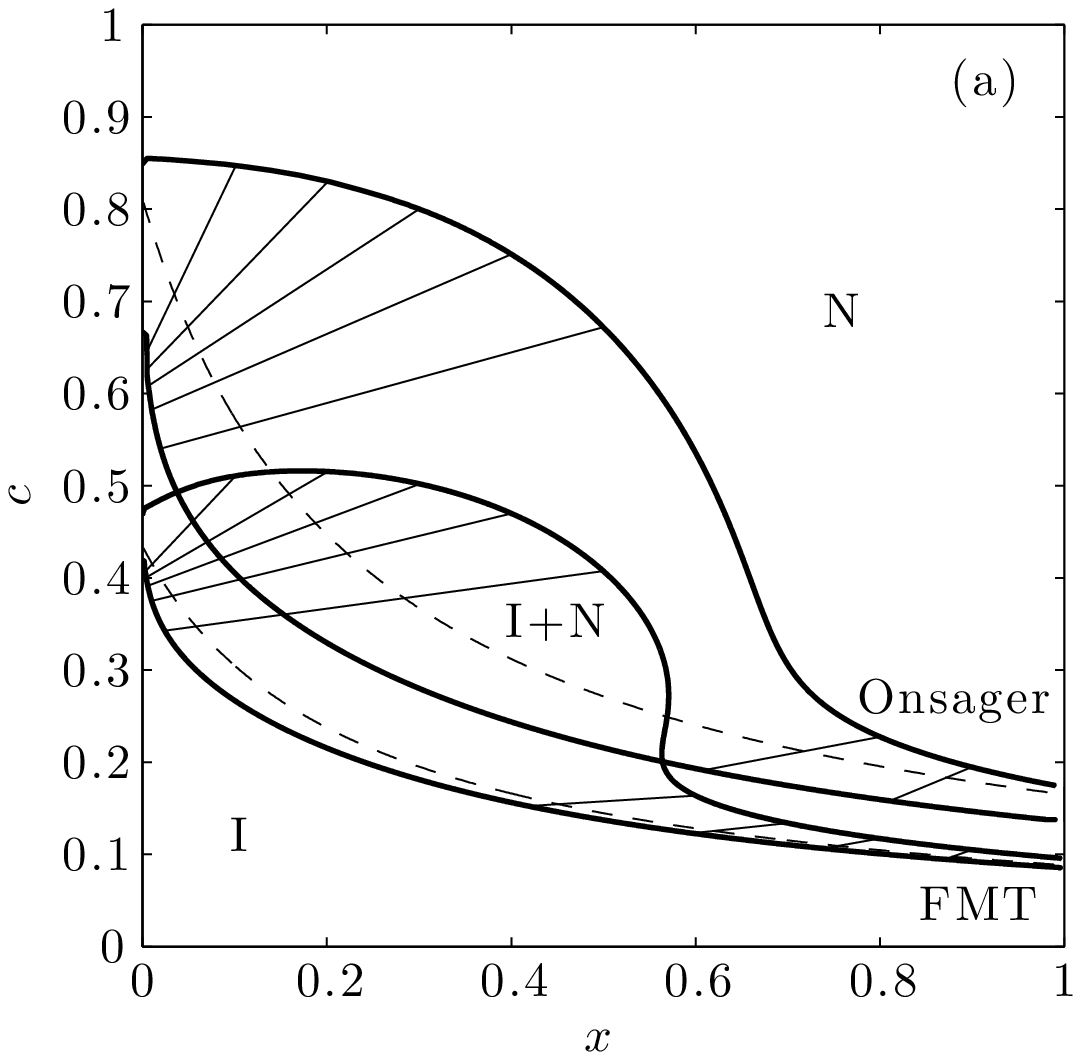}}
\subfigure{\includegraphics[width=0.45\textwidth,clip]{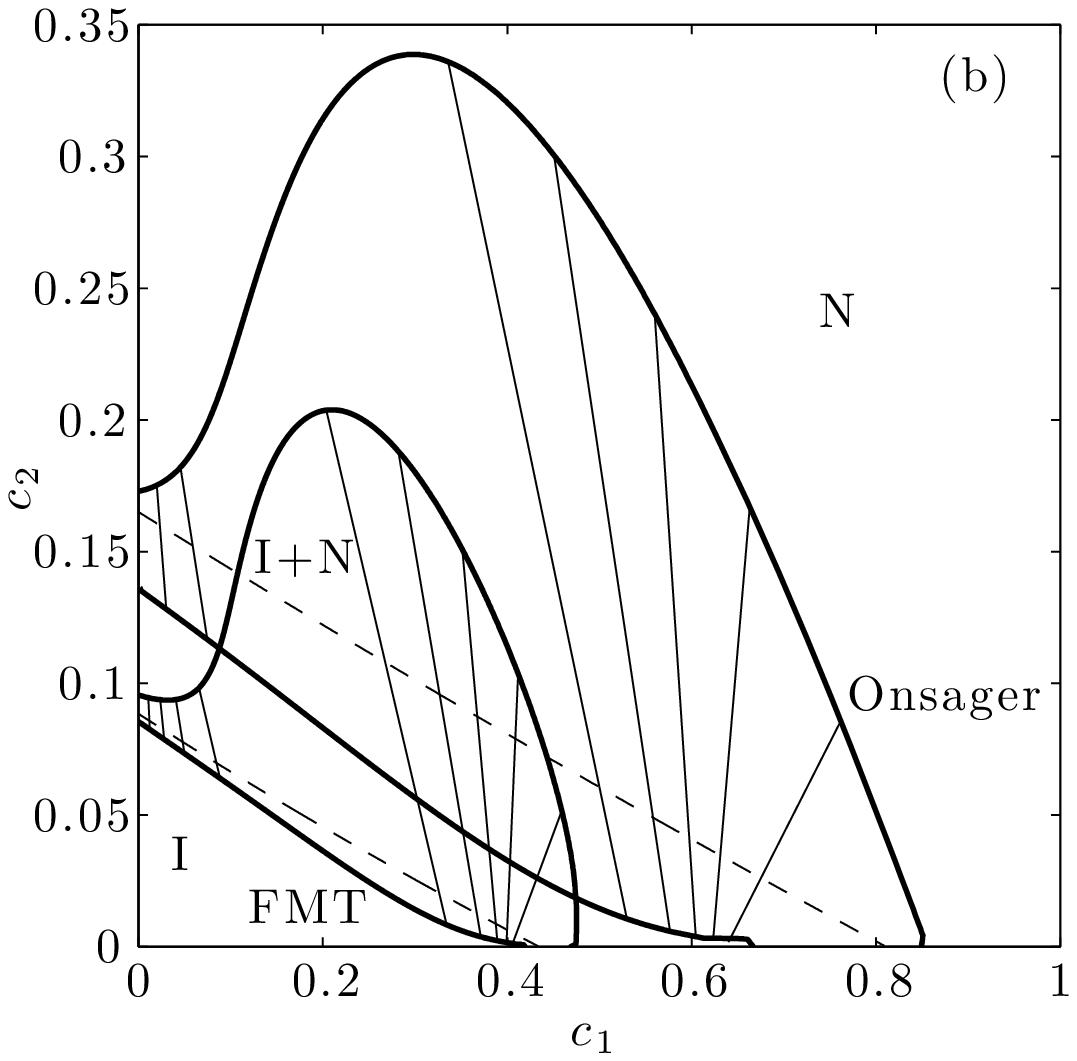}}
\subfigure{\includegraphics[width=0.45\textwidth,clip]{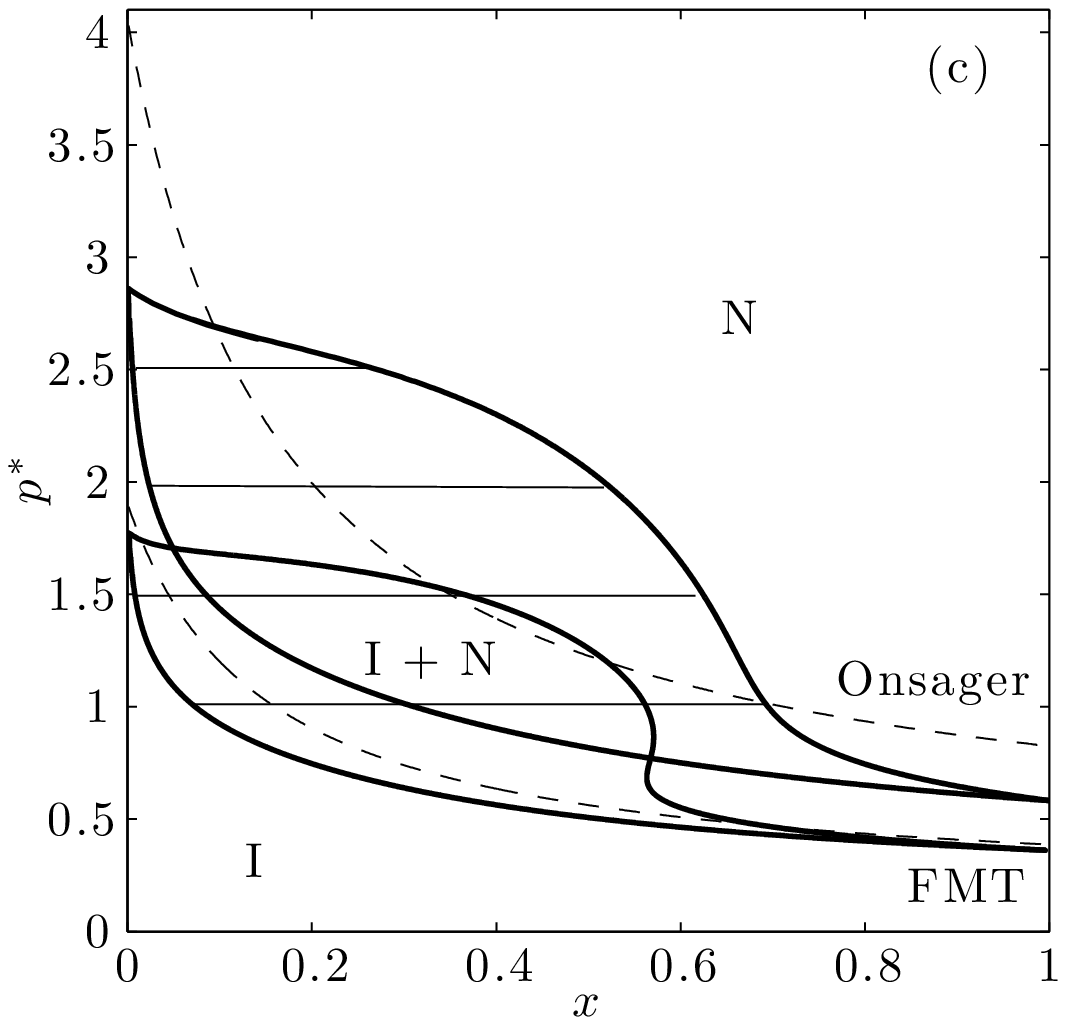}}
\subfigure{\includegraphics[width=0.45\textwidth,clip]{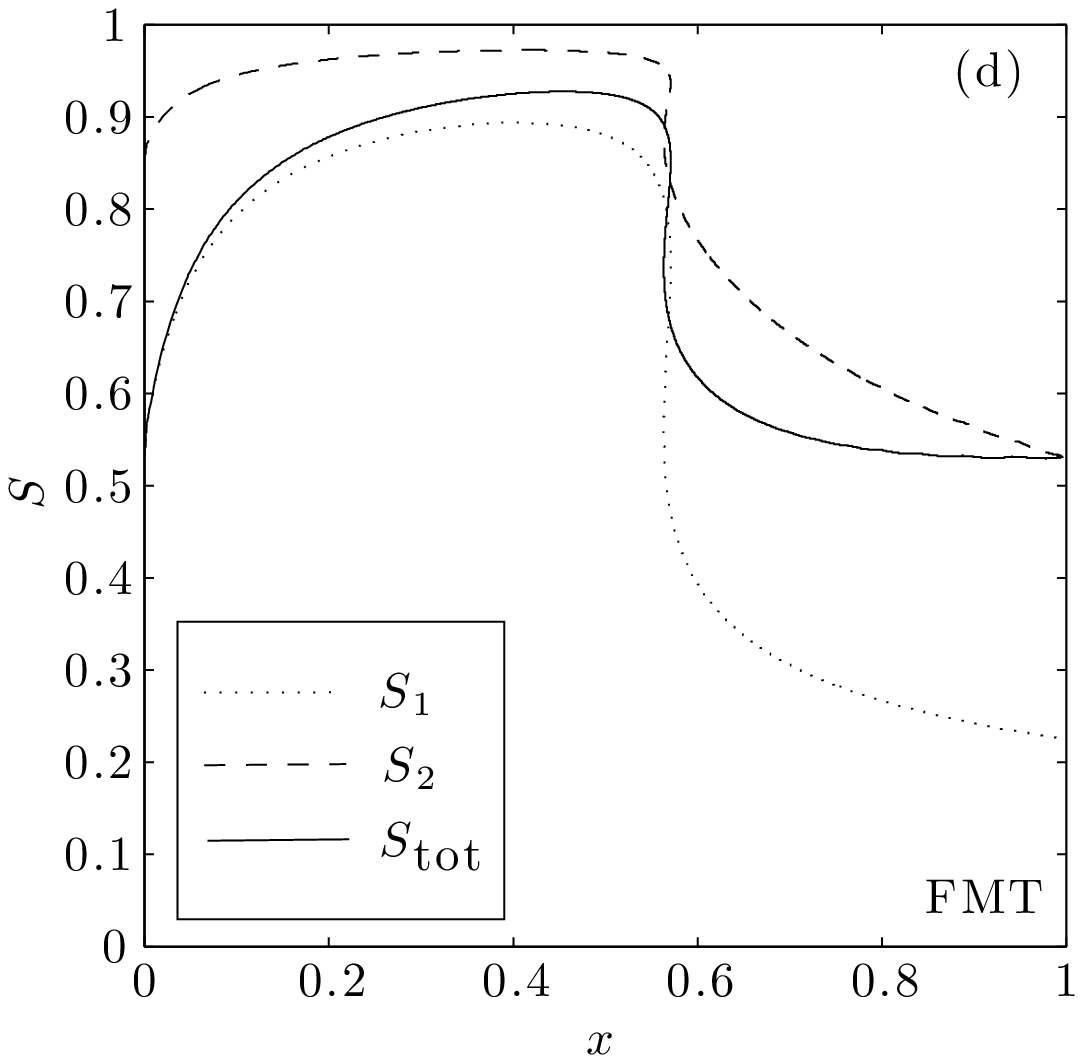}}
\caption{\label{fig:3}  Results for $\lambda=1.7$. The notation is the same as in Fig.~1. In (c) for FMT, if one keeps a constant fluid composition, for example at around $x=0.59$ but \textit{increases} the pressure from $p^{*}=0$ to $p^{*}=1$ the state changes from \textit{I} $\rightarrow$ \textit{I}+\textit{N$_{2}$} $\rightarrow$ \textit{N$_{2}$} $\rightarrow$ \textit{I}+\textit{N$_{2}$} $\rightarrow$ \textit{N$_{2}$}. In (d) we present the order parameters only for FMT.}
\label{figure3} 
\end{figure}

\begin{figure}
\centering 
\subfigure{\includegraphics[width=0.45\textwidth,clip]{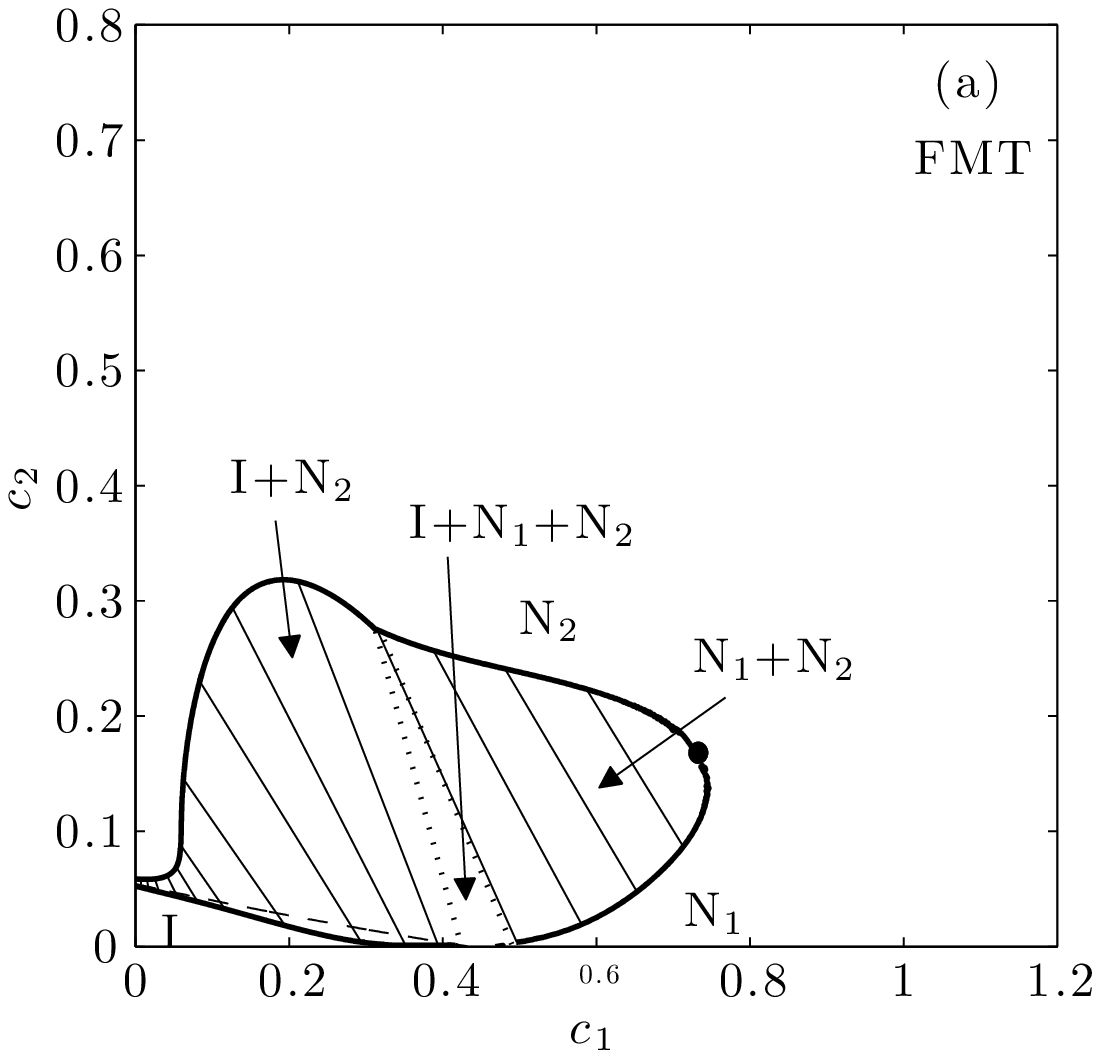}}
\subfigure{\includegraphics[width=0.45\textwidth,clip]{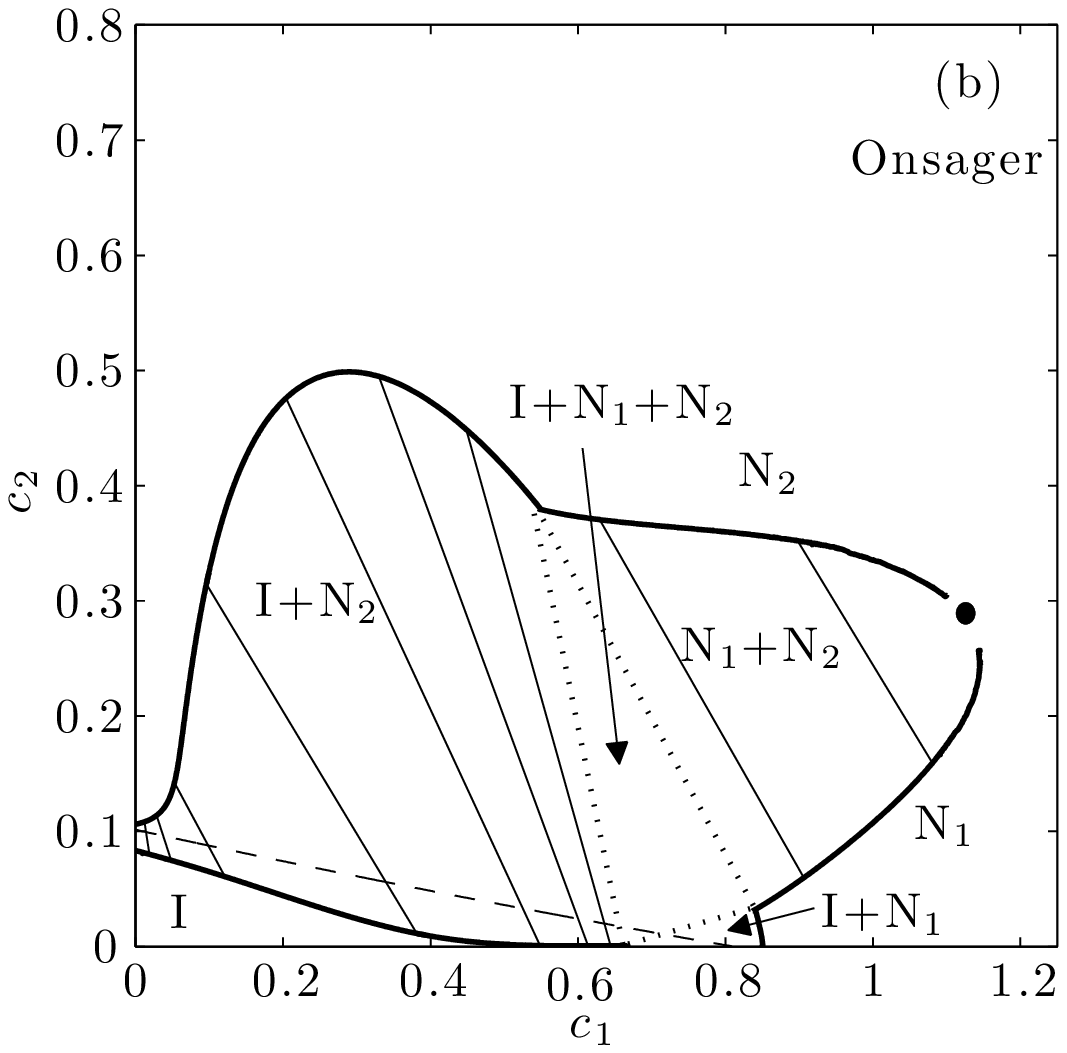}}
\subfigure{\includegraphics[width=0.45\textwidth,clip]{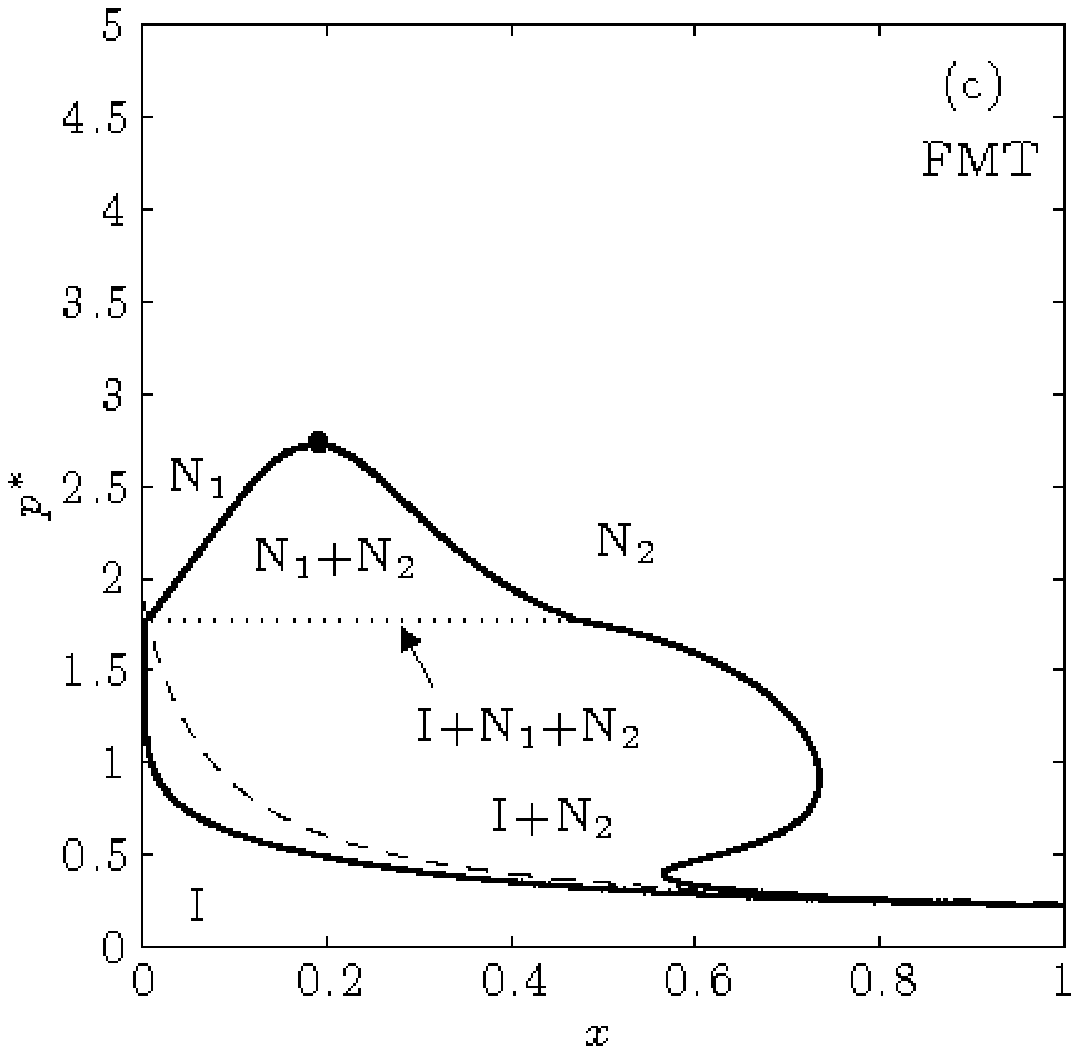}}
\subfigure{\includegraphics[width=0.45\textwidth,clip]{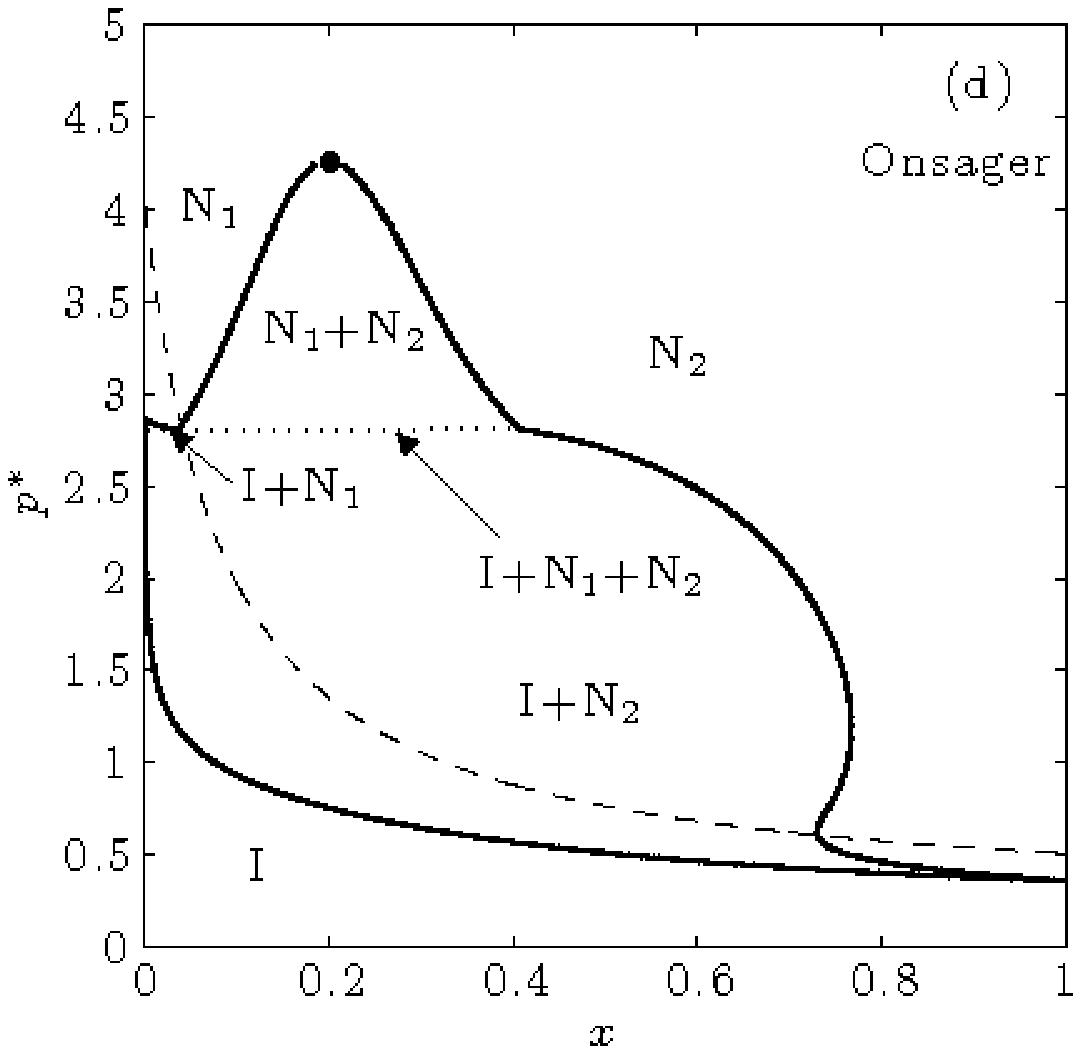}}
\caption{\label{fig:4} Phase diagrams for $\lambda=2$. (a) shows the phase behaviour in the $(c_{1},c_{2})$ representation for Onsager theory and (b) for FMT. (c) shows the phase behaviour in the $(x,p^{*})$ representation for Onsager theory and (b) for FMT. The topology is the same for Onsager theory and FMT. Solid lines denote the binodals and the thinner solid lines denote tie lines. As well as \textit{I}-\textit{N} coexistence, there is now coexistence between two nematic phases; the \textit{N}-\textit{N} phase ends in a critical point  (depicted as a shaded circle) in both Onsager theory and FMT, though the critical point is at lower densities for FMT. The dashed lines denote the \textit{I}-\textit{N} spinodals but we do not show the location of the \textit{N}-\textit{N} spinodals on these phase diagrams. There is also \textit{I}-\textit{N}-\textit{N} coexistence. In (a) and (b) this triple point is represented by a region bounded by dotted lines, the vertices of which are the pure phases. In (c) and (d) this region is collapsed onto a triple line, shown here with a dotted line. }
\label{fig:figure4}  
\end{figure}

\begin{figure}
\centering 
\subfigure{\includegraphics[width=0.45\textwidth,clip]{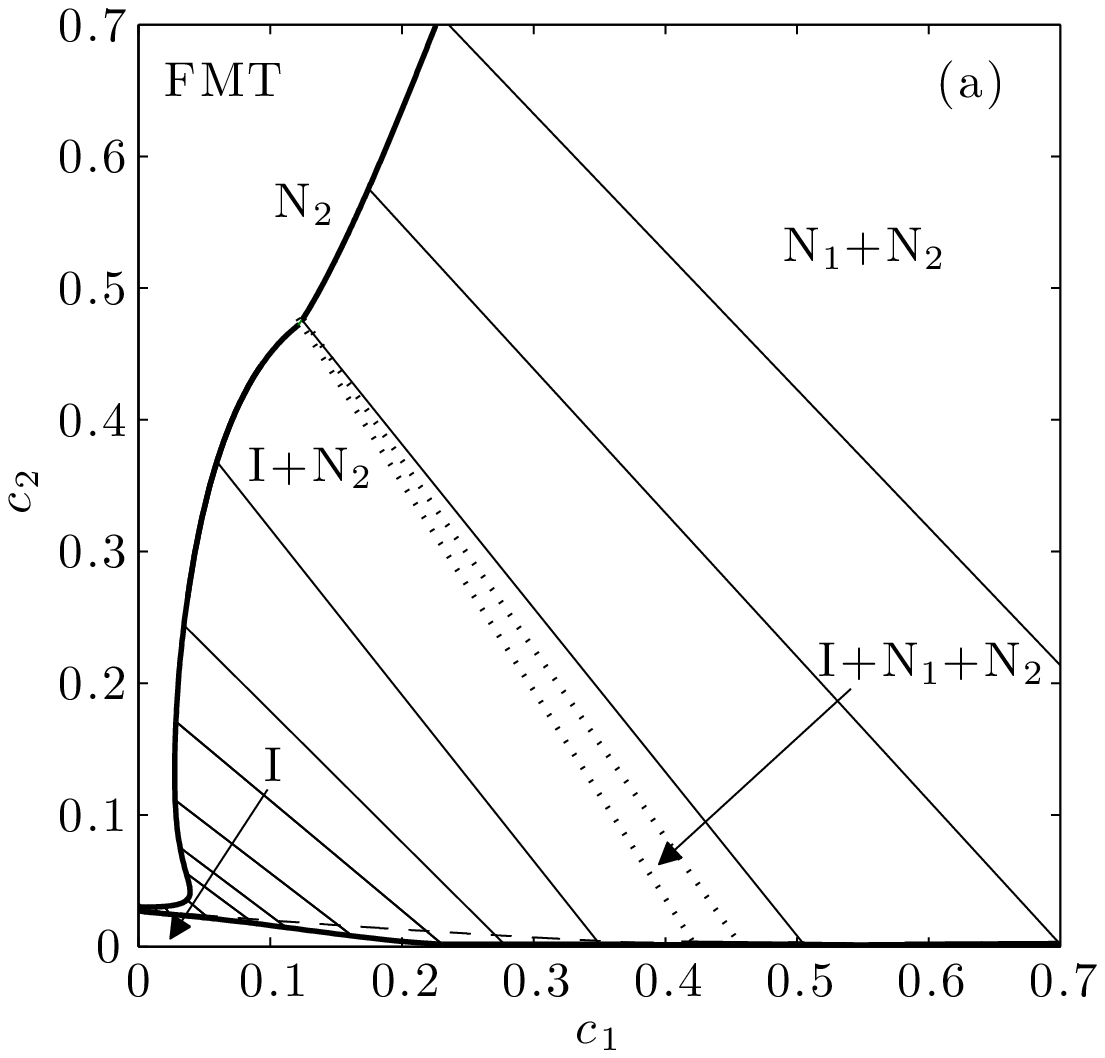}}
\subfigure{\includegraphics[width=0.45\textwidth,clip]{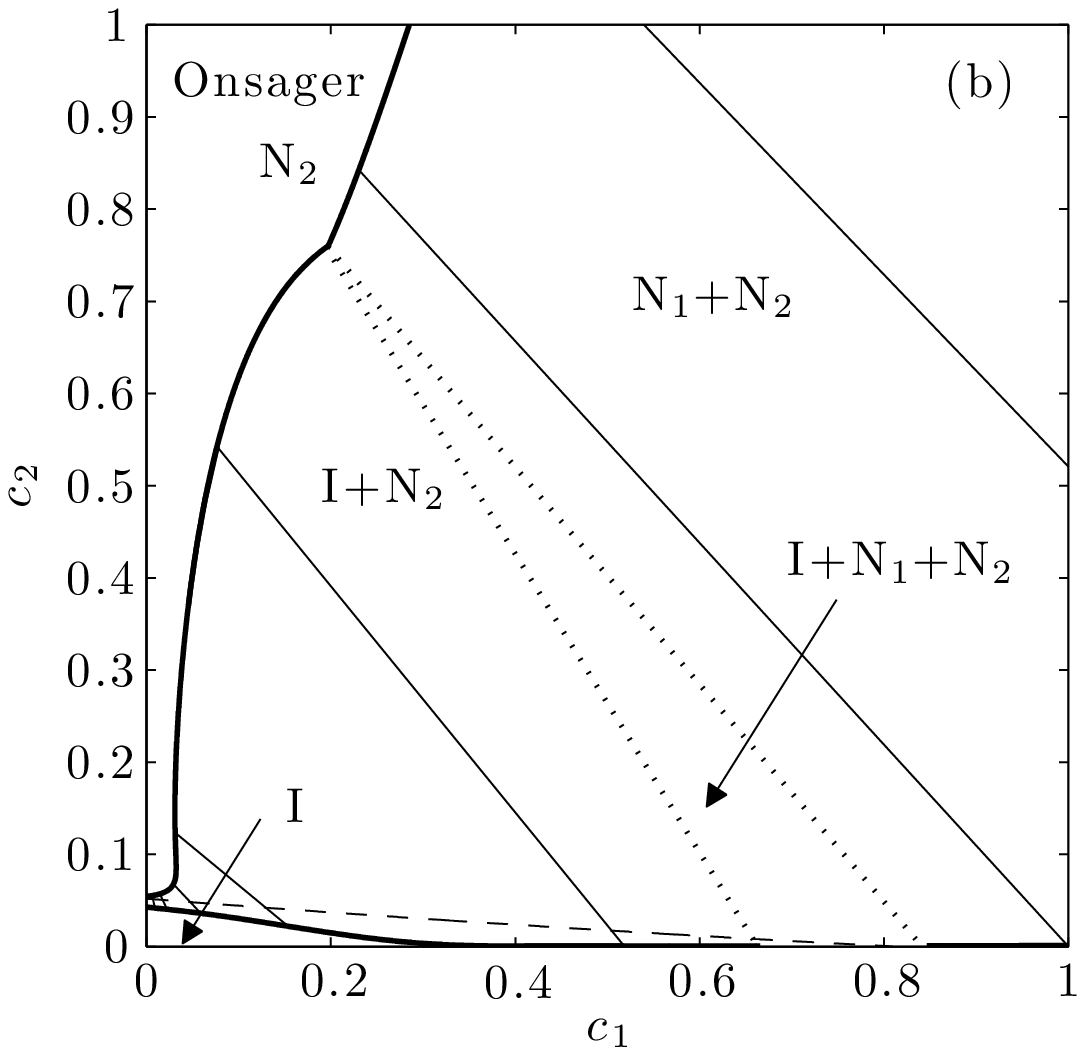}}
\subfigure{\includegraphics[width=0.45\textwidth,clip]{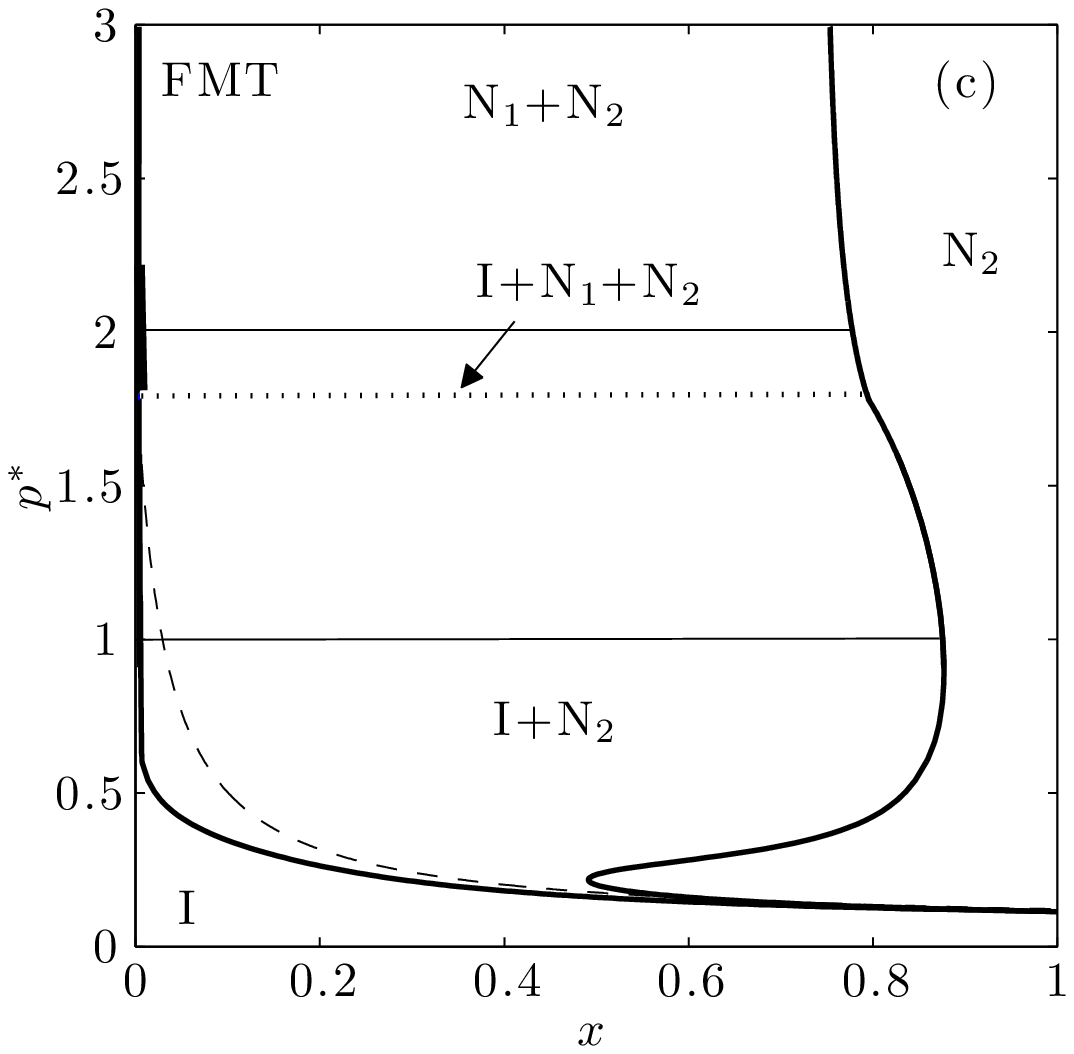}}
\subfigure{\includegraphics[width=0.45\textwidth,clip]{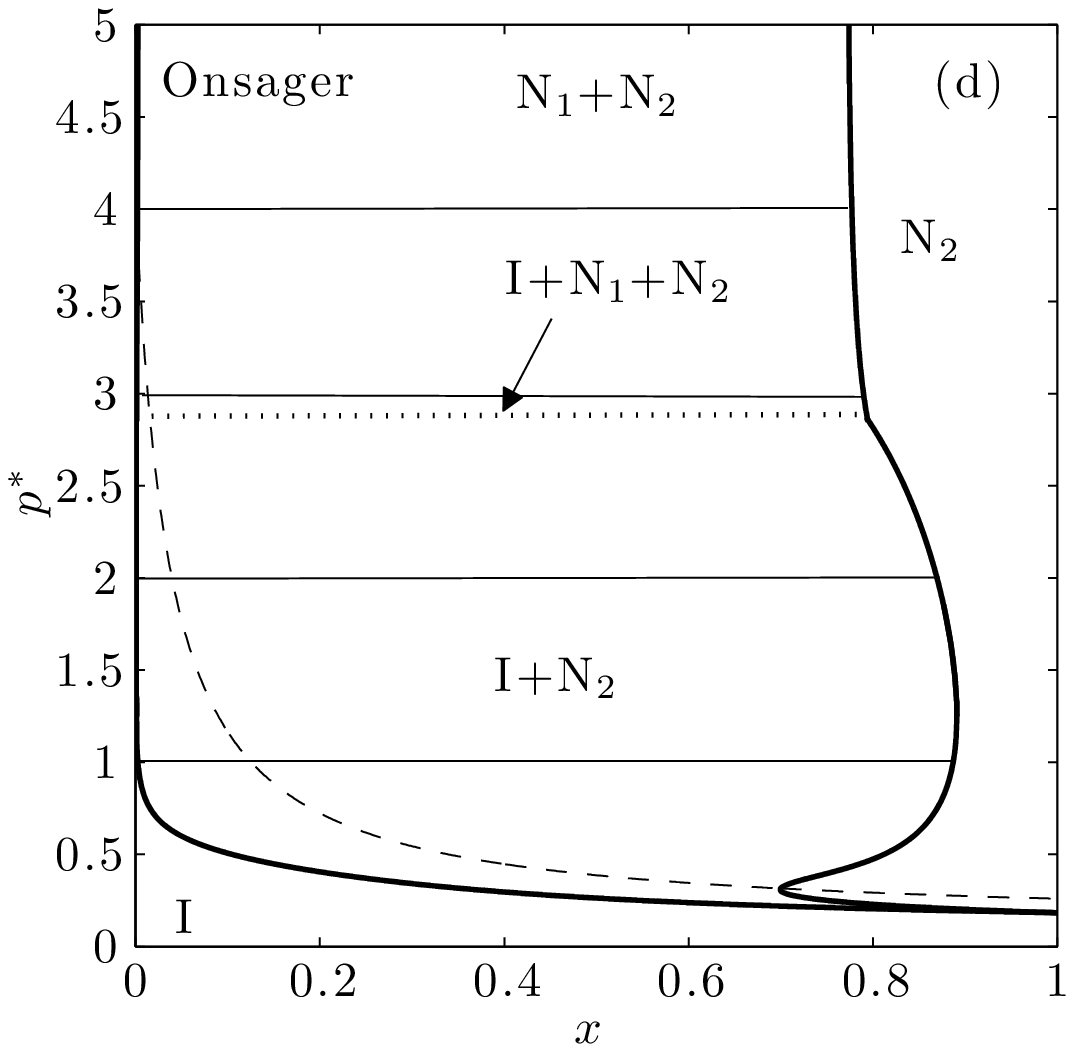}}
\caption{\label{fig:5} Phase diagrams for $\lambda=2.5$. The notation is the same as in Fig.~\ref{fig:4}. The \textit{N}-\textit{N} coexistence does not end in a critical point in these phase diagrams. }
\label{fig:figure5} 
\end{figure}

\begin{figure}
\centering 
\subfigure{\includegraphics[width=0.45\textwidth,clip]{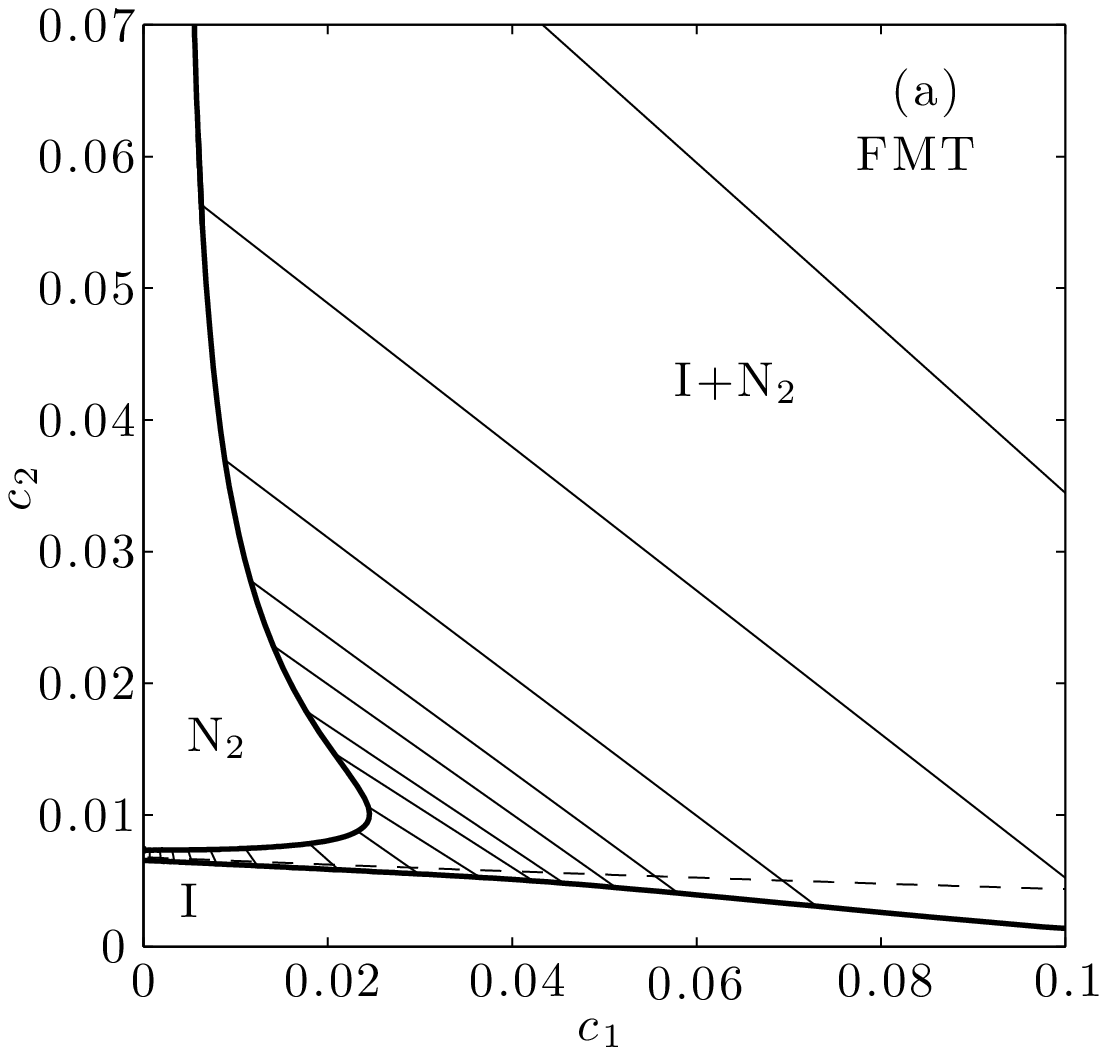}}
\subfigure{\includegraphics[width=0.45\textwidth,clip]{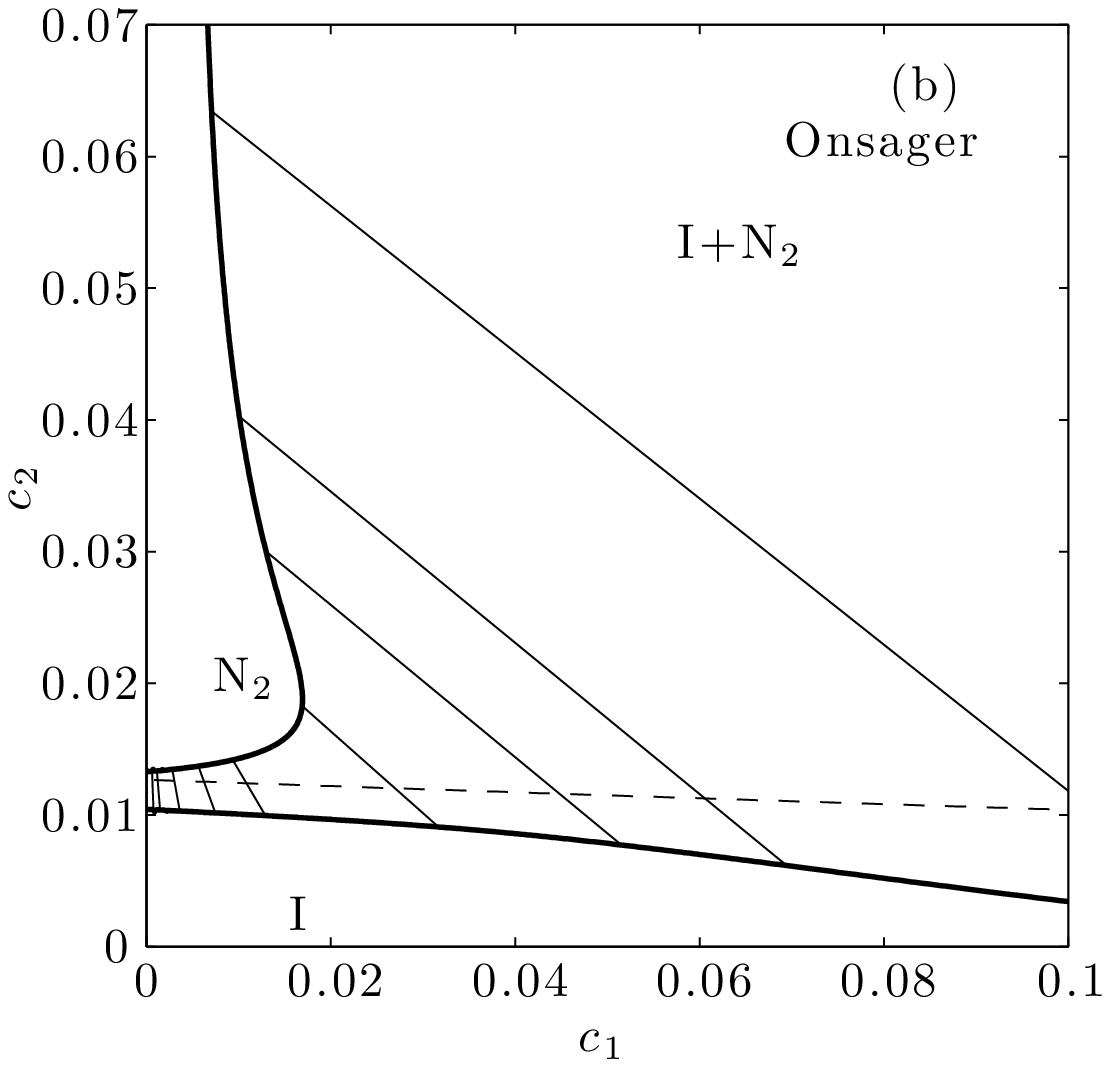}}
\subfigure{\includegraphics[width=0.45\textwidth,clip]{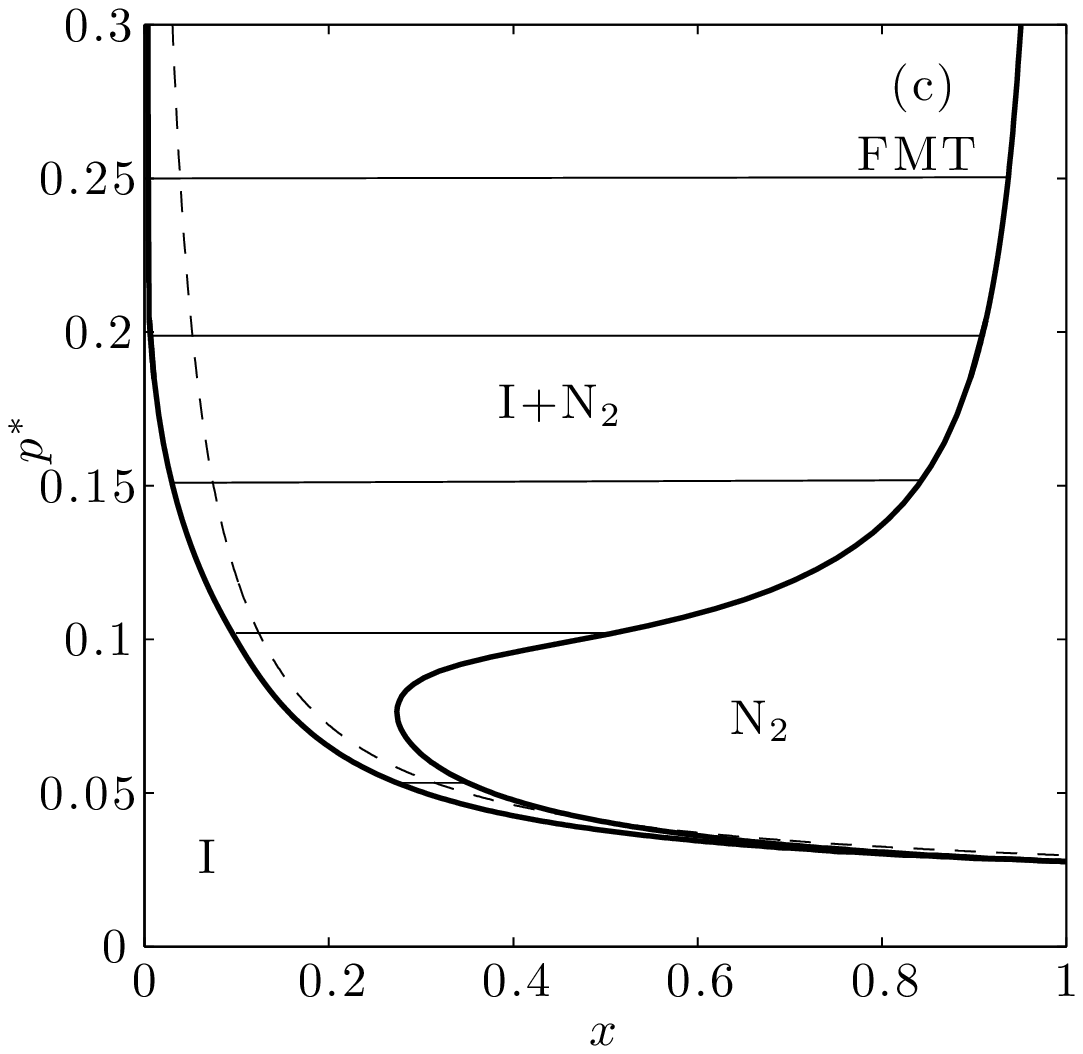}}
\subfigure{\includegraphics[width=0.45\textwidth,clip]{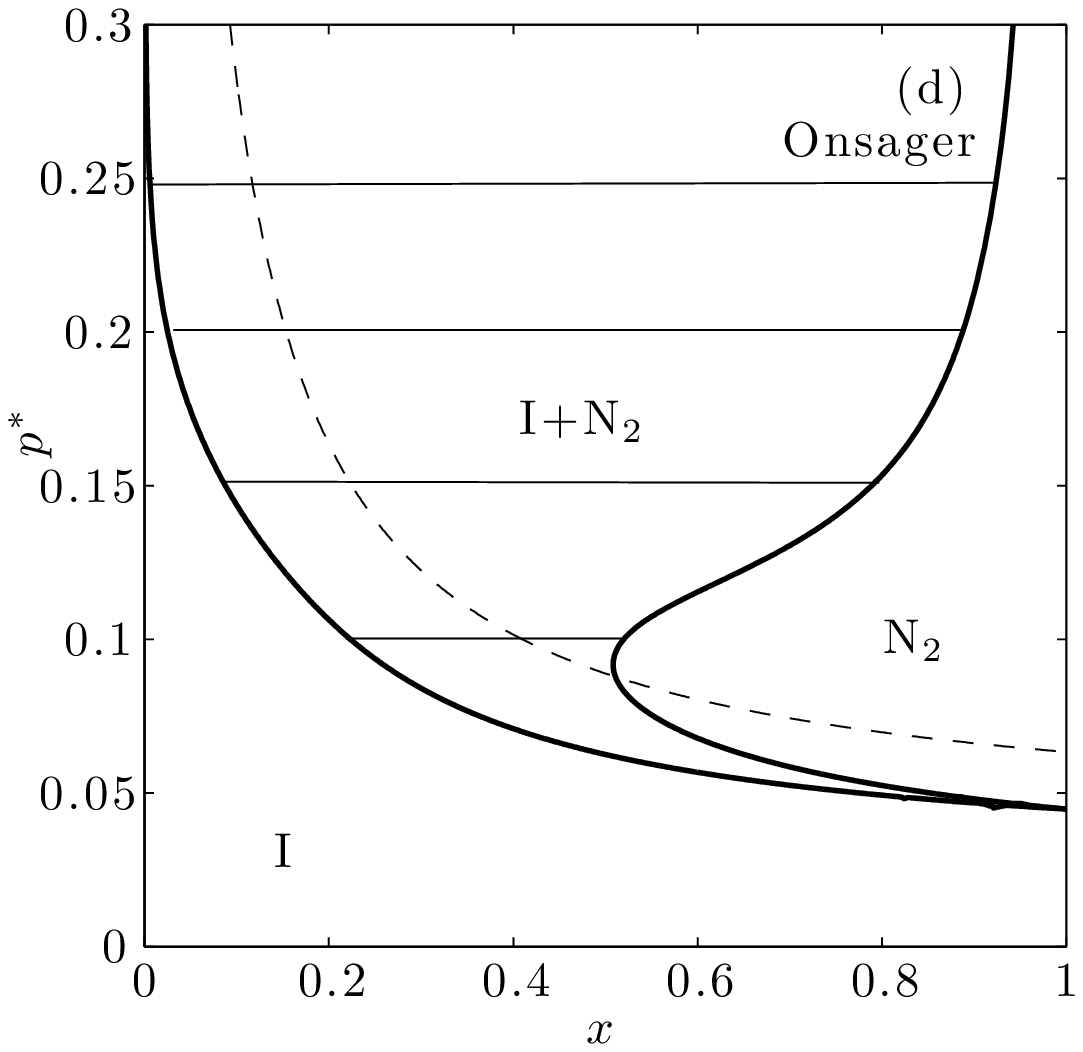}}
\caption{\label{fig:6} Phase diagrams for $\lambda=4$. The notation is the same as in Fig.~\ref{fig:4}. We have focused on the lower part of the phase diagram as this is where the most interesting phase behaviour occurs. The \textit{N}-\textit{N} phase separation continues to high densities beyond the scale of these plots, with a huge immiscibility gap. }
\label{fig:figure6}  
\end{figure}

\begin{figure}
\centering 
\subfigure{\includegraphics[width=0.45\textwidth,clip]{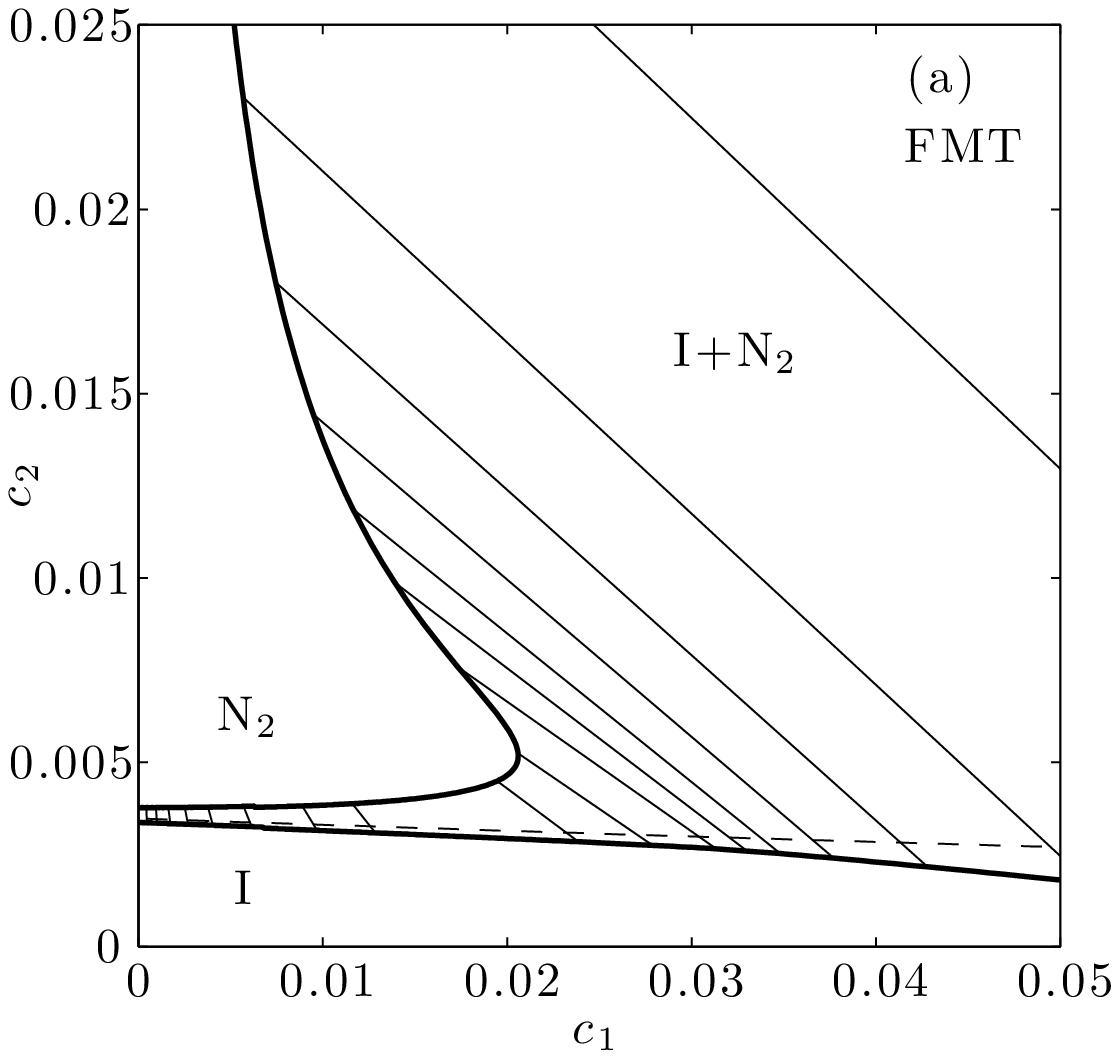}}
\subfigure{\includegraphics[width=0.45\textwidth,clip]{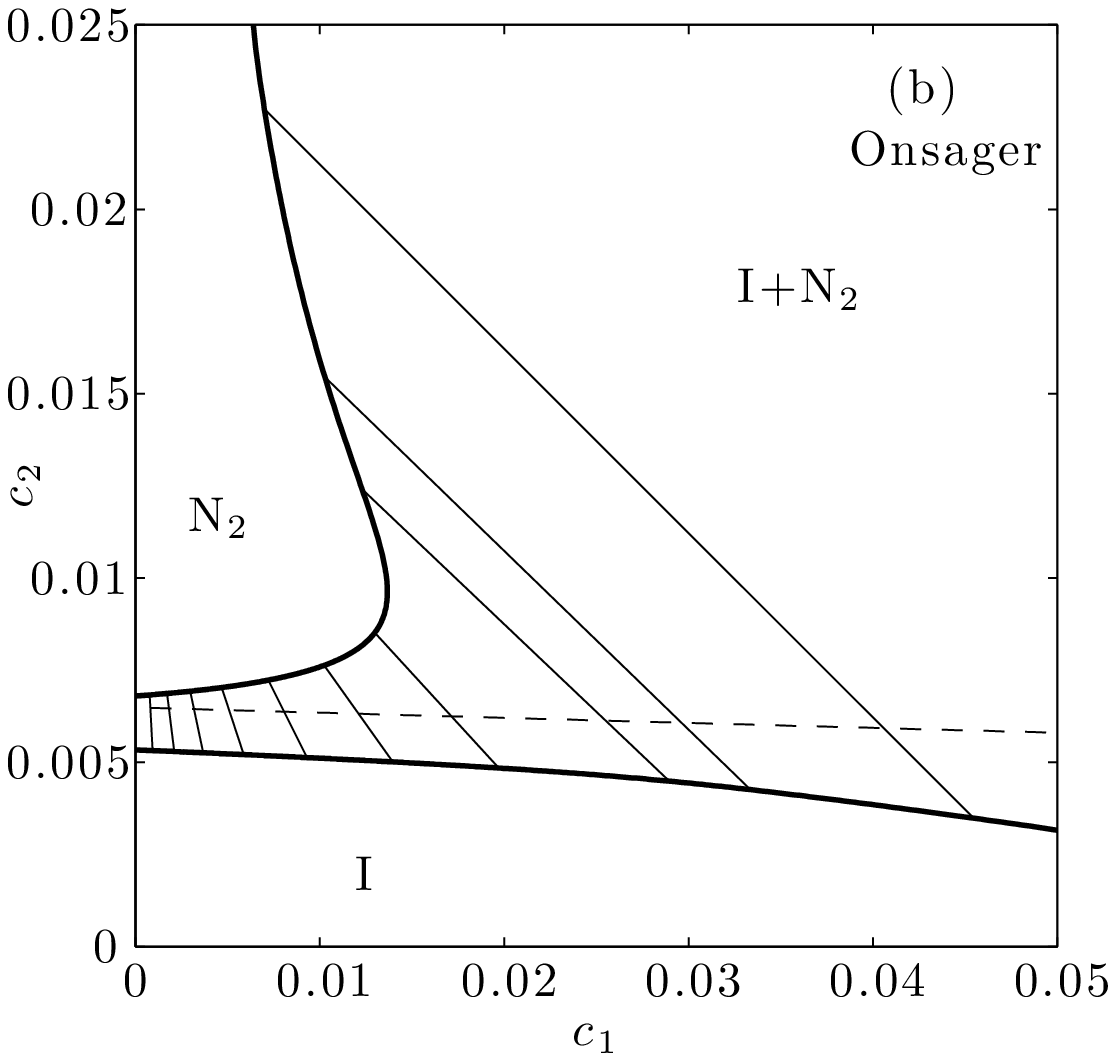}}
\subfigure{\includegraphics[width=0.45\textwidth,clip]{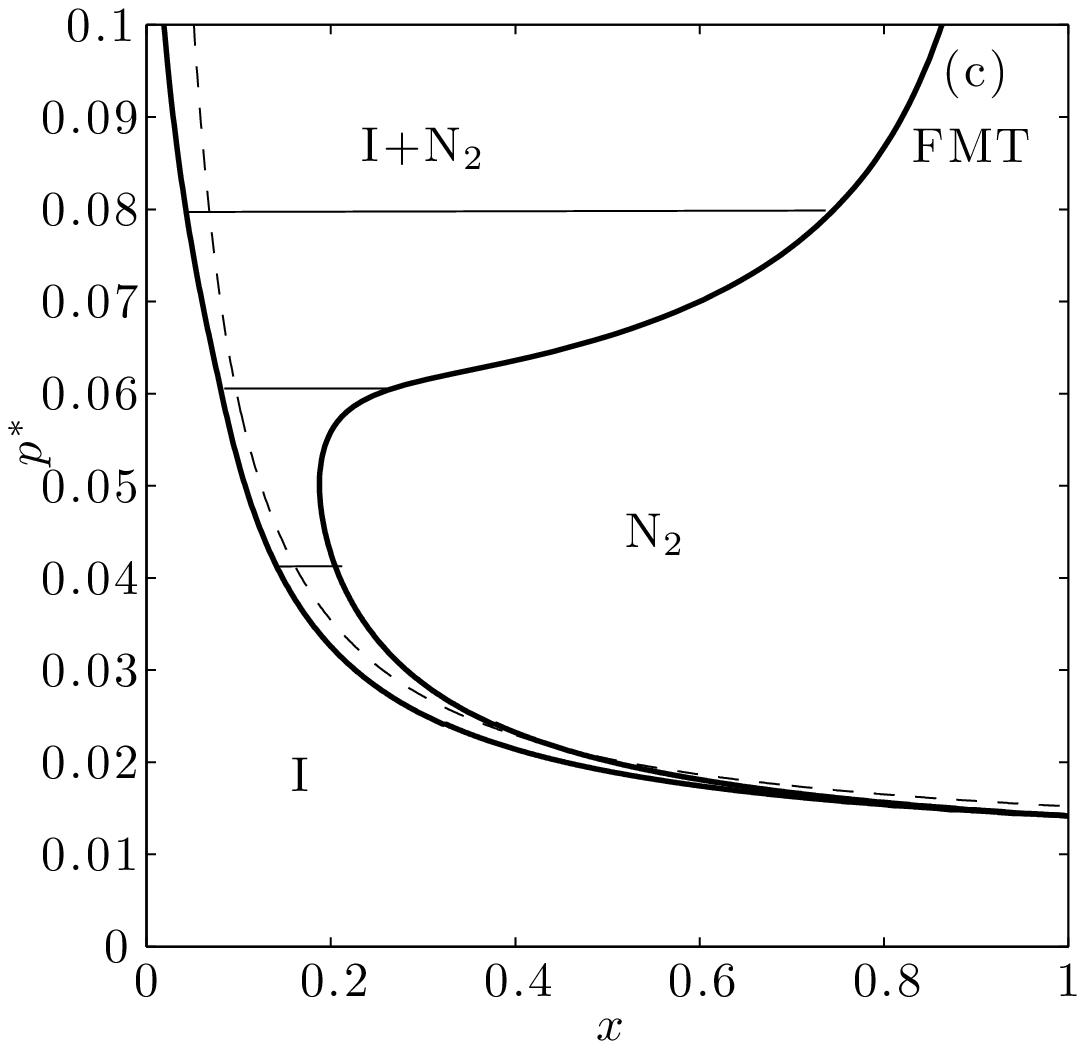}}
\subfigure{\includegraphics[width=0.45\textwidth,clip]{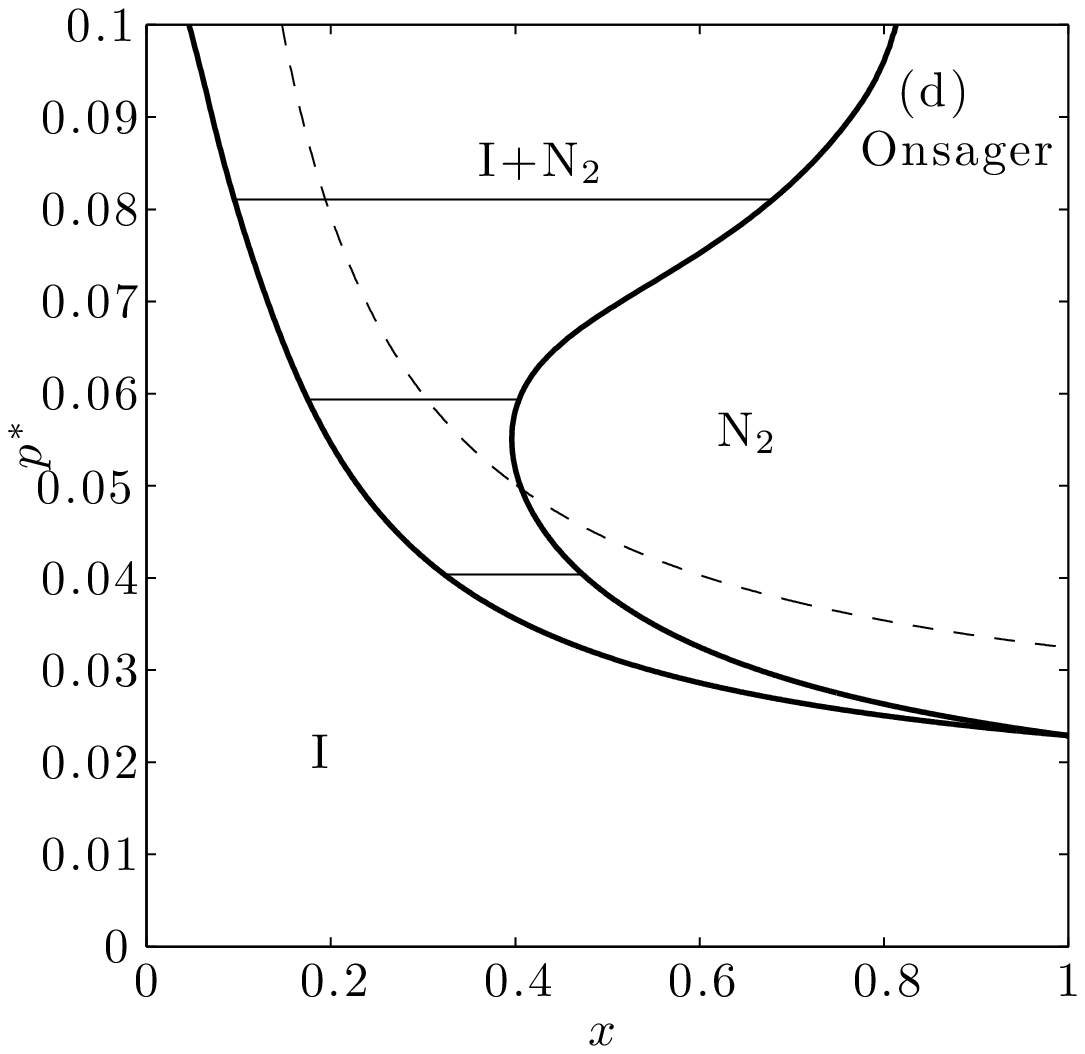}}
\caption{\label{fig:7} Phase diagrams for $\lambda=5$. The notation is the same as in Fig.~\ref{fig:4}. We focus on the re-entrant behaviour in this figure. The general trend of the re-entrant bend moving to lower composition continues here once again. }
\label{fig:figure7} 
\end{figure}

\begin{figure}
\centering 
\subfigure{\includegraphics[width=0.45\textwidth,clip]{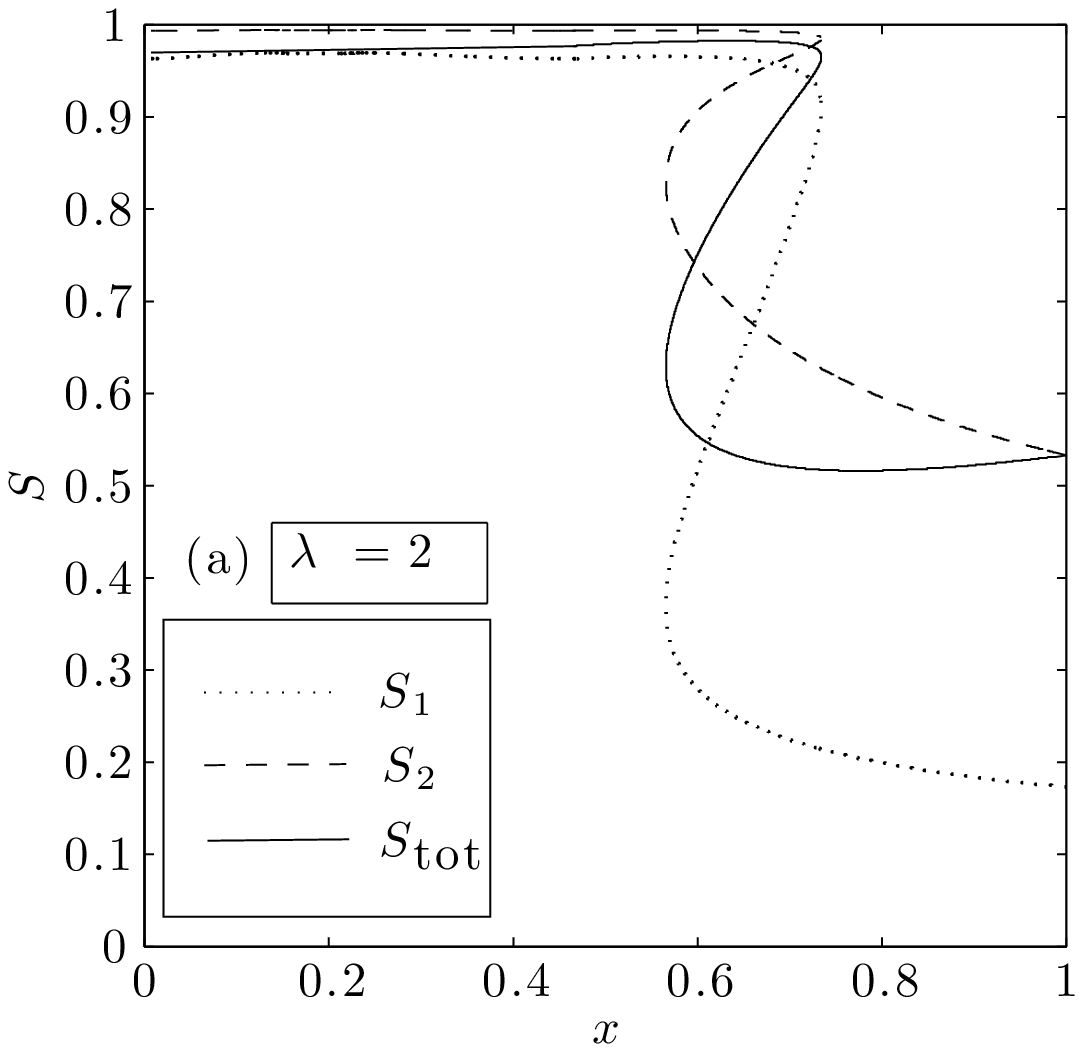}}
\subfigure{\includegraphics[width=0.45\textwidth,clip]{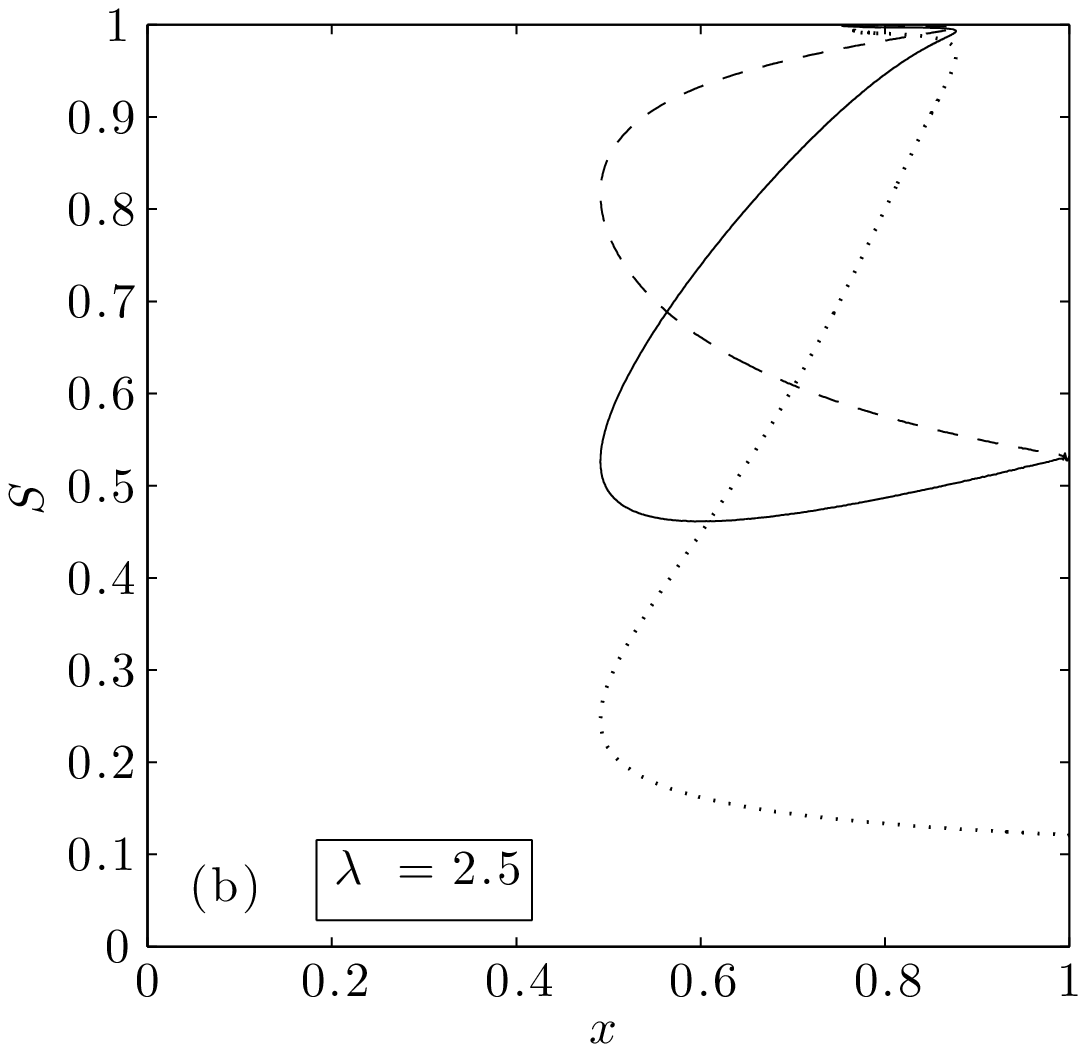}}
\subfigure{\includegraphics[width=0.45\textwidth,clip]{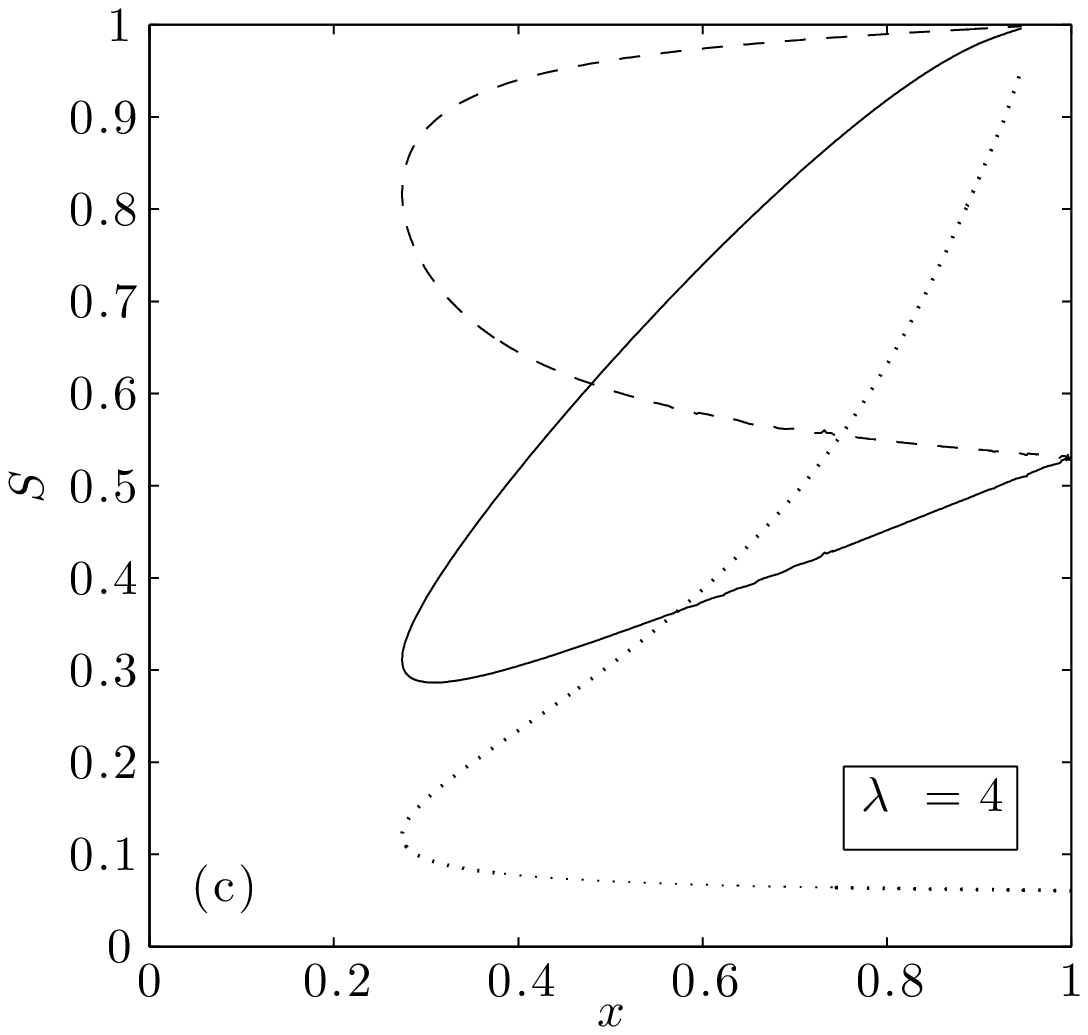}}
\subfigure{\includegraphics[width=0.45\textwidth,clip]{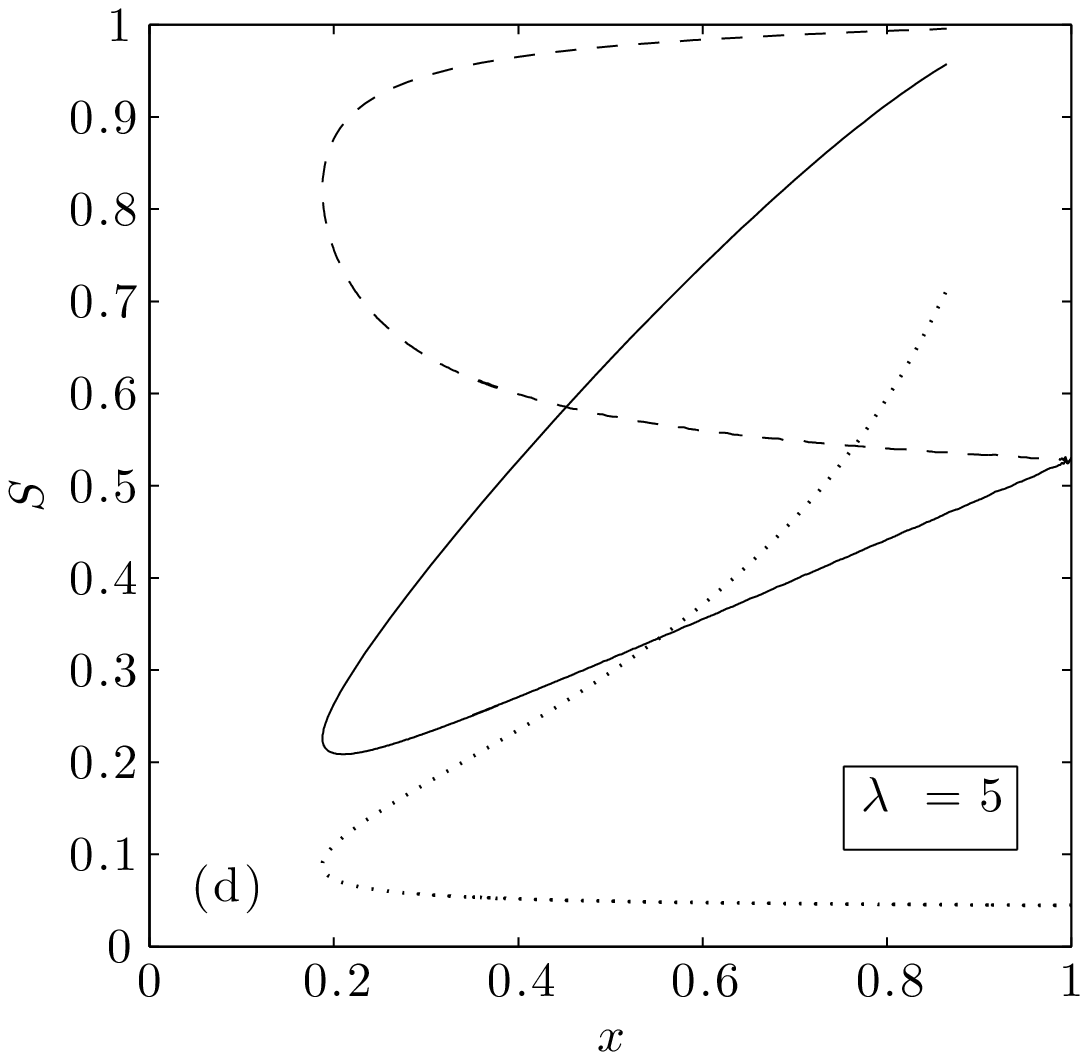}}
\caption{\label{fig:8} The nematic order parameters along the nematic binodal for $\lambda = 2, 2.5, 4$ and $5$ using FMT. The dotted curves indicate $S_{1}$, the dashed curves indicate $S_{2}$ and the solid curves indicate $S_{\textrm{tot}}$. }
\label{fig:figure7} 
\end{figure}

\end{widetext}


\begin{thebibliography}{10}

\bibitem{vanderkooij.f.m:2000.a}
F.~M. van~der Kooij, K. Kassapidou, and H.~N.~W. Lekkerkerker, Nature {\bf
  406},  868  (2000).

\bibitem{dijkstra.m:1995.a}
M. Dijkstra, J.~P. Hansen, and P.~A. Madden, Phys. Rev. Lett. {\bf 75},  2236
  (1995).

\bibitem{dijkstra.m:1997.a}
M. Dijkstra, J.~P. Hansen, and P.~A. Madden, Phys. Rev. E {\bf 55},  3044
  (1997).

\bibitem{pizzey.c:2004.a}
C. Pizzey, S. Klein, E. Leach, J.~S. van Duijneveldt, and R.~M. Richardson, J. Phys.\ Condens.\ Matter {\bf 16},  2479  (2004).

\bibitem{vanduijneveldt.js:2005.a}
J.~S. van Duijneveldt, S. Klein, E. Leach, C. Pizzey, and R.~M. Richardson, J. Phys.\ Condens.\ Matter {\bf 17},  2255  (2005).

\bibitem{leach.esh:2005.a}
E.~S.~H. Leach and J.~S. van Duijneveldt, Langmuir {\bf 21},  3821  (2005).

\bibitem{mourad.m.c.d.:2008.b}
M.~C.~D. Mourad, E.~J. Devid, M.~M. van Schooneveld, S. Vonk, and H.~N.~W.
  Lekkerkerker, J. Phys.\ Chem.\ B {\bf 112},  10142  (2008).

\bibitem{maurice_thesis}
M.~C.~D. Mourad, {\em Liquid Crystal Phases of Colloidal Platelets and their
  Use as Nanocomposite Templates} (PhD Thesis, Utrecht University, 2009).

\bibitem{maitland.gc:2000.a}
G.~C. Maitland, Curr.\ Opin.\ Colloid Interface Sci. {\bf 51},  301  (2000).

\bibitem{mason.tg:2002.a}
T.~G. Mason, Phys. Rev. E {\bf 66},  060402(R)  (2002).

\bibitem{majumdar.a:2007.a}
A. Majumdar, C.~J.~P. Newton, J.~M. Robbins, and M. Zyskin, Phys. Rev. E {\bf
  75},  051703  (2007).

\bibitem{chandrasekhar:1990.a}
S. Chandrasekhar and G.~S. Ranganath, Rep. Prog. Phys. {\bf 53},  57  (1990).

\bibitem{kumar:2004.a}
S. Kumar, Liquid Crystals {\bf 31},  1037  (2004).

\bibitem{bushby:2002.a}
R.~J. Bushby and O.~R. Lozman, Curr. Op. Coll. Interf. Sci. {\bf 7},  343
  (2002).

\bibitem{onsager.l:1949.a}
L. Onsager, Ann.\ N. Y. Acad.\ Sci. {\bf 51},  627  (1949).

\bibitem{Hansen06}
J.~P. Hansen and I.~R. McDonald, {\em Theory of Simple Liquids}, 3rd ed.
  (Academic Press, London, 2006).

\bibitem{gray.cg:1984.a}
C.~G. Gray and K.~E. Gubbins, {\em Theory of molecular fluids. 1. Fundamentals}
  (Clarendon Press, Oxford, 1984).

\bibitem{forsyth.pa:1977.a}
P.~A. {Forsyth Jr.}, S. Mar\u{c}elja, and D.~J. Mitchell, J. Chem. Soc. Faraday
  Trans. II {\bf 73},  84  (1977).

\bibitem{forsyth.pa:1978.a}
P.~A. {Forsyth Jr.}, S. Mar\u{c}elja, and D.~J. Mitchell, Adv. Coll. Interf.
  Sci. {\bf 9},  37  (1978).

\bibitem{eppenga.r:1982.a}
D. Frenkel and R. Eppenga, Phys. Rev. Lett. {\bf 49},  1089  (1982).

\bibitem{veerman.jac:1992.a}
J.~A.~C. Veerman and D. Frenkel, Phys. Rev. A {\bf 45},  5632  (1992).

\bibitem{fartaria.rps:2009.a}
R. Fartaria and M.~B. Sweatman, Chem. Phys. Lett. {\bf 478},  150  (2009).

\bibitem{bates.ma:1999.a}
M.~A. Bates, J.\ Chem.\ Phys. {\bf 110},  6553  (1999).

\bibitem{bates.ma:1999.b}
M.~A. Bates, J. Phys. Chem. B {\bf 111},  4  (1999).

\bibitem{reich.h:2007.a}
H. Reich, M. Dijkstra, R. van Roij, and M. Schmidt, J. Phys.\ Chem.\ B {\bf
  111},  7825  (2007).

\bibitem{cheung.dl:2008.e}
D.~L. Cheung, L. Anton, M.~P. Allen, and A.~J. Masters, Phys. Rev. E {\bf 77},
  011202  (2008).

\bibitem{vanroij.r:1998.a}
R. van Roij, B. Mulder, and M. Dijkstra, Physica A {\bf 261},  374  (1998).

\bibitem{lekkerkerker.hnw:1984.a}
H.~N.~W. Lekkerkerker, P. Coulon, R. van~der Haegen, and R. Deblieck, J.\
  Chem.\ Phys. {\bf 80},  3427  (1984).

\bibitem{varga.s:2003.a}
S. Varga, A. Galindo, and G. Jackson, J.\ Chem.\ Phys. {\bf 80},  3427  (1984).

\bibitem{varga.s:2005.a}
S. Varga, K. Purdy, A. Galindo, S. Fraden, and G. Jackson, Phys. Rev. E {\bf
  72},  051704  .

\bibitem{wensink.hh:2001.a}
H.~H. Wensink, G.~J. Vroege, and H.~N.~W. Lekkerkerker, J. Phys. Chem. B {\bf
  105},  10610  (2001).

\bibitem{parsons.jd:1979.a}
J.~D. Parsons, Phys. Rev. A {\bf 19},  1225  (1979).

\bibitem{lee.sd:1987.a}
S.~D. Lee, J. Chem. Phys. {\bf 87},  4972  (1987).

\bibitem{lee.sd:1989.a}
S.~D. Lee, J. Chem. Phys. {\bf 89},  7036  (1989).

\bibitem{bier.m:2004.a}
M. Bier, L. Harnau, and S. Dietrich, Phys. Rev. E {\bf 69},  021506  (2004).

\bibitem{harnau.l:2002.c}
L. Harnau and S. Dietrich, Phys. Rev. E {\bf 66},  051702  (2002).

\bibitem{harnau.l:2002.d}
L. Harnau, D. Rowan, and J.~P. Hansen, J. Chem. Phys. {\bf 117},  11359
  (2002).

\bibitem{harnau.l:2008.f}
L. Harnau, Mol. Phys. {\bf 106},  1977  (2008).

\bibitem{verhoeff.a.a:2009.a}
A.~A. Verhoeff, H.~H. Wensink, M. Vis, G. Jackson, and H.~N.~W. Lekkerkerker,
  J. Phys. Chem. B {\bf 113},  13476  (2009).

\bibitem{wensink.hh:2004.c}
H.~H. Wensink, Phys. Rev. Lett. {\bf 93},  157801  (2004).

\bibitem{evans.r:1979.a}
R. Evans, Adv. Phys. {\bf 28},  143  (1979).

\bibitem{rosenfeld.y:1989.a}
Y. Rosenfeld, Phys. Rev. Lett. {\bf 63},  980  (1989).

\bibitem{rosenfeld.y:1997.a}
Y. Rosenfeld, M. Schmidt, H. L{\"o}wen, and P. Tarazona, Phys. Rev. E {\bf 55},
   4245  (1997).

\bibitem{rosenfeld.y:1994.a}
Y. Rosenfeld, Phys. Rev. E {\bf 50},  R3318  (1994).

\bibitem{rosenfeld.y:1995.a}
Y. Rosenfeld, Mol. Phys. {\bf 86},  637  (1995).

\bibitem{cinacchi.g:2002.a}
G. Cinacchi and F. Schmid, J. Phys.\ Condens.\ Matter {\bf 14},  12223
  (2002).

\bibitem{hansen-goos.h:2009.a}
H. Hansen-Goos and K.~R. Mecke, Phys. Rev. Lett. {\bf 102},  018302  (2009).

\bibitem{reich.h:2007.b}
H. Reich and M. Schmidt, J. Phys.\ Condens.\ Matter {\bf 19},  326103  (2007).

\bibitem{hendrik_communication}
H. Reich, private communication.

\bibitem{fmt_note}
In Ref.~\cite{esztermann.a:2006.a} where the FMT for pure platelets (as the
  appropriate limit of a ternary mixture of platelets, rods and spheres) was
  developed, the values $c_{I}=0.418$, $c_{N}=0.46$ and $S_{N}=0.492$ were
  obtained using a numerically less accurate method. Hence the value of $S_{N}$
  differs from that of Ref.~\cite{reich.h:2007.b} and the present study.

\bibitem{vanderbeek.d:2006.a}
D. van~der Beek, H. Reich, P. van~der Schoot, M. Dijkstra, T. Schilling, R.
  Vink, M. Schmidt, R. van Roij, and H.~N.~W. Lekkerkerker, Phys. Rev. Lett.
  {\bf 97},  087801  (2006).

\bibitem{esztermann.a:2006.a}
A. Esztermann, H. Reich, and M. Schmidt, Phys. Rev. E  011409  (2006).

\bibitem{vanroij.r:1996.a}
R. van Roij and B. Mulder, Phys. Rev. E {\bf 56},  6430  (1996).

\bibitem{Sollich_communication}
A similar form was independently obtained by P. Sollich, private communication.

\bibitem{barker.ja:1976.a}
J.~A. Barker and D. Henderson, Rev. Mod. Phys. {\bf 48},  587  (1976).

\bibitem{cuesta.ja:2002.a}
J.~A. Cuesta, Y. Martinez-Raton, and P. Tarazona, J.\ Phys.:\ Condens.\ Matter
  {\bf 14},  11965  (2002).

\bibitem{tarazona.p:1997.a}
P. Tarazona and Y. Rosenfeld, Phys. Rev. E {\bf 55},  R4873  (1997).

\bibitem{herzfeld84}
J. Herzfeld, A. Berger, and J. Wingate, Macromolecules {\bf 17},  1718  (1984).

\bibitem{vanroij.r.:2005.d}
R. van Roij, Eur. Phys. J. E {\bf 26},  S57  (2005).

\bibitem{press.wh:2007.a}
W.~H. Press, S.~A. Teukolsky, W.~T. Vetterling, and B.~P. Flannery, {\em
  Numerical Recipes}, 3rd ed. (Cambridge University Press, Cambridge, 2007).

\bibitem{eppenga.r:1984.a}
R. Eppenga and D. Frenkel, Molec.\ Phys. {\bf 52},  1303  (1984).

\bibitem{kayser.rf:1978}
R.~F. Kayser and H.~J. Ravech\'e, Phys. Rev. A {\bf 17},  2067  (1978).

\bibitem{mulder89}
B. Mulder, Phys. Rev. A {\bf 39},  360  (1989).

\bibitem{wensink.hh:2001.b}
H.~H. Wensink, G.~J. Vroege, and H.~N.~W. Lekkerkerker, J. Chem. Phys. {\bf
  115},  7319  (2001).

\bibitem{wensink.hh:2004.a}
H.~H. Wensink and G.~J. Vroege, J. Phys.\ Condens.\ Matter {\bf 16},  S2015
  (2004).

\bibitem{bates.m:1998.a}
M.~A. Bates and D. Frenkel, Phys. Rev. E {\bf 57},  4824  (1998).

\bibitem{wensink.h:2009.a}
H.~H. Wensink and H.~N.~W. Lekkerkerker, Mol. Phys. {\bf 107},  2111  (2009).

\bibitem{varga.s.:2000.a}
S. Varga and I. Szalai, Phys.\ Chem.\ Chem.\ Phys. {\bf 2},  1955  (2000).

\bibitem{dobra.s:2006.a}
S. Dobra, I. Szalai, and S. Varga, J.\ Chem.\ Phys. {\bf 125},  074907  (2006).

\bibitem{vanderkooij.f.m.:2000.b}
F.~M. van~der Kooij and H.~N.~W. Lekkerkerker, Phys. Rev. Lett. {\bf 84},  781
  (2000).

\bibitem{varga.s.:2002.a}
S. Varga, A. Galindo, and G. Jackson, J.\ Chem.\ Phys. {\bf 117},  7207
  (2002).

\bibitem{varga.s.:2002.b}
S. Varga, A. Galindo, and G. Jackson, Phys. Rev. E {\bf 66},  011707  (2002).

\bibitem{varga.s.:2002.c}
S. Varga, A. Galindo, and G. Jackson, Mol.\ Phys.\ Phys. {\bf 101},  817
  (2002).

\bibitem{wensink.h.h.:2002.b}
H.~H. Wensink, G.~J. Vroege, and H.~N.~W. Lekkerkerker, Phys. Rev. E {\bf 66},
  041704  .

\bibitem{speranza.a:2002.a}
A. Speranza and P. Sollich, J.\ Chem.\ Phys. {\bf 117},  5421  (2002).

\bibitem{speranza.a:2003.a}
A. Speranza and P. Sollich, J.\ Chem.\ Phys. {\bf 110},  5213  (2003).

\bibitem{speranza.a:2003.b}
A. Speranza and Sollich, Phys. Rev. E {\bf 67},  061702  (2003).

\bibitem{wensink.h.h:2002.a}
H.~H. Wensink and G.~J. Vroege, Phys. Rev. E {\bf 67},  031716  (2002).

\bibitem{abramowitz.m:1965.a}
M. Abramowitz and I.~A. Stegun, {\em Handbook of Mathematical Functions} (Dover
  Publications Inc., New York, 1965).

\end{thebibliography}
\end{document}